\documentclass[aps,prd,notitlepage,showpacs,nofootinbib,superscriptaddress]{revtex4-1}

\usepackage{amsmath, amsthm, amssymb}
\usepackage{mathtools}
\usepackage{bm}
\usepackage{mathrsfs}
\usepackage{relsize}

\usepackage{epsfig,color,graphicx}
\usepackage{xcolor}
\usepackage{subfigure}

\colorlet{darkred}{red!80!black}
\colorlet{darkgreen}{green!50!black}
\colorlet{darkblue}{blue!50!black}

\usepackage{soul}
\usepackage{cancel}

\usepackage[pdftex,hypertexnames=false,linktocpage=true]{hyperref}
\hypersetup{colorlinks=true,linkcolor=blue,anchorcolor=blue,citecolor=darkgreen,filecolor=blue,urlcolor=blue,bookmarksnumbered=true,
pdfview=FitB
}

\mathchardef\-="2D

\newcommand{\half}[1][1] {\mathsmaller{\frac{#1}{2}}}

\newcommand{\cd}{\makebox[0.08cm]{$\cdot$}}
\newcommand{\bg}[1]{\mbox{\boldmath $#1$}}

\newcommand{\ket}[1] {{\left.|#1\right>}}

\DeclareMathAlphabet{\mathpzc}{OT1}{pzc}{m}{it}

\begin{document}

\title{Nonperturbative solution of scalar Yukawa model in two- and three-body Fock space truncations}

\author{Vladimir~A.~Karmanov}
\affiliation{Lebedev Physical Institute, Leninsky Prospekt 53, 119991 Moscow, Russia}

\author{Yang~Li}
\affiliation{Department of Physics and Astronomy, Iowa State University, Ames, IA 50011, USA}

\author{Alexander~V.~Smirnov}
\affiliation{JSC All-Russian Research Institute for Nuclear Power
Plant Operations, 25 Ferganskaya St., 109507 Moscow,
Russia}

\author{James~P.~Vary}
\affiliation{Department of Physics and Astronomy, Iowa State University, Ames, IA 50011, USA}

\date{\today}

\begin{abstract}
The Light-Front Tamm-Dancoff method of finding the nonperturbative
solutions in field theory is based on the Fock decomposition of
the state vector, complemented with the sector-dependent
nonperturbative renormalization scheme. We show in detail how to
implement the renormalization procedure and to solve the simplest
nontrivial example of the scalar Yukawa model in the two- and
three-body Fock space truncations incorporating scalar ``nucleon''
and one or two scalar ``pions''.
\end{abstract}
\maketitle


\section{Introduction}
\label{sect1}


Light-Front Tamm-Dancoff method is a promising nonperturbative
Hamiltonian approach to quantum field theories~\cite{Perry90}. It
is based on the Fock decomposition of the state vector, which
schematically reads
\begin{equation}
\phi(p)=\psi_1\ket{1}+\psi_2\ket{2}+\psi_3\ket{3}+ \ldots,
\label{Focks}
\end{equation}
where $p$ is the total four-momentum of the physical system
considered, $|n\rangle$ represents a state with the fixed number
$n$ of particles (the $n$-body Fock sector, $n=1,\,2,\,3,\ldots$),
and the coefficients $\psi_n$ are relativistic wave functions (or
Fock components). The interaction between constituents, generally
speaking, does not conserve the number and type of particles, so
that the state vector is a mixture of an infinite number of Fock
sectors. Light-Front Dynamics (LFD) proposed by Dirac~\cite{Dirac}
represents an effective formalism to calculate state vectors in
Fock space. LFD defines the state vector on a null plane, also
known as a light front. In covariant notations, this plane is
given by the equation $\omega\cd x=0$, where $\omega$ is a null
four-vector, $\omega^2=0$ (see,  e.g., Ref.~\cite{Carbonell98} for
a review). It is traditional to choose the light front to be $x^+
\equiv t + z = 0$, corresponding to $\omega=(1, 0, 0, -1)$
\cite{Brodsky98,Coester92}. The
state vector of a physical particle can be obtained by
diagonalizing the light-front Hamilton operator which is the
minus-component of the four-momentum operator:
%
\begin{equation}
\label{eqn:Hamiltonian_eigenvalue_equation}
\hat{P}^-\phi(p)=p^-\phi(p).
\end{equation}
%
The symbol ``hat'' hereafter indicates that the corresponding
quantity is an operator. The standard LFD minus-, plus-, and transverse components of the
four-momentum are, respectively, $p^- \equiv p^0-p^3=
({p}^2_\perp+M^2)/p^+$, $p^+\equiv p^0+p^3$, ${\bf
p}_{\perp}\equiv (p^1,p^2)$, and $M$ is the mass of the physical
system considered.  The eigenvector $\phi(p)$ can be used to
calculate observables, such, e.g., as the electromagnetic form
factors. The light-front Tamm-Dancoff method does not rely on the
expansion in powers of coupling constants and thus is
nonperturbative in nature. Wave functions obtained in this process
provide direct information on the structure of the
system~\cite{Carbonell98}. The light-front Hamiltonian approach
also enjoys some other advantages that makes it particularly
appealing as an alternative method to nonperturbative Lagrangian
approaches such as Lattice gauge theory~\cite{Brodsky98}.

In practical calculations however one can not retain the whole
(infinite) set of the Fock sectors and one has to truncate the Fock
decomposition of the state vector by omitting Fock sectors which
contain more than a finite number $N$ of constituents. We will
refer to such an approximation as the Fock space truncation of
order $N$, {or, equivalently, the $N$-body truncation}. In
truncated Fock space, the Hamiltonian eigenvalue
equation~(\ref{eqn:Hamiltonian_eigenvalue_equation}) reduces to a
finite system of coupled linear integral equations for the wave
functions $\psi_1,\,\psi_2,\ldots ,\,\psi_N$. It is convenient to
represent this equation in a diagrammatic form by using the LFD
graph techniques~\cite{Carbonell98}. Fock space truncation means
that one should neglect all diagrams containing more than $N$
particles in intermediate states.

Quantum field theory suffers from divergences, with no exception
for LFD. As a consequence, they appear in the eigenvalue problem
Eq.~(\ref{eqn:Hamiltonian_eigenvalue_equation}) as well.
Regularization and renormalization have to be carried out wherein
the bare coupling and bare masses, or the corresponding
counterterms, are fixed via the physical coupling and physical
masses. The divergences are then absorbed into the counterterms
which are not observable. In nonperturbative approaches such as
the light-front Tamm-Dancoff method, the renormalization, of
course, is also nonperturbative. A particular challenge faced in
the light-front Tamm-Dancoff method is how to guarantee the exact
cancellation of the divergences. In perturbation theory, the
divergences are canceled order-by-order in the coupling constant
$g$. If some perturbative diagrams of a certain order are absent,
the cancellation of divergences of that order may be destroyed.
Such a situation takes place, when calculating the state
vector in truncated Fock space. Indeed, the light-front
Tamm-Dancoff method sums over an infinite number of diagrams with
no more than $N$ intermediate particles, while all diagrams with
$(N+1)$ and more intermediate particles are omitted. Consider the
perturbative expansion of any calculated observable. Since the
light-front Tamm-Dancoff method is nonperturbative, this expansion
contains contributions of all orders in $g$   but not an exhaustive
set in a given order (say, in the order $n$). The contributions of
the order $g^n$    corresponding to $(N+1)$ and more intermediate
particles are absent because of truncation (do not confuse here
the order $n$ of perturbative expansion with the Fock space
truncation of the order $N$). Starting with some finite order $n$
of perturbative expansion, we would see that divergences are not
canceled, because a part of the divergent contributions related to the
omitted diagrams is missed. The reason is that diagrams which are
of the same order in $g$ may correspond to different Fock sectors.
Since higher Fock sectors are excluded from consideration, we
inevitably omit a part of (divergent) contributions needed to
cancel those coming from the Fock sectors involved. As a
consequence, the cancellation of divergences may not occur when
following the standard renormalization procedure.

Fock sector-dependent renormalization (FSDR) was
proposed~\cite{Perry90} and systematically developed~\cite{Karmanov08} to address this issue. While in
perturbation theory the counterterms are determined order-by-order
in the coupling constant, in the FSDR scheme the counterterms are
determined sector-by-sector in Fock space expansion. That is, we
first find the counterterms in the leading, e.g., two-body, Fock
space truncation. They provide renormalization and cancellation of
infinities in the leading Fock sector. However, they are not
sufficient to cancel infinities in the three-body
(next-to-leading) sector truncation,  as it contains both the two-
and three-body intermediate states. The three-body intermediate
states require new counterterms --- the three-body counterterms,
which are found from the renormalization performed within the
three-body Fock space truncation. The same procedure is continued
in the four-body and higher order truncations.

Strict mathematical proof that this procedure eliminates
infinities is complicated by the nonperturbative nature of the
equations and does not yet exist. However, the validity of FSDR is
strongly supported by numerical calculations. For instance, in
Ref.~\cite{Karmanov12} the FSDR scheme was applied to the coupling
constant and fermion mass renormalization in the Yukawa model up
to the three-body (one fermion plus two scalar bosons) truncation.
Numerical calculations of renormalized observables demonstrated
their good stability with the increase of the regularization
parameters --- the Pauli-Villars (PV) masses. In
Refs.~\cite{Li15a, Li15b}, very good stability of calculated
observables was found in the scalar Yukawa model up to the
four-body truncation (one heavy scalar boson plus three light
scalar bosons). These highly nontrivial numerical calculations
provide good arguments in favor of FSDR as an effective method of
nonperturbative renormalization and show a prospect for a broader
range of its applications.

Recent studies of the scalar Yukawa model~\cite{Li15a, Li15b} also
give one more dimension of support for the light-front
Tamm-Dancoff method equipped with the FSDR {scheme}. Comparison of
the electromagnetic form factors obtained successively within
two-, three-, and four-body truncations shows their rather fast
convergence with respect to the order of truncation. This result
indicates that, at least in the given model, the four-body
truncation almost saturates the state vector and the calculated
value of the electromagnetic form factor is already close to the
exact one.

Originally, the FSDR scheme was formulated on the basis of the
``true'' Yukawa model with a spin-1/2 fermion~\cite{Karmanov08}.
Meanwhile, renormalization of a theory of particles with spin in
LFD encounters many technical difficulties having no direct
relation to FSDR (more complicated spin structure of wave
functions, appearance of additional counterterms depending on the
light front orientation, sensitivity of results to the choice of
regularization, etc.) The complexity of attendant mathematical
derivations conceals, to some extent, the basic ideas of FSDR,
which are rather general and applicable to a variety of realistic
quantum field theories. For this reason, in the present paper we
give a detailed exposition how to apply FSDR {scheme} in practice,
using the scalar Yukawa model in truncated Fock
space. This allows us to illustrate the FSDR method in a simple but
nontrivial example. We will present in detail the solution of the
scalar Yukawa model in the two- and three-body truncations.
Another purpose of the paper concerns the following. According to
the FSDR {scheme}, in recent studies of the scalar Yukawa
model~\cite{Li15a, Li15b} in the four-body truncation, the values
of the bare coupling constant and the heavy boson mass counterterm
from the three-body truncation were used. However, the details of
their derivation were omitted. This paper serves to fill the gap.
The bare coupling constant and the mass counterterm obtained below
can also be used in the future for solving a relativistic bound state
problem up to four-body truncation (two heavy plus two light
scalar bosons).

Note that the renormalized scalar Yukawa model in the three-body
truncation was also studied in Ref.~\cite{Bernard01}, but without
reference to FSDR. Though such an approach led to acceptable
results for the particular model and the particular order of
truncation, it does not seem universal from the point of view of
divergence cancellation, in contrast to FSDR.

The paper is organized as follows. We start in Sec.~\ref{sect2a}
with a brief description of the scalar Yukawa model. In
Sec.~\ref{sect2} a general equation for the state vector in LFD is
formulated. In Sec.~\ref{sect3} we expose the main features of
{FSDR}. Solutions for the state vector in the
scalar Yukawa model are found in the two- and three-body
truncations in Secs.~\ref{sec5} and~\ref{sec6}, respectively. In
Sec.~\ref{ff} we calculate an observable quantity
--- the scalar heavy boson electromagnetic form factor
--- in the two- and three-body truncations, successively.
In Sec.~\ref{g03x} we discuss the properties of the bare coupling
constant determined by the renormalization. Sec.~\ref{concl}
contains concluding remarks.


\section{Scalar Yukawa model}
\label{sect2a}

\begin{figure}
 \centering
 \includegraphics[width=0.15\textwidth]{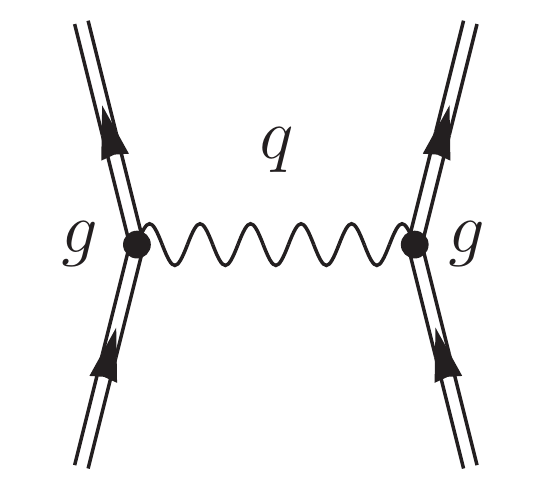}
 \caption{
Amplitude of elastic scattering of two scalar
nucleons (double solid lines) near the scalar pion (wavy line)
pole $q^2=\mu^2$. In the vicinity of this pole the amplitude has
the form 
$\mathcal M_{\chi\chi\to{}\chi\chi} = {-g^2}/{(q^2-\mu^2+i0)}$.}
 \label{fig:scattering}
\end{figure}

We consider {an electrically charged heavy} scalar boson ($\chi$)
with the physical mass $m$, dressed by lighter neutral scalar
bosons ($\varphi$) with the physical mass $\mu$. To mimic somehow
real nucleon-pion physics, we tentatively assign them,
respectively, the nucleon and pion masses\footnote{The masses are in GeV. However in
this model, only the ratio $\mu/m$ matters, and we will suppress all units.}, $m=0.94$, $\mu =
0.14$, and will call them scalar nucleon and scalar pion,
omitting sometimes the word ``scalar'', for shortness. The
corresponding Lagrangian reads
\begin{equation}
\label{Lagr}
 \mathscr L =
 \partial_\nu \chi^\dagger \partial^\nu \chi - m^2 \chi^\dagger \chi + \half \partial_\nu \varphi \partial^\nu \varphi - \half
\mu^2 \varphi^2
+ g_0\chi^\dagger \chi \varphi + \delta m^2 \chi^\dagger \chi,
\end{equation}
where the bare coupling constant $g_0$ and the nucleon mass
counterterm $\delta m^2$ are renormalization constants to be
determined by the renormalization procedure. We denote the
physical coupling constant as $g$ which is
found from typical scattering experiments,
e.g., by the analytic continuation of the measured
two scalar nucleon elastic scattering amplitude, as a function of
the momentum transfer square, to the scalar pion pole in the
nonphysical kinematical region (see Fig.~\ref{fig:scattering}).
For convenience, we introduce a
dimensionless coupling constant
\begin{equation}
\label{alpha} \alpha \equiv \frac{g^2}{16\pi m^2},
\end{equation}
which appears as the coupling constant of the nonrelativistic
Yukawa potential {$U(r)=-\alpha e^{-\mu r}/r$}
between two scalar nucleons. The electromagnetic interaction is
not explicitly included into the Lagrangian~(\ref{Lagr}) because
it is assumed much weaker than the interaction between scalar
nucleons and pions. We will need it only for the calculation of
the nucleon electromagnetic form factor, where it will be taken
into account perturbatively. In contrast to the electromagnetic
fine structure constant $e^2$, the coupling constant $\alpha$ is
not implied to be small and no expansions in it are used.

To regularize the theory, we introduce a PV scalar pion field
$\varphi_\textsc{pv}$ with a large mass $\mu_\textsc{pv} \gg
m,\,\mu$. The PV pion field is enough to regularize rather weak
(logarithmic) divergences which appear in the scalar Yukawa model,
i.e., there is no need to introduce an analogous PV nucleon field.
Since PV fields have negative metric, the Lagrangian becomes
\begin{equation}
 \mathscr L = \partial_\nu \chi^\dagger \partial^\nu \chi - m^2 \chi^\dagger \chi + \delta m^2 \chi^\dagger \chi
+ \half \sum_{j=0}^1 (-1)^j \big[ \partial_\nu \varphi_j
\partial^\nu \varphi_j - \mu^2_j \varphi^2_j \big] + \sum_{j=0}^1
g_0\chi^\dagger \chi \varphi_j , \label{LagrPV}
\end{equation}
where the index $j$ denotes a type of particle: the values $j=0$
and $j=1$ correspond, respectively, to the physical and PV scalar
pion fields, $\mu_0 \equiv \mu$, $\mu_1 \equiv \mu_\textsc{pv}$.
Similar procedure was used in Ref.~\cite{Hiller98}.

Our main goal is to calculate the state vector $\phi(p)$ of the
scalar nucleon. Then it can be used for calculating
observables. The Fock space generated by the Lagrangian~(\ref{Lagr})
embraces all Fock sectors composed of scalar nucleons,
antinucleons, and pions. Each Fock sector contains one nucleon
plus an arbitrary number of nucleon-antinucleon pairs plus arbitrary
number of pions. It is known however that the contribution from
the nucleon-antinucleon loops causes the instability of the
vacuum~\cite{Baym60,gross-tjon}. We therefore truncate away all
Fock sectors with antinucleons and construct a truncated Fock space
from a set of Fock sectors with one scalar nucleon and increasing
number of scalar pions. This procedure, however, comes with a
penalty, as we will discuss below.

The introduction of PV scalar pions into the Lagrangian~(\ref{LagrPV})
extends the Fock space, which impacts the rule of particle counting
inside Fock sectors. We postulate that PV scalar pions come to the
theory on equal grounds with the physical ones. This means that
any pion is counted as one particle, regardless to its type.


\section{State vector in Light-Front dynamics}
\label{sect2}

The explicitly covariant form
of LFD, as a more general approach mentioned in the
Introduction, has many technical advantages in comparison with
its noncovariant forms \cite{Carbonell98}.
In particular, the four-vector $\omega$ serves as an indicator of
possible dependence of calculated results on the light front
orientation. This is especially important in approximate
nonperturbative calculations, where such dependence may appear in
calculated observables due to rotational symmetry breaking. For
particles with spin, covariant LFD facilitates studying the spin
structure of scattering amplitudes. In spite of these merits, for
the case of scalar particles, these different forms of LFD are almost equivalent,
even from the technical point of view. For this reason, we will
not distinguish them below and, retaining in some instances the
four-vector $\omega$ in explicit form, we will assume that it has
definite components $(1,0,0,-1)$. If so, we have $\omega^+=0$,
${\bg \omega}_{\perp}={\bf 0}$, $\omega^-=2$, and $\omega\cd
a=a^+$ for an arbitrary four-vector $a$. 

In LFD the state vector of a physical state is a solution of the
eigenvalue equation~(\ref{eqn:Hamiltonian_eigenvalue_equation})
which can be written in an invariant form:
\begin{equation}
\label{Eqmain}
\hat P^2\phi(p)= M^2\phi(p),
\end{equation}
where $\hat P^2 = \hat P^+ \hat P^- - {\hat{P}}^2_\perp$. The
plus- and transverse components of the momentum operator in LFD
do not contain the interaction; so they can be substituted,
respectively, by the $p^+$ and ${\bf p}_{\perp}$ components of the
total four-momentum $p$. The interaction is only contained in the
{minus-component of the momentum operator which can be represented
as a sum of the free and interacting parts:
${\hat{P}}^-={\hat{P}}^-_0+\hat{P}_{\text{int}}^-$. The
interacting part, in its turn, tightly relates to the} light-front
interaction Hamiltonian $\mathcal H_\text{int}(x)$:
\begin{equation}
 \hat{P}^-_\text{int} = 2\int   \mathcal H_\text{int}(x)\delta(\omega\cdot x)d^4 x =
 2\int_{{-\infty}}^{{+\infty}}
\widetilde{\mathcal H}_\text{int}(\omega \tau)\frac{d \tau}{2\pi},
\end{equation}
where $\widetilde{\mathcal H}_\text{int}$ is a Fourier transform
of the interaction Hamiltonian:
\begin{equation}
\widetilde{\mathcal H}_\text{int}(\omega \tau)=\int \mathcal
H_\text{int}(x)e^{-i(\omega\cd x)\tau}d^4x.
\end{equation}
In covariant form, the four-momentum operator can be written as
\begin{equation}
\label{Pcovar} \hat P^{\nu}=\hat
P_0^{\nu}+\omega^{\nu}\int_{{-\infty}}^{{+\infty}}\widetilde{\mathcal
H}_\text{int}(\omega \tau)\frac{d \tau}{2\pi}.
\end{equation}
Since $\omega^2=0$, we have $\omega\cd \hat P=\omega\cd \hat
P_0=p^+$. In Ref.~\cite{Bernard01} it was
proven that the operators $\omega\cd \hat P_0$
and $\hat P_\text{int}^-$ commute. We thus get
\begin{equation}
\label{Psquared} \hat P^2=\hat
P_0^2+2p^+\int_{{-\infty}}^{{+\infty}}\widetilde{\mathcal
H}_\text{int}(\omega \tau)\frac{d \tau}{2\pi}.
\end{equation}
Substituting this result into Eq.~(\ref{Eqmain}), we finally
obtain~\cite{Bernard01}
\begin{equation}
\label{eqphi} \left[\hat
P_0^2-M^2\right]\phi(p)=-2p^+\int_{{-\infty}}^{{+\infty}}\widetilde{\mathcal
H}_\text{int}(\omega \tau)\frac{d \tau}{2\pi}\,\phi(p).
\end{equation}

The interaction Hamiltonian can be derived from the corresponding
Lagrangian. We need the Hamiltonian in the interaction
representation, i.e., that expressed through the free fields. For
particles with spin or if the interaction depends on field
derivatives the procedure may be, generally speaking, very
nontrivial. The reason is that in LFD some of the equations of
motion for field components are not dynamical equations but
constraints. Exclusion of the non-dynamical degrees of freedom give
rise to specific (contact) terms in the Hamiltonian. This point is
explained in more detail in Ref.~\cite{Karmanov04}. Fortunately,
all that does not concern the case of scalar Yukawa model we
consider here, because each scalar field has only one component.
If so, one can simply identify the Hamiltonian with the
interaction part of the Lagrangian taken with the opposite sign:
\begin{equation}
 \mathcal H_\text{int}(x) = - g_0 \, \chi^\dagger \chi \varphi
-\delta m^2 \chi^\dagger \chi. \label{Hamiltonian}
\end{equation}
To avoid overload with notations, we do not show explicitly the
contribution of PV particles. They can be introduced later
directly in the equations for the Fock components.

To solve Eq.~(\ref{eqphi}), we make use of the Fock decomposition
of the state vector $\phi(p)$, as given schematically by
Eq.~(\ref{Focks}). We define the $n$-body Fock sector as a state
containing one free scalar nucleon with the four-momentum $k_1$
plus $(n-1)$ free scalar pions with the four-momenta $k_2,\ldots,
k_n$. This state is obtained by acting with the corresponding creation
operators on the vacuum:
\begin{equation}
\label{nFock} |n\rangle=\hat{a}^{\dag}({\bf
k}_1)\hat{c}^{\dag}({\bf k}_2)\ldots \hat{c}^{\dag}({\bf
k}_n)|0\rangle.
\end{equation}
The creation operators satisfy the standard commutation relation
$[\hat{a}({\bf k}),\hat{a}^{\dag}({\bf k}')]=\delta^{(3)}({\bf
k}-{\bf k}')$ (for $\hat{c}$ and $\hat{c}^{\dag}$ analogously).
Due to the interaction, the total four-momentum $p$ of the
physical nucleon is not equal to the sum of the constituent
four-momenta: $k_1+\ldots +k_n\neq p$, i.e., momentum
conservation is violated. Within LFD, only plus- and transverse
components of the total four-momentum are conserved:
\begin{equation}
\label{momconsLFD} k_1^++\ldots +k_n^+=p^+,\,\,\,\,\,\,{\bf
k}_{1\perp}+\ldots +{\bf k}_{n\perp}={\bf p}_{\perp}.
\end{equation}
In the following, we will set ${\bf p}_{\perp}={\bf 0}$. This can
be safely done due to the invariance of LFD with respect to
transverse boosts. Using the four-vector $\omega$ introduced
above, the relations~(\ref{momconsLFD}) can be written in an
explicitly covariant form which looks like the momentum
conservation law:
\begin{equation}
\label{momconsLFDcov} k_1+\ldots +k_n=p+\omega\tau_n.
\end{equation}
The scalar parameter $\tau_n$ (the off-shell light-front energy)
can be expressed through the particle momenta by squaring both
sides of Eq.~(\ref{momconsLFDcov}):
\begin{equation}
\label{taun} \tau_n=\frac{s_n-M^2}{2p^+},
\end{equation}
where $s_n$ is the invariant mass squared of the $n$-body Fock
sector:
\begin{equation} \label{sn}
s_n \equiv (k_1 + \ldots + k_n)^2.
\end{equation}
By definition, $s_1=m^2$.
Note that $s_n$ is an eigenvalue of the free four-momentum
operator squared $\hat P^2_0 = p^+\hat P^-_0 - p^2_\perp$:
\begin{equation}
\label{P20eigenvalue} \hat P^2_0 |n\rangle = s_n |n\rangle.
\end{equation}

The Fock decomposition of the physical scalar
nucleon state vector can be written as~\cite{Bernard01}:
\begin{equation}\label{eqn:Fock_representation}
\phi(p)=\sum_{n=1}^\infty \frac{2p^+(2\pi)^{3/2}}{(n-1)!}\int d
\tau_n\left(\prod_{i=1}^n
\frac{d^3k_i}{(2\pi)^{3/2}\sqrt{2\varepsilon_{k_i}}}\right)\,\psi_n(k_1,\ldots
k_n;p)\,\delta^{(4)}(k_1+\ldots+k_n-p-\omega\tau_n)\, |n\rangle,
\end{equation}
 where $\varepsilon_{k_i}=\sqrt{{\bf k}_i^2+m_i^2}$ and $m_i$ is
the mass of the $i$-th constituent. All the four-momenta are on
their mass shells, $k_i^2 = m_i^2$. The
combinatorial factor $1/(n-1)!$ takes into account the identity of
scalar pions. The Dirac's delta-function accounts for the
four-momentum conservation law~(\ref{momconsLFDcov}). Note that
Eq.~(\ref{eqn:Fock_representation}) may be considered as an exact
definition of the light-front wave functions $\psi_n$.

The state vector satisfies the normalization
condition
\begin{equation}
{ \phi^{\dag}(p')\phi(p)=2\varepsilon_{p}\delta^{(3)}({\bf p}-{\bf
p}')}
\end{equation}
which reduces to
\begin{equation}
\sum_{n=1}^\infty I_n=1,
\label{def:Fock_sector_norms}
\end{equation}
where
\begin{equation}
\label{Innorm} I_n=\frac{2p^+}{(2\pi)^{3(n-1)}(n-1)!}\int
d\tau_n\left(\prod_{i=1}^n\frac{d^3k_{i}}{2\varepsilon_{k_i}}\right)
|\psi_n(k_1,\ldots,k_n;p)|^2\delta^{(4)}(k_1+\ldots
+k_n-p-\omega\tau_n)
\end{equation}
is the $n$-body Fock sector contribution to the full norm equal to
unity. By its physical sense, $I_n$ is the probability that the
physical state appears in the $n$-body Fock sector.

\begin{figure}
\centering
\includegraphics[width=0.25\textwidth]{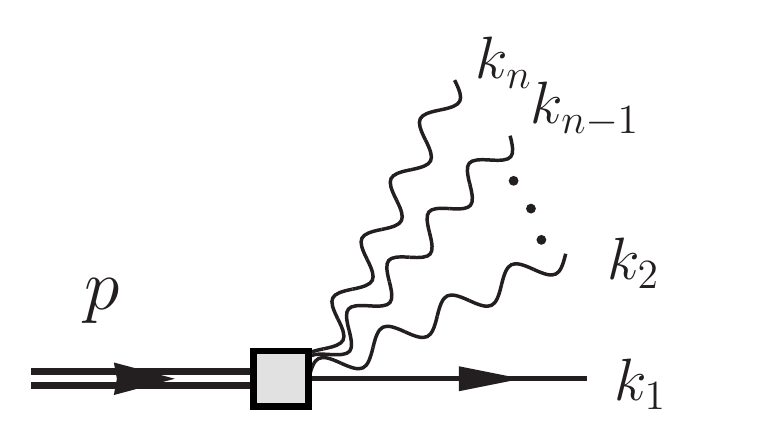}
\caption{Diagram for the $n$-body vertex function $\Gamma_n(k_1,
k_2, \ldots, k_n; p)$. The scalar pions are represented by the
wavy lines. The constituent and physical scalar nucleons are
represented by the single and double straight lines, respectively.
}
\label{fig:vertex_diagram}
\end{figure}

It is useful to introduce the light-front vertex functions
$\Gamma_n$ related to the wave functions by
\begin{equation} \label{def:vertex_function}
 \Gamma_n \equiv (s_n - M^2) \psi_n
\end{equation}
and the new state vector
\begin{equation}
\label{SVG} {\cal G}(p)=2p^+\hat{\tau}\phi(p),
\end{equation}
where the operator $\hat{\tau}$ acting on each Fock component
$\psi_n$ yields $\tau_n\psi_n$. ${\cal G}(p)$ has the same Fock
decomposition~(\ref{eqn:Fock_representation}), changing the wave
functions $\psi_n$ by the corresponding vertex functions
$\Gamma_n$. Using Eqs.~(\ref{P20eigenvalue}) and~(\ref{taun}),
and the definition~(\ref{def:vertex_function}), the main dynamical
equation~{(\ref{eqphi})} for the state vector can
be rewritten as~\cite{Bernard01}
\begin{equation}
\label{EqGamma} {\cal
G}(p)=\frac{1}{2\pi}\int_{{-\infty}}^{{+\infty}}
\left[-\widetilde{\mathcal H}_\text{int}(\omega
\tau)\right]\,\frac{d\tau}{\tau}\,{\cal G}(p).
\end{equation}
The vertex function {$\Gamma_n$} is closely related to the full
transition amplitude~\cite{Hiller98}. This connection allows us to
represent the system of equations for the vertex functions using
the light-front time-ordered diagrams via the so-called covariant
{LFD} graphical rules
\cite{Carbonell98}. An $n$-body vertex diagram is shown in
Fig.~\ref{fig:vertex_diagram}.

For practical applications, it is convenient to transform the
dependence of the wave and vertex functions on the constituent
four-momenta $k_1,\ldots ,k_n$ into their dependence on the light-front variables which are the transverse momenta ${\bf
k}_{i\perp}$ and the longitudinal momentum fractions $x_i\equiv
k_i^+/p^+$ ($i=1,\ldots,n$). The $n$ pairs of the arguments $({\bf
k}_{i\perp},x_i)$ are constrained by the conditions
\begin{equation}
\label{lfvar} \sum_{i=1}^n x_i = 1, \quad \sum_{i=1}^n {\bf
k}_{i\perp}= {\bf 0},
\end{equation}
directly following from {Eqs.~(\ref{momconsLFD})}. We thus have
$(n-1)$ pairs of independent kinematical variables $({\bf
k}_{i\perp},x_i)$ in the $n$-body Fock sector. The invariant mass
squared $s_n$ of the $n$-body Fock sector is expressed through the
light-front variables as
\begin{equation}
\label{snlf}
s_n =\sum_{i=1}^n \frac{k^2_{i\perp}+m^2_i}{x_i}.
\end{equation}
The dependence of the wave and vertex functions on the total
four-momentum $p$ reduces to their dependence on $p^2=M^2$. It is
convenient to {exclude, by means of Eqs.~(\ref{lfvar}), the scalar
nucleon momenta ${\bf k}_{1\perp}$ and $x_1$ and to} choose the
scalar pion momenta as a set of independent variables. We thus
write
\begin{equation}
\Gamma_n = \Gamma_n( {\bf k}_{2\perp}, x_{2}, \ldots, {\bf
k}_{n\perp}, x_{n}; M^2)
\label{Gamman}
\end{equation}
and analogously for $\psi_n$. For simplicity, we will further
suppress the dependence of Fock components on $M^2$ for the
physical particle ($M^2=m^2$), whenever there is no danger of
confusion.


\section{Fock sector dependent renormalization}
\label{sect3}

\begin{figure*}
\includegraphics[width=.225\textwidth]{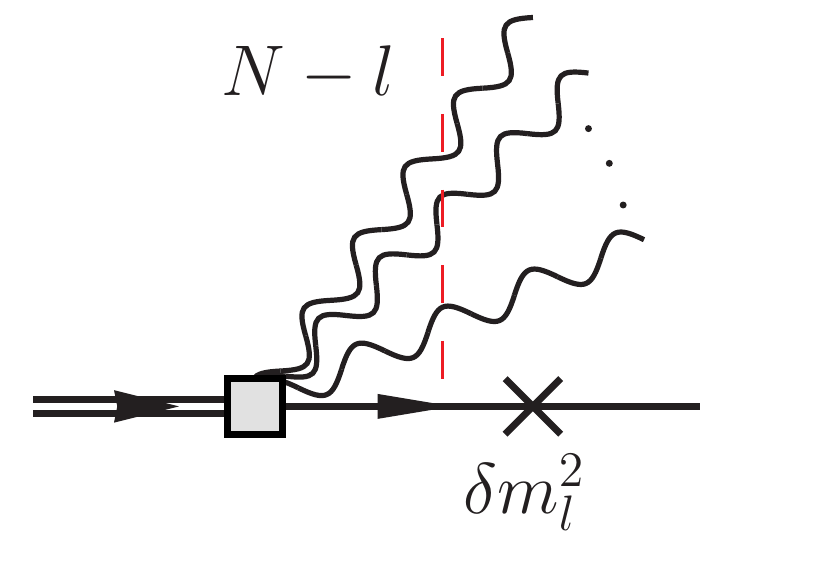}
\includegraphics[width=.45\textwidth]{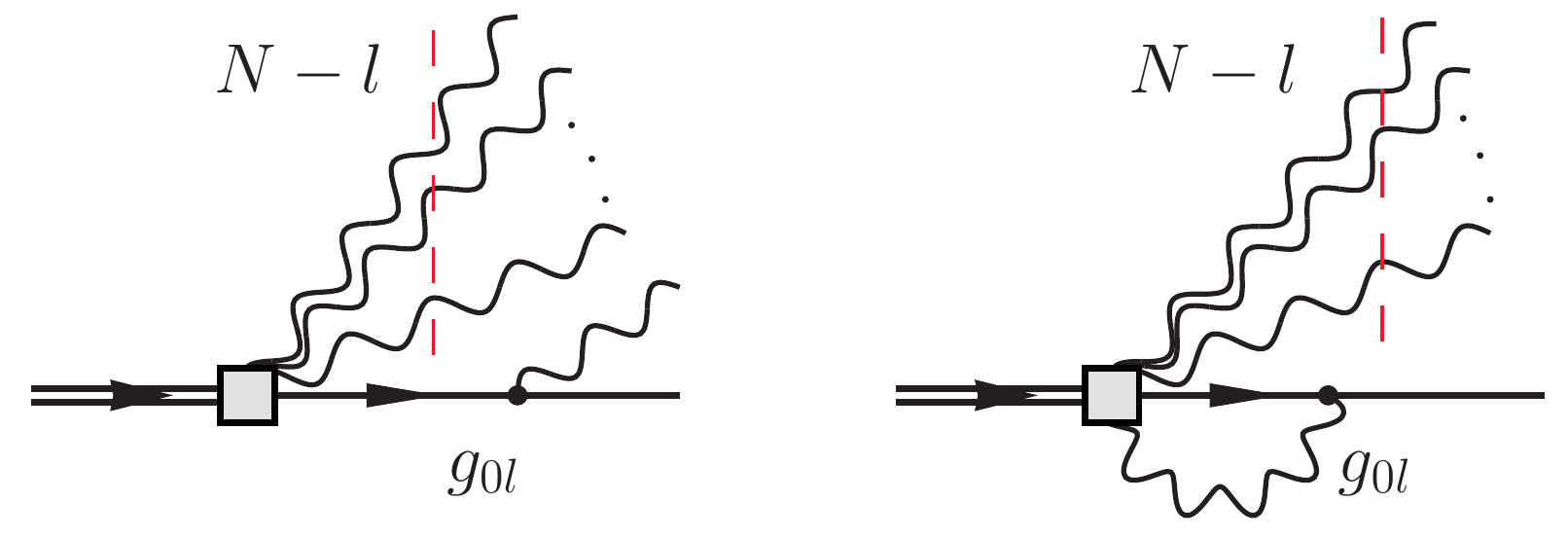}
\caption{Assignment of Fock sector dependent {bare parameters}.
Here $ N$ is the maximal number of particles allowed by the
truncation [one scalar nucleon plus $(N-1)$ scalar pions] and $N-l$
is the number of {pion} spectators which are intersected by the
dashed line.} \label{fig:Fock_sector_dependence}
\end{figure*}

Fock sector dependent renormalization (FSDR) is a systematic
scheme to renormalize light-front Hamiltonian field theory in
truncated Fock space~\cite{Karmanov08}. In this approach, the bare
parameters (i.e., the full set of parameters entering into the
interaction Hamiltonian and used for renormalization, such as bare
coupling constants, bare masses, various counterterms, etc.)
explicitly depend on the Fock sector, where they appear in the
equations for the Fock components. For example, instead of the
unique bare coupling constant $g_0$ one should assign to each
interaction vertex in light-front diagrams the factor $g_{0l}$,
where the index $l$ equals the difference between the order of
Fock space truncation $N$ and the total number of other particles
``in flight'' at the instant which corresponds to the given
vertex. The same concerns the mass counterterm $\delta m^2$.
Actually, one has to deal with a whole series of bare coupling
constants and mass counterterms being different for different Fock
sectors:
\begin{equation}
\label{bareq} g_{0} \; \to \; g_{0l}, \quad \delta m^2 \; \to \;
\delta m^2_{l}, \quad (l=1,2,\ldots, N).
\end{equation}
In the general case, $l=(N-n_s)$, where $n_s$ is the number of
pion spectators. The assignment of the Fock sector dependence is
illustrated in Fig.~\ref{fig:Fock_sector_dependence}.

These sector dependent bare parameters can be determined
successively, by increasing the order of truncation
$N$. The trivial case $N=1$ yields $g_{01} = 0$ and $\delta
m^2_{1} = 0$, since the only particle allowed is the scalar
nucleon with no interactions and mass renormalization. Then,
$g_{02}$ and $\delta m_2^2$ are determined in the two-body
truncation ($N=2$), where the state vector is a superposition of
the single scalar nucleon and one scalar nucleon plus one scalar
pion Fock sectors. $g_{03}$ and $\delta m_3^2$ are determined in
the three-body {($N=3$)} truncation, where the scalar nucleon plus
two scalar pions Fock sector is added. The bare parameters
$g_{02}$ and $\delta m_2^2$ appearing in this approximation as
well are used untouched, as they have been found from the $N=2$
case. The process repeats, until one's desired Fock sector
truncation is reached. Therefore, in order to find the state
vector for the $N$-body truncation, one has to  solve first the
two-, three-, ..., $(N-1)$-body problems. Below, to distinguish
from each other the same quantities calculated in different
approximations, we will supply the former ones by the superscript
``$(N)$'' indicating the order of Fock space truncation. Thus
$\Gamma_n^{(N)}$ means the $n$-body vertex function found within
the $N$-body Fock space truncation, $I_n^{(N)}$ stands for the
$n$-body Fock sector norm obtained in the same approximation, etc.

The bare parameters relate to the physical ones by the
renormalization conditions. The scalar nucleon mass counterterm
is determined from the requirement that the physical and
constituent nucleon masses coincide, i.e., $M=m$. In other words,
one demands that the interaction does not change the nucleon mass.
The bare coupling constant is obtained from the standard condition
that the ``dressed'' two-body on-energy-shell vertex function turns
into the physical coupling constant~(see, e.g.,
Ref.~\cite{Peskin95}):
\begin{equation}
\label{rencg}
\sqrt{Z_{\chi}}\,\widetilde{\Gamma}_2^{\text{on-shell}}\sqrt{Z_{\chi}}\sqrt{Z_{\varphi}}=g,
\end{equation}
where $Z$'s are the so-called field strength renormalization
factors taking into account ``radiative'' corrections to the
two-body vertex external legs and $\widetilde{\Gamma}$ denotes the
two-body vertex amputated from all radiative corrections to its
external legs. The factor $Z_\chi$ tightly relates to the
corresponding scalar nucleon self-energy $\Sigma(p^2)$ by
\begin{equation}\label{def:Z_factor}
 Z_\chi = \left[1{-}\Sigma'(m^2)\right]^{-1},
\end{equation}
where the prime means the derivative
\begin{equation}
\label{deriv} \Sigma'(m^2)\equiv\frac{\partial}{\partial
p^2}\Sigma(p^2)\Big|_{p^2=m^2}.
\end{equation}
The self-energy is given by a sum of amplitudes of all irreducible
diagrams with one-body initial and final states. For the scalar
pion factor $Z_\varphi$ a formula analogous to
Eq.~(\ref{def:Z_factor}) can be written down.

Note that the factorization of the ``dressed'' vertex into a product
of the ``bare'' vertex $\widetilde{\Gamma}_2$ and the external leg
factors $\sqrt{Z}$'s, which appears automatically in the
four-dimensional Feynman approach, is a very nontrivial fact in
the framework of LFD. First, such a factorization in LFD takes
place on the energy shell only, while in the Feynman case it holds
for the off-mass-shell vertex as well. Second, the factorization
may be destroyed by approximations, e.g., the Fock space
truncation. Fortunately, the LFD two-body vertex function enters
into the renormalization condition~(\ref{rencg}) just being taken
on the energy shell, where it coincides with the corresponding
Feynman on-mass-shell two-body vertex. In addition, we do not
consider here antinucleon contributions to the state vector, that
leaves scalar pion a point-like particle, so that
$Z_{\varphi}\equiv 1$. Under these conditions, one can safely
accept Eq.~(\ref{rencg}) as a starting point of the bare coupling
constant renormalization, even in truncated Fock space. Below we
relate $\widetilde{\Gamma}_2$ to the previously introduced
two-body vertex function $\Gamma_2$.

The condition that the nucleon-pion state is on the energy shell
means that the constituent four-momenta satisfy the conservation
law $k_1+k_2=p$ and, hence, $s_2=(k_1+k_2)^2=m^2$. Going over to
the light-front variables, we have $x_1+x_2=1$ and ${\bf
k}_{1\perp}+{\bf k}_{2\perp}={\bf 0}$. So, the two-body vertex
function depends on the two variables which we denote as
\begin{equation}
\label{kpx} k_{\perp}\equiv |{\bf k}_{2\perp}|=|{\bf
k}_{1\perp}|,\quad x\equiv x_2=1-x_1.
\end{equation}
The invariant two-body mass squared in terms of these variables has the form
\begin{equation}
\label{s2}
s_2=\frac{k_{\perp}^2+\mu^2}{x}+\frac{k_{\perp}^2+m^2}{1-x}.
\end{equation}
On the energy shell, where $s_2=m^2$, we get
\begin{equation}
\label{kstar} k_{\perp}=k_{\perp}^*(x)\equiv
i\sqrt{m^2x^2+\mu^2(1-x)}
\end{equation}
(the choice of the sign,
$+i\sqrt{\vphantom{m}\ldots}$ or $-i\sqrt{\vphantom{m}\ldots}$, is
not important, since the two-body vertex function depends in fact
on $k_{\perp}^2$) and
\begin{equation}
\label{G2x}
\widetilde{\Gamma}_2^{\text{on-shell}}=\widetilde{\Gamma}_2(s_2=m^2)=\widetilde{\Gamma}_2(k_{\perp}^*(x),x).
\end{equation}
Since the field strength renormalization factors are constants
(i.e., they do not depend on any kinematical variables), the
renormalization condition~(\ref{rencg}) implies that the two-body
vertex function taken on the energy shell must turn into a
constant too. This is indeed so in perturbation theory. It would
be true in exact nonperturbative calculations, if they were
possible. In approximate nonperturbative approach however such a
property is not automatically guaranteed and the calculated
two-body vertex keeps $x$-dependence even on the energy shell. If
so, one may consider Eq.~(\ref{rencg}) to be true for some particular
value of $x=x^*$ only, choosing $x^*$ at our own
will~\cite{Karmanov10, Mathiot11}. An evident flaw here is the
dependence of calculated observables on the extra nonphysical
parameter $x^*$. Since there are not any strict arguments in favor
of some preset value $x^*$, whether this dependence is
weak or not is a matter of chance. An alternative way proposed in
Ref.~\cite{Karmanov12} seems more justified. It demands
Eq.~(\ref{rencg}) to be true {\em for all} $0\leq x \leq 1$, but
admits $x$-dependence of the bare parameters, uniquely determined
directly from the system of equations for the Fock components. Now
the nonphysical dependence of the on-energy-shell two-body vertex
function on kinematical variables, caused by the Fock space
truncation, shifts to unobserved quantities, while the
renormalization condition~(\ref{rencg}) becomes fully
self-consistent. One may also expect that this method improves the
stability of calculated observables as a function of the
regularization parameters (PV masses), as, e.g., the calculations
of the spin-1/2 fermion anomalous magnetic moment in the Yukawa
model, obtained in Ref.~{\cite{Karmanov12}}, show.

We emphasize: by making a truncation, we approximate the initial
field-theoretical Hamiltonian by a matrix of finite dimension (in
terms of the particle number), acting in Fock space. This is the
reason, why in ``new'' dynamics the on-shell two-body vertex function~(\ref{G2x})
calculated with constant bare parameters acquires dependence on
the variable $x$. Assuming appropriate $x$-dependence of the bare coupling
constant which implicitly enters into
$\widetilde{\Gamma}_2^{\text{on-shell}}$ allows the latter to be
$x$-independent. So, $x$-dependence of bare parameters compensates,
to some extent, the effect of missed (because of the Fock space
truncation) contributions. In Sec.~\ref{g03x} below,
by using an example, we will demonstrate
explicitly that after
taking into account the contribution eliminated by
truncation
the $x$-dependence of the
on-shell two-body vertex function does
completely disappear.

Upon adoption within a truncated Fock space, the general renormalization
condition~(\ref{rencg}) should be reformulated according to the
FSDR requirements. The factor $\sqrt{Z_{\chi}}$ in front of
$\widetilde{\Gamma}_2^{\text{on-shell}}$ comes from the ``dressing''
of a single scalar nucleon line. For the Fock space truncation of
order $N$ it should be thus substituted by
$\sqrt{Z_{\chi}^{(N)}}$. The analogous factor behind
$\widetilde{\Gamma}_2^{\text{on-shell}}$ corresponds to the
``dressing'' of a scalar nucleon line in the two-body (nucleon plus
pion) state. Since the total number of particles in any Fock
sector can not exceed $N$, this factor should be calculated in the
lower $(N-1)$ approximation. Taking into account that
$\sqrt{Z_{\varphi}}\equiv 1$, we obtain
\begin{equation}
\label{rencgN1}
\sqrt{Z_{\chi}^{(N)}}\,\widetilde{\Gamma}^{{(N)}}_2(s_2=m^2)\sqrt{Z_{\chi}^{(N-1)}}=g.
\end{equation}

To make practical use of Eq.~(\ref{rencgN1}) one should relate the
``amputated'' two-body vertex $\widetilde{\Gamma}_2$ to the
previously introduced two-body Fock component $\Gamma_2$. This
relation has the form~\cite{Karmanov10}\footnote{Though in
Ref.~\cite{Karmanov10} the Yukawa model with a spin-1/2 ``nucleon''
was considered, some results obtained there are rather general and
can be applied to the scalar case as well.}
\begin{equation}
\label{G2G2t} \Gamma_2^{{(N)}}(s_2=m^2)=
\sqrt{I_1^{(N)}}\widetilde{\Gamma}_2^{{(N)}}(s_2=m^2)Z_{\chi}^{(N-1)},
\end{equation}
where $I_1^{(N)}$ is the one-body Fock sector normalization
integral in the $N$-body truncated Fock space. Its calculation
according to Eq.~(\ref{Innorm}) yields
$I_1^{(N)}=|\psi_1^{(N)}|^2$. Note that the normalization
condition~(\ref{def:Fock_sector_norms}) for the state vector now
acquires the form
\begin{equation}
\label{normN}
\sum_{n=1}^NI_n^{(N)}=1.
\end{equation}
In terms of the light-front variables, $I_n^{(N)}$ is expressed
through the corresponding vertex function as
\begin{widetext}
\begin{equation}
\label{Inexpl}
I_n^{(N)}=\frac{2}{(2\pi)^{3(n-1)}(n-1)!}\int\prod_{i=1}^n
\frac{d^2k_{i\perp}dx_i}{2x_i}\,\left[\frac{\Gamma_n^{(N)}}{s_n-M^2}\right]^2\,
\delta^{(2)}\left(\sum_{i=1}^n{\bf
k}_{i\perp}\right)\delta\left(\sum_{i=1}^nx_i-1\right).
\end{equation}
\end{widetext}
In Ref.~\cite{Karmanov10} it was proven that the field strength
renormalization factor for a spin-1/2 fermion exactly coincides with
the corresponding one-body normalization integral. The proof can
be easily reduced to the scalar case. Applying
this result to the quantities in truncated Fock
space means
\begin{equation}
\label{ZI}
Z_{\chi}^{(N)}=I_1^{(N)}.
\end{equation}
Combining Eqs.~(\ref{rencgN1}), (\ref{G2G2t}), and~(\ref{ZI})
together gives the final form of the renormalization condition for the bare coupling constant:
\begin{equation}
\label{rencgN2}
\Gamma_2^{(N)}(s_2=m^2)=g\sqrt{I_1^{(N-1)}}.
\end{equation}

On introducing PV particles, one has to supply the vertex
functions with additional indices pointing out the types of scalar
pions in the corresponding Fock sectors. We will denote the pion
type by the superscript $j_l$, $l=2,\,3,\ldots ,\,n$:
$$
\Gamma_n^{(N)}({\bf k}_{2\perp},x_2,\ldots ,{\bf k}_{n\perp},x_n)\to
\Gamma_n^{(N)j_2\ldots j_n}({\bf k}_{2\perp},x_2,\ldots ,{\bf k}_{n\perp},x_n).
$$
According to the notations accepted in Sec.~\ref{sect2a}, $j_l=0$
stands for a physical pion, while $j_l=1$ corresponds to a PV one.
The renormalization condition~(\ref{rencgN2}) is imposed on the
physical component $\Gamma_2^{(N)j_{2}=0}$ of the two-body vertex.


\section{Scalar nucleon state vector in the two-body ($N=2$) truncation}
\label{sec5}



\subsection{Equations for the Fock components and their solution}
\label{sec51}

\begin{figure*}
\centering
\includegraphics[width=.65\textwidth]{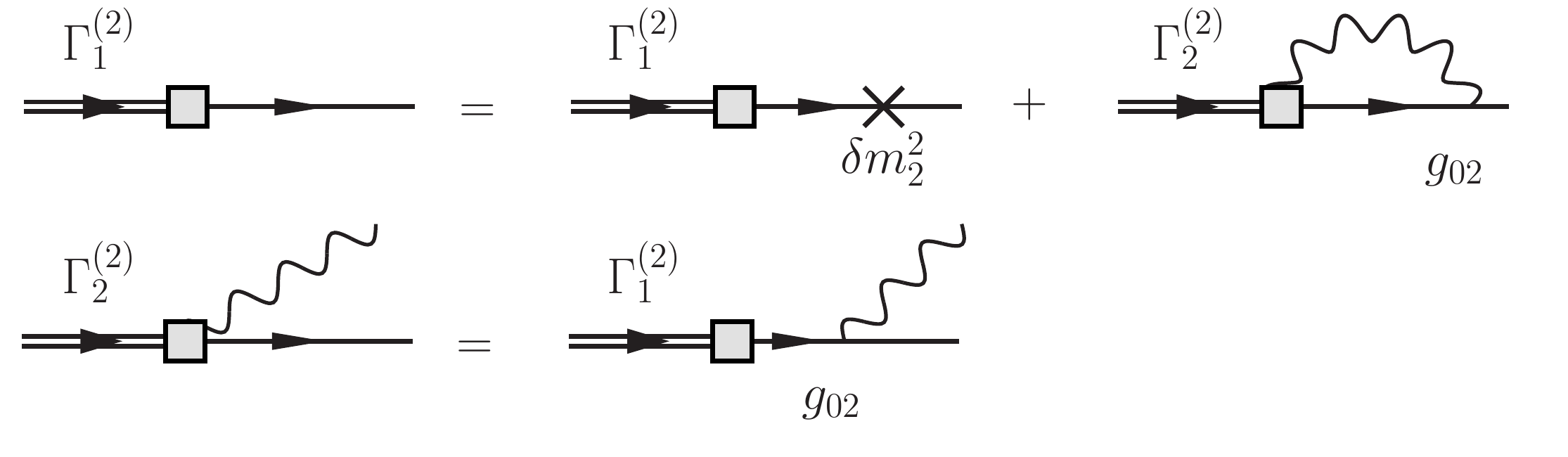}
\caption{System of equations for the vertex functions in the
two-body truncation.}\label{fig1}
\end{figure*}
\begin{figure}
\centering
\includegraphics[width=0.225\textwidth]{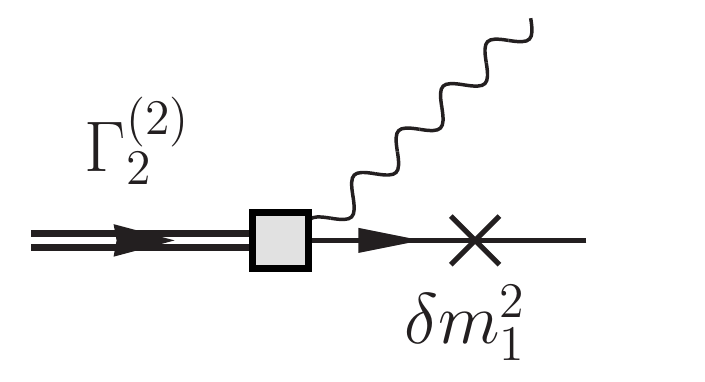}
\caption{Contribution from the mass counterterm, which is absent
in the two-body truncation.} \label{dm12}
\end{figure}

In the two-body truncation, we keep up to two particles (one scalar nucleon plus one scalar pion) in
the Fock space.
The system of equations for the vertex functions,
obtained from the general equation~(\ref{EqGamma}) for the state vector,  is shown
graphically in Fig.~\ref{fig1}.
The rules of the LFD graph techniques are exposed,
in covariant form, e.g., in Ref.~\cite{Carbonell98}. Applying
them to the system of equations considered, one gets
\begin{widetext}
\begin{align}
\Gamma_1^{(2)} \; = \;\; &  \delta m^2_2 \frac{\Gamma^{(2)}_1}{m^2-M^2} + g_{02}\sum_{j=0}^1 (-1)^j \int_0^1\frac{d x}{2x(1-x)}
\int\frac{d^2k_\perp}{(2\pi)^3} \frac{\Gamma_2^{(2)j}(k_\perp, x)}{
s_2^j-M^2},
 \label{eqn:SoE_N2n1} \\
\Gamma^{(2)j}_2(k_\perp, x) \; = \;\; &g_{02}
\frac{\Gamma^{(2)}_1}{m^2-M^2},  \label{eqn:SoE_N2n2}
\end{align}
\end{widetext}
where
\begin{equation}
\label{s2j}
s_2^j=\frac{k^2_\perp+\mu^2_j}{x}+\frac{k^2_\perp+m^2}{1-x}
\end{equation}
is the invariant mass squared of the two-body state made from the
one scalar nucleon and one scalar pion of the $j$-th type [cf.
with Eq.~(\ref{s2})]. The arguments of the two-body vertex
function are defined by Eqs.~(\ref{kpx}). The factor $(-1)^j$
takes into account the negative norm of the PV scalar pion. Note
that $\Gamma^{(2)}_1/(m^2-M^2) = \psi_1^{(2)}$ is a constant in
the sense that it does not depend on kinematical variables. The
bare parameters are assigned to the vertices of the diagrams,
according to the FSDR requirements. This is the reason why
Eq.~(\ref{eqn:SoE_N2n2}) does not contain, on its right-hand side,
a contribution from the scalar nucleon mass counterterm. In
principle, one should add such a contribution (it is shown in
Fig.~\ref{dm12}), because it is generated by the interaction
Hamiltonian~(\ref{Hamiltonian}). At the same time, within the
two-body truncation, one has to assign the factor $\delta m_1^2$
to the corresponding vertex given by the mass counterterm, since
there is already one scalar pion in flight in the two-body state.
Due to the fact that $\delta m_1^2=0$, the diagram in
Fig.~\ref{dm12} does not contribute to Eq.~(\ref{eqn:SoE_N2n2}).

In the limit $M\to m$ the one-body vertex function
$\Gamma_1^{(2)}\sim (m^2-M^2)\to 0$, while $\psi_1^{(2)}$ has a
constant value determined from the normalization
condition~(\ref{normN}) for the state vector. The system of
equations~(\ref{eqn:SoE_N2n1}) and~(\ref{eqn:SoE_N2n2})
thus reduces to
\begin{widetext}
\begin{align}
0 \; = \;\; &  \delta m^2_2 \psi_1^{(2)} \
{-} g_{02}^2\bar{\Sigma}^{(2)}(m^2)\psi_1^{(2)},
 \label{eqn:SoE_N2n1a} \\
\Gamma^{(2)j}_2(k_\perp, x) \; = \;\; &g_{02} \psi_1^{(2)},
\label{eqn:SoE_N2n2a}
\end{align}
\end{widetext}
where $\bar{\Sigma}^{(2)}$ is nothing but the scalar nucleon
self-energy in the two-body approximation, $\Sigma^{(2)}$,
amputated from the coupling constant squared. For an arbitrary
value of its argument $p^2$, this function is given by
\begin{equation}
\label{SE} \bar{\Sigma}^{(2)}(p^2)={-}\sum_{j=0}^1
(-1)^j \int_0^1\frac{d x}{2x(1-x)} \int\frac{d^2k_\perp}{(2\pi)^3}
\frac{1}{s_2^j-p^2}.
\end{equation}
By definition,
$\bar{\Sigma}^{(2)}(p^2)=\Sigma^{(2)}(p^2)/g_{02}^2$. It enters
into Eq.~(\ref{eqn:SoE_N2n1a}) at $p^2=m^2$.
When the PV scalar pion mass $\mu_1$ tends to
infinity, $\bar{\Sigma}^{(2)}(p^2)$ diverges like $\log
(\mu_1/m)$. The function $\bar{\Sigma}^{(2)}(p^2)$ is calculated
in an explicit form in Appendix~\ref{SE2b}.

Equation~(\ref{eqn:SoE_N2n1a}) determines the mass counterterm:
\begin{equation}
\label{dm2g02} \delta m_2^2=
{g_{02}^2\bar{\Sigma}^{(2)}(m^2)},
\end{equation}
while $\psi_1^{(2)}$ still remains a free constant. $\delta m_2^2$
is not immediately needed for the two-body truncation and will be
analyzed later. The two-body vertex function, as follows from
Eq.~(\ref{eqn:SoE_N2n2a}), is a constant too: it depends neither
on kinematical variables nor on the index $j$. This fact is a
direct consequence of the two-body Fock space truncation and,
generally speaking, it does not hold in higher order truncations.
The renormalization condition~(\ref{rencgN2}) at $N=2$ reads
simply
\begin{equation}
\label{rcN2} \Gamma_2^{(2)j=0}(k^*_{\perp}(x),x)=g,
\end{equation}
where we have used $I_1^{(1)} = 1$ (free theory).
Since $\Gamma_2^{(2)j}(k_{\perp},x)\equiv
\Gamma_2^{(2)}$ is a constant, one gets
\begin{equation}
\label{eqn:Gamma2_N2}
 \Gamma_2^{{(2)}} = g.
\end{equation}
The one-body wave function $\psi_1^{(2)}$ is now
defined by the state vector normalization:
\begin{equation}
\label{psi1a} \psi_1^{(2)}=\sqrt{I_1^{(2)}}=\sqrt{1-I_2^{(2)}},
\end{equation}
where
\begin{eqnarray}
\label{I2n}
I_2^{(2)}&=&{\sum_{j=0}^1(-1)^j
\int_0^1\frac{d x}{2x(1-x)}\int \frac{d^2k_\perp}{(2\pi)^3}
\left[\frac{\Gamma_2^{(2)}}{s_2^j-m^2}\right]^2}\nonumber\\
&=& {\frac{g^2}{16\pi^2m^2}} \int_0^1 d x
\bigg[\frac{x(1-x)}{(1-x)\mu^2/m^2+x^2} -
\frac{x(1-x)}{(1-x)\mu^2_{1}/m^2+x^2}\bigg].
\end{eqnarray}
The two-body Fock sector norm $I_2^{(2)}$ (and, hence,
$I_1^{(2)}$) is finite even after removing the ultraviolet regulator, i.e., at
$\mu_1\to\infty$. It is convenient to introduce the two-body norm
$\bar{I}_2^{(2)}$ amputated from the coupling constant squared. By
definition, $\bar{I}_2^{(2)}=I_2^{(2)}/g^2$. It does not depend on
$g$. Note that the following identity is valid:
\begin{equation}
\label{Sigma2I2}
\bar{I}_2^{(2)}={-}{\bar{\Sigma}^{(2)}}{'}(m^2).
\end{equation}
This result can be checked by differentiating the right-hand side of
Eq.~(\ref{SE}) and comparing the result with the right-hand side
of Eq.~(\ref{I2n}). The derivative ${\bar{\Sigma}^{(2)}}{'}(m^2)$ is
calculated analytically in Appendix~\ref{SE2b}. Now we obtain for
the one-body wave function
\begin{equation}
\label{psi12f} \psi_1^{(2)}=\sqrt{1-g^2\bar{I}_2^{(2)}}.
\end{equation}
Eqs.~(\ref{eqn:Gamma2_N2}) and~(\ref{psi12f}) determine the
normalized (and renormalized) Fock components of the scalar
nucleon state vector in the two-body truncation.

The two-body wave function is
\begin{equation}
\begin{split}
 \psi_2^{(2)j}(k_\perp, x)=\frac{\Gamma_2^{(2)}}{s_2^j-m^2}
=
\frac{gx(1-x)}{k_{\perp}^2+\mu_j^2(1-x)+m^2x^2}.
\end{split}
\end{equation}
In contrast to the vertex function, it depends on
both kinematical variables and on the index $j$.


\subsection{Renormalization Parameters}
\label{sect52}


To fix all the renormalization parameters, one
should relate the bare coupling constant $g_{02}$ with the
physical one. Once
the Fock components are available, the
relation desired can be obtained from
Eq.~(\ref{eqn:SoE_N2n2a}):
\begin{equation}
\label{g02g}
g_{02}=\frac{g}{\psi_1^{(2)}} = \frac{g}{\sqrt{
        1-{g^2\bar{I}_2^{(2)}}}}.
\end{equation}
Substituting Eq.~(\ref{g02g}) into
Eq.~(\ref{dm2g02}), we find the mass counterterm:
\begin{equation}
\label{dm2f} \delta
m_2^2=
{\frac{g^2\bar{\Sigma}^{(2)}(m^2)}{1-g^2{\bar{I}_2^{(2)}}}}.
\end{equation}
Then,
the field strength renormalization
factor defined by Eq.~(\ref{def:Z_factor}),
\begin{equation}
\label{eqn:Z_N2}
 Z^{(2)}_\chi={\left[1
{-}g_{02}^2{\bar{\Sigma}^{(2)}}{'}(m^2)\right]^{-1} =
1
{+}g^2{\bar{\Sigma}^{(2)}}{'}(m^2)=1-I_2^{(2)}}=I_1^{(2)},
\end{equation}
as expected. In the limit of infinite PV scalar pion mass $\mu_1$
the quantities $g_{02}$ and $Z_{\chi}^{{(2)}}$ tend
to finite values, while $\delta m_2^2$ diverges logarithmically,
like the self-energy $\bar{\Sigma}^{(2)}$. The bare parameters
$g_{02}$ and $\delta m_2^2$ defined by Eqs.~(\ref{g02g})
and~(\ref{dm2f}), respectively, will be used as an input in the
next order ($N=3$) approximation.


\subsection{Critical coupling associated with the Landau pole}
\label{sect4b}


From Eq.~(\ref{g02g}) it is seen that $g_{02}^2$
considered as a function of $g^2$ becomes singular at
$g^2=1/\bar{I}_2^{(2)}$. A
similar singularity arises in the bare coupling of QED and is
called the Landau pole\footnote{In QED, due to
the Ward Identity, the renormalization of the
charge entirely comes from the vacuum polarization. Our case is
different: we exclude the vacuum
polarization and, because of the
Fock space truncation, the coupling
constant renormalization
is fully caused by the self-energy correction.}.
The critical coupling constant
$\alpha=\alpha_\textsc{l}$ associated with the Landau pole is
determined by
\begin{equation}
\label{alphaL1} \alpha_\textsc{l}^{-1} =
{16\pi m^2\bar{I}_2^{(2)}=\frac{1}{\pi}}
\int_0^1 d x \bigg[\frac{x(1-x)}{(1-x)\mu^2/m^2+x^2} -
\frac{x(1-x)}{(1-x)\mu^2_{1}/m^2+x^2}\bigg],
\end{equation}
where $\alpha$ relates to $g$ by
Eq.~(\ref{alpha}). If $\alpha>\alpha_\textsc{l}$, the bare
coupling constant $g_{02}$ becomes imaginary. In principle, one
can always adjust the PV scalar pion mass $\mu_1$ to make
$\alpha_\textsc{l}$ large enough for the mathematical
self-consistency of the model. From physical considerations
however it is evident that one has to take $\mu_1\gg m$ to claim
that the renormalization procedure allows one to eliminate the
regularization parameters. In the limit $\mu_1\to \infty$
Eq.~(\ref{alphaL1}) reduces to
\begin{equation}
\label{alphaL2} \alpha_\textsc{l}=\pi \left[
\frac{\xi(3-\xi^2)}{\sqrt{4-\xi^2}}\,\arctan\left(\frac{\sqrt{4-\xi^2}}{\xi}\right)-1+(1-\xi^2)\log\frac{1}{\xi}\right]^{-1},
\end{equation}
where $\xi=\mu/m$. For $\mu/m = 0.14 / 0.94$, $\alpha_\textsc{l}
\simeq 2.630$.
Above the critical coupling, the scalar Yukawa
theory becomes ill-defined. At the same time, the threshold of the
coupling constant may not be apparent in calculated observables within
the two-body truncation, which are well-defined for arbitrary
strong coupling. The critical coupling~(\ref{alphaL2}) however
brings real restrictions on admitted values of $\alpha$ in
the three-body truncation, where the renormalized Fock components
do not exist at $\alpha>\alpha_\textsc{l}$. We will discuss these
points in more detail below.


\section{Scalar nucleon state vector in the three-body ($N=3$) truncation}
\label{sec6}



\subsection{Equations for the Fock components and their solution}
\label{sec61}


%
%
%
\begin{figure*}
\centering
\includegraphics[width=0.7\textwidth]{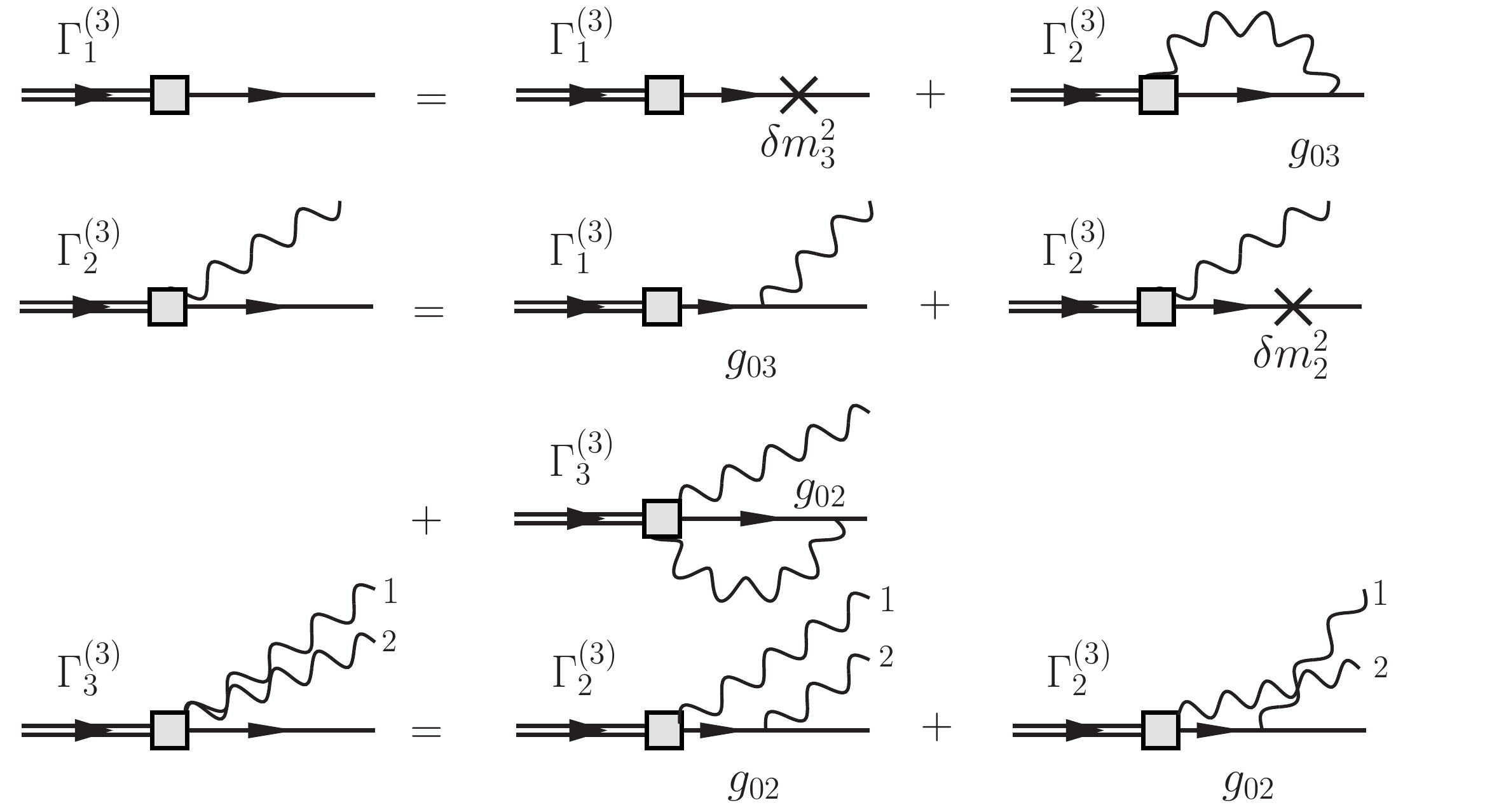}
\caption{System of equations for the Fock components in the
three-body truncation.}
\label{fig4}
\end{figure*}
The system of equations for the vertex functions in the three-body
Fock space truncation ($N=3$) is graphically shown in
Fig.~\ref{fig4}. It differs from that in the $N=2$ case by the
presence of three-body intermediate states which complicate the
equations to some extent. According to the FSDR
rules~\cite{Karmanov08}, the elementary interaction vertices
inside full three-body states, (i.e., the vertices appearing
simultaneously with a scalar pion spectator), contain the bare
coupling constant $g_{02}$ or the mass
counterterm $\delta m^2_2$. The interaction vertices with no
pion spectator above them correspond to the factors $g_{03}$ or
$\delta m_3^2$. The appearance, in different intermediate states,
of the sector dependent bare coupling constants, either $g_{02}$
or $g_{03}$, and the mass counterterms, either $\delta m_2^2$ or
$\delta m_3^2$, is the very essence of the sector dependent
renormalization scheme.

When we solve the problem in the three-body truncation, the values $g_{02}$ and  $\delta m^2_2$ are
assumed to be known --- they were obtained in the
two-body truncation [see Eqs.~(\ref{g02g}) and~(\ref{dm2f})].
The new renormalization parameters $g_{03}$ and $\delta m_3^2$ will be
found by applying the renormalization conditions
again. So, in the framework of FSDR, the refinement of these
quantities from sector to sector is analogous to their refinement,
from order to order, in perturbation theory. As explained in
Sec.~\ref{sect3}, we need the sector dependent renormalization
scheme in order to eliminate divergences for any given truncation.

Applying the rules of the LFD graph techniques, we cast the system
of equations for the vertex functions in the three-body truncation
in an analytical form:
\begin{align}
\Gamma_1^{(3)}  \; = \;\; &   \frac{\delta m^2_3\,\Gamma^{(3)}_1}{m^2-M^2} + g_{03}\sum_{j=0}^1 (-1)^j \int_0^1\frac{d x}{2x(1-x)}
\int\frac{d^2k_\perp}{(2\pi)^3}  \frac{\Gamma_2^{(3)j}(k_\perp, x)}{{{s}_2^j}-M^2},
\label{eqn:SoE_N3n1} \\
\Gamma^{(3)j}_2(k_\perp, x)    \; = \;\;   & \frac{g_{03}\,\Gamma^{(2)}_1}{m^2-M^2}
+ \frac{\delta m_2^2\,\Gamma^{(3)j}_2(k_\perp, x)}{{(1-x)}({s_2^j}-M^2)} \nonumber \\
+&  g_{02}\sum_{\mathclap{j'=0}}^1(-1)^{j'}\int\limits_0^{\mathclap{1-x}} \frac{d x'}{2x'(1-x-x')} \int
\frac{d^2k'_\perp}{(2\pi)^3} \frac{\Gamma_3^{(3)jj'}(k_\perp, x, k'_\perp, x')}{{s_3^{jj'}}-M^2}, \label{eqn:SoE_N3n2} \\
\Gamma_3^{(3){jj'}}({k_{\perp}, x, k'_{\perp}, x'})  \; = \;\; &
\frac{g_{02}\,\Gamma_2^{(3){j}}({k_{\perp}, x})}{(1-{x})({s_2^j} -M^2)}
+ \frac{g_{02}\,\Gamma_2^{(3){j}}({k'_{\perp}, x'})}{(1-{x'})({{s'}_2^{j'}} -M^2)},
\label{eqn:SoE_N3n3}
\end{align}
where $s_2^j$ is defined by Eq.~(\ref{s2j}), ${s'}_2^{j'}$ is given by the same formula, changing $k_{\perp}\to k'_{\perp}$, $x\to x'$, and
$j\to j'$, and
\begin{equation}
\label{s3jjp}
s_{3}^{jj'}=\frac{k^2_\perp+\mu^2_j}{x}+\frac{{ k'}^2_\perp+\mu^2_{j'}}{x'}+\frac{({\bf k}_\perp+{\bf k}'_\perp)^2+m^2}{1-x-x'}
\end{equation}
is the three-body invariant mass squared.

As before, the mass eigenvalue $M$ is implied to be identical to the physical
nucleon mass $m$, i.e., the limit $M\to m$ should
be taken in Eqs.~(\ref{eqn:SoE_N3n1})--(\ref{eqn:SoE_N3n3}). The
three-body vertex function $\Gamma_3^{{(3)}}$ is
expressed through the two-body vertex. Therefore, it can be
excluded by substituting Eq.~(\ref{eqn:SoE_N3n3}) into
Eq.~(\ref{eqn:SoE_N3n2}).
The corresponding analytical expression reads
\begin{multline}
\label{eqn:SoE_N3} \bigg[1 -
\frac{g_{02}^2\bar{\Sigma}^{(2)}(\ell^2)
{-}\delta m_2^2}{\ell^2-m^2}\bigg]
\Gamma^{(3)j}_2({k}_{\perp},x)= g_{03}\psi_1^{(3)} +
g_{02}^2\sum_{j'=0}^1 (-1)^{j'}
\int\limits_0^{\mathclap{1-x}} \frac{d x'}{2x'(1-x')(1-x-x')} \int\frac{d^2k'_{\perp}}{(2\pi)^3} \\
\times
\frac{\Gamma^{(3)j'}_2(k'_{\perp},x')}{({s'}_2^{j'}-m^2)(s_3^{jj'}-m^2)}.
\end{multline}
where $\ell^2 = m^2-(1-x)({s_2^j-m^2})$. The term proportional to
the self-energy $\bar{\Sigma}^{(2)}(\ell^2)$ is generated by the
substitution of the first addendum on the right-hand side of
Eq.~(\ref{eqn:SoE_N3n3}) into the integral term of
Eq.~(\ref{eqn:SoE_N3n2}). Indeed, the result of this substitution
has the form
\begin{equation}
\label{add1}
\frac{g_{02}^2\Gamma_2^{(3)j}(k_{\perp},x)}{(1-x)(s_2^j-m^2)}\,
\sum_{j'=0}^1(-1)^{j'}\int_0^{1-x}
\frac{dx'}{2x'(1-x-x')}\int\frac{d^2k_{\perp}'}{(2\pi)^3}\,\frac{1}{s_3^{jj'}-m^2}.
\end{equation}
Making the sequential change of the integration variables $x'\to
(1-x)x'$ and then ${\bf k}'_{\perp}\to {\bf k}_{\perp}'-x'{\bf
k}_{\perp}$, we can cast the expression~(\ref{add1}) in the form
\begin{equation}
\label{add2}
\frac{g_{02}^2\Gamma_2^{(3)j}(k_{\perp},x)}{(1-x)(s_2^j-m^2)}\,
\sum_{j'=0}^1(-1)^{j'}\int_0^{1}
\frac{dx'}{2x'(1-x')}\int\frac{d^2k_{\perp}'}{(2\pi)^3}\,
\frac{1}{{s'}_2^{j'}-{\ell^2}}=\frac{g_{02}^2\Gamma_2^{(3)j}(k_{\perp},x)
\bar{\Sigma}^{(2)}(\ell^2)}{{\ell^2}-m^2},
\end{equation}
as follows from Eq.~(\ref{SE}) and the definition of the quantity
$\ell^2$. The latter is nothing else than the square of the
off-shell four-momentum of the constituent scalar nucleon in the
two-body state: $\ell^2=(p-k_2)^2$, where $k_2$ is the scalar pion
spectator four-momentum. Note that both the self-energy
$g_{02}^2\bar{\Sigma}^{(2)}(\ell^2)$ and the mass counterterm
$\delta m_2^2$ diverge logarithmically at large mass $\mu_1$ of
the PV scalar pion, but their combination
\begin{equation}
\label{cancel} g_{02}^2\bar{\Sigma}^{(2)}(\ell^2)
{-}\delta
m_2^2=g_{02}^2\left[\bar{\Sigma}^{(2)}(\ell^2)-\bar{\Sigma}^{(2)}(m^2)\right]
\end{equation}
entering into Eq.~(\ref{eqn:SoE_N3}) is finite in this limit,
provided $\ell^2$ is of order of physical masses squared. This
cancellation of divergent terms is just an important feature of FSDR. The
equation~(\ref{eqn:SoE_N3}) determining the two-body vertex
function $\Gamma^{(3)j}_2(k_{\perp},x)$ in the three-body
truncation is a three-body counterpart of
Eq.~(\ref{eqn:SoE_N2n2a}). When $k_{\perp}\to\infty$, it turns
into $\Gamma^{(3)j}_2 \to g_{03}\psi_1^{{(3)}}$, which differs
from $\Gamma_2^{(2)j}$ by the replacement of the index pointing
out the order of truncation.

\begin{figure*}
\centering
\includegraphics[width=0.7\textwidth]{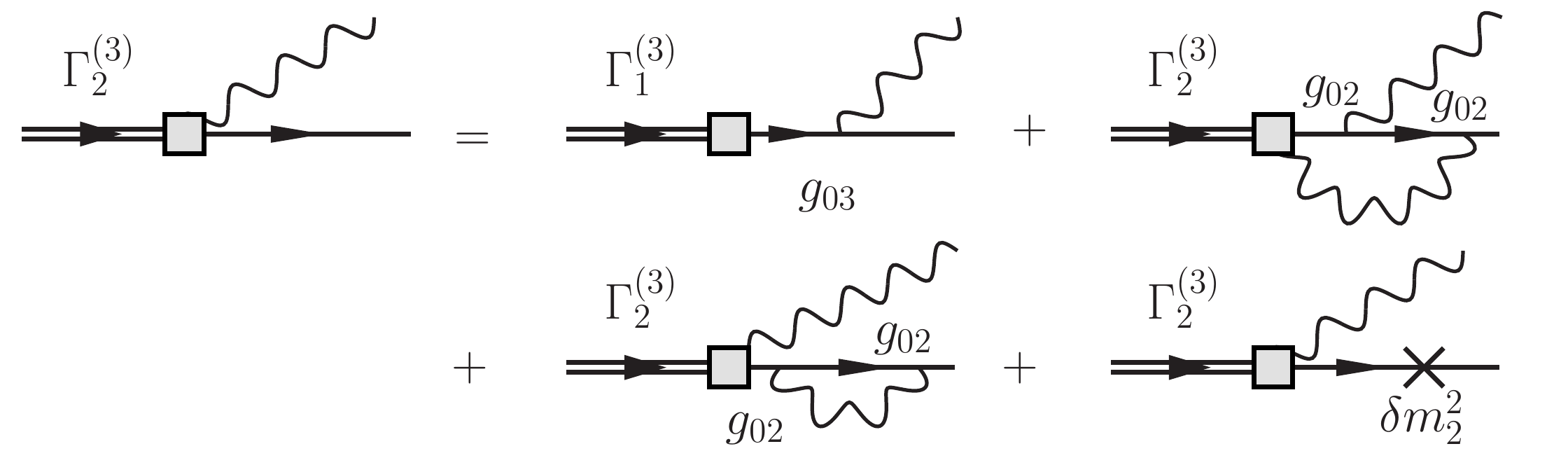}
\caption{Equation for the two-body component {after
the exclusion of the three-body component}.}\label{fig5}
\end{figure*}

The substitution of Eq.~(\ref{eqn:SoE_N3n3}) into
Eq.~(\ref{eqn:SoE_N3n2}), which has been done analytically, could
be realized diagrammatically as well. In such a way, we would
obtain the graphical equation for the two-body vertex function,
shown in Fig.~\ref{fig5}. Using the LFD graph techniques rules
leads to the same analytical equation~(\ref{eqn:SoE_N3}).

Now we make use of Eqs.~(\ref{g02g}), (\ref{Sigma2I2}),
and~(\ref{dm2f}) in order to get rid of the second order bare
parameters $g_{02}$ and $\delta m_2^2$ in Eq.~(\ref{eqn:SoE_N3}).
After simple transformations, we arrive at the following equation
\begin{eqnarray}
\left[1-\frac{g^2\bar{\Sigma}_r^{(2)}(\ell^2)}{\ell^2-m^2}\right]
\Gamma_2^{(3)j}(k_{\perp},x)&=&g_{03}\psi_1^{(3)}\left[1-g^2\bar{I}_2^{(2)}\right]\nonumber\\
&&+\frac{g^2}{8\pi^2}\sum_{j'=0}^1
(-1)^{j'}\int_0^{1-x}dx'\int_0^{\infty}dk_{\perp}'\,k_{\perp}'\,V^{jj'}(k_{\perp},x,k_{\perp}',x')
\Gamma_2^{(3)j'}(k_{\perp}',x'),\nonumber\\
\label{G2f}
\end{eqnarray}
where
\begin{equation}
\label{SEr}
\bar{\Sigma}_r^{(2)}(\ell^2)=\bar{\Sigma}^{(2)}(\ell^2)-\bar{\Sigma}^{(2)}(m^2)-\bar{\Sigma}^{(2)}{'}(m^2)(\ell^2-m^2)
\end{equation}
is the renormalized scalar nucleon self-energy in the two-body
truncation, its argument
\begin{equation}
\label{ell}
\ell^2=-\frac{k_{\perp}^2}{x}+(1-x)m^2-\mu_j^2\left(\frac{1-x}{x}\right),
\end{equation}
and
\begin{eqnarray}
V^{jj'}(k_{\perp},x,k_{\perp}',x')&=&\frac{1}{2\pi x'(1-x')(1-x-x')({s'}_2^{j'}-m^2)}\int_0^{2\pi}\frac{d\phi'}{s_3^{jj'}-m^2}\nonumber\\
&=&\frac{1}{{k'}^2_{\perp}+\mu_{j'}^2(1-x')+m^2{x'}^2}\nonumber\\
&&\times
\left[(1-x-x')^2\left(\frac{k_{\perp}^2+\mu_j^2}{x}+\frac{{k'}^2_{\perp}+\mu_{j'}^2}{x'}
+\frac{k_{\perp}^2+{k'}^2_{\perp}+m^2}{1-x-x'}-m^2\right)^2-4k_{\perp}^2{k'}^2_{\perp}\right]^{-1/2}.
\label{Vker}
\end{eqnarray}
The integration over the azimuthal angle
$\phi'$
has been done analytically by
using the formula
\begin{equation}
 \int\limits_0^{2\pi}  \frac{d\phi'}{A + B \cos\phi'} = \frac{2\pi}{\sqrt{A^2-B^2}},
\qquad (A^2>B^2).
\end{equation}

Eq.~(\ref{G2f}) contains the undefined bare coupling constant
$g_{03}$. To fix it, one should apply the renormalization
condition~(\ref{rencgN2}) which now becomes
\begin{equation}
\label{rcN3}
\Gamma_2^{(3)j=0}(k^*_{\perp}(x),x)=g\sqrt{I_1^{(2)}}=g\sqrt{1-g^2\bar{I}_2^{(2)}}
\end{equation}
with $k^*_{\perp}(x)$ given by Eq.~(\ref{kstar}). We thus set
$k_{\perp}=k^*_{\perp}(x)$ and $j=0$ on both sides of
Eq.~(\ref{G2f}) and demand the condition~(\ref{rcN3}) to hold for
arbitrary $0\leq x\leq 1$. The argument of the self-energy
$\ell^2$ turns into $m^2$ at the renormalization point. Taking
into account that
$$
\bar{\Sigma}_r^{(2)}(\ell^2)\stackrel{\ell^2\to
m^2}{\sim}(\ell^2-m^2)^2,
$$
we get
\begin{eqnarray}
g_{03}\psi_1^{(3)}&=&\left[1-g^2\bar{I}_2^{(2)}\right]^{-1}\nonumber\\
&&\times\left[g\sqrt{1-g^2\bar{I}_2^{(2)}}-\frac{g^2}{8\pi^2}\sum_{j'=0}^1
(-1)^{j'}\int_0^{1-x}dx'\int_0^{\infty}dk_{\perp}'\,k_{\perp}'\,V^{0j'}(k^*_{\perp}(x),x,k_{\perp}',x')
\Gamma_2^{(3)j'}(k_{\perp}',x')\right]. \label{g03psi1}
\end{eqnarray}
An immediate observation is that the right-hand side of
Eq.~(\ref{g03psi1}) depends on the longitudinal momentum fraction
of the scalar pion $x$. Therefore, we allow $g_{03}$ to depend on
$x$ in order to satisfy the condition~Eq.~(\ref{rcN3}) for any
value of $x$ \cite{Karmanov12} (see the detailed discussion below, in
Sec.~\ref{g03x}). Substituting the combination
$g_{03}\psi_1^{(3)}$ back into Eq.~(\ref{G2f}), we find a closed
renormalized equation for the two-body vertex function:
\begin{eqnarray}
\left[1-\frac{g^2\bar{\Sigma}_r^{(2)}(\ell^2)}{\ell^2-m^2}\right]
\Gamma_2^{(3)j}(k_{\perp},x)&=&g\sqrt{1-g^2\bar{I}_2^{(2)}}\nonumber\\
&&+\frac{g^2}{8\pi^2}\sum_{j'=0}^1
(-1)^{j'}\int_0^{1-x}dx'\int_0^{\infty}dk_{\perp}'\,k_{\perp}'\,\Delta
V^{jj'}(k_{\perp},x,k_{\perp}',x')
\Gamma_2^{(3)j'}(k_{\perp}',x'),\nonumber\\
\label{G2ff}
\end{eqnarray}
where
\begin{equation}
\label{DV} \Delta
V^{jj'}(k_{\perp},x,k_{\perp}',x')=V^{jj'}(k_{\perp},x,k_{\perp}',x')-V^{0j'}(k^*_{\perp}(x),x,k_{\perp}',x').
\end{equation}
In fact, Eq.~(\ref{G2ff}) is a system of two inhomogeneous linear
integral equations for
the two components of $\Gamma_2^{(3)j}$ (i.e., those with $j=0$
and $j=1$).
These equations are fully nonperturbative. On solving them, we
obtain a properly normalized two-body vertex function
$\Gamma_2^{(3)j}(k_{\perp},x)$. Eq.~(\ref{eqn:SoE_N3n3}) taken for
$M=m$ uniquely determines the three-body vertex function
$\Gamma_3^{(3)jj'}(k_\perp,x,k'_\perp,x')$ in terms of the
two-body vertex function. The one-body wave function $\psi_1^{(3)}$ is then
found from the normalization condition for the whole state vector:
\begin{equation}
\label{psi13}
\psi_1^{(3)}=\sqrt{I_1^{(3)}}=\sqrt{1-I_2^{(3)}-I_3^{(3)}},
\end{equation}
where the two- and three-body Fock sector norms are calculated
according to Eq.~(\ref{Inexpl}) with $N=3$, taking into account PV
particle contributions:
\begin{eqnarray}
\label{I23}
 I_2^{(3)}&=& \int\limits_0^1 \frac{d x}{2x(1-x)}
\int \frac{d^2k_\perp}{(2\pi)^3} \sum_{j=0}^1(-1)^j \left[\frac{\Gamma^{(3)j}_2(k_{\perp},x)}{s_2^j-m^2}\right]^2, \\
\label{I33}
 I_3^{(3)}&=& \frac{1}{2}\int\limits_0^1 \frac{d x}{2x} \int \frac{d^2k_{\perp}}{(2\pi)^3} \int\limits_0^{1-x}\frac
{d x'}{2x'(1-x-x')}  \int \frac{d^2k'_{\perp}}{(2\pi)^3}
\sum_{j,j'=0}^1(-1)^{j+j'}
\left[\frac{\Gamma^{(3)jj'}_3(k_{\perp},x,
k'_{\perp},x')}{s_3^{jj'}-m^2}\right]^2.
\end{eqnarray}

We emphasize that all Fock components of the
scalar nucleon state vector in the three-body truncation can be
calculated without any reference to Eq.~(\ref{eqn:SoE_N3n1}) which
determines the mass counterterm $\delta m_3^2$. Together with the
bare coupling constant $g_{03}$, it will be needed in higher order
($N\geq 4$) truncations only. This feature reflects a general
property of FSDR: the highest order bare parameters $g_{0N}$ and
$\delta m_N^2$ found in the $N$-body truncation are actually
needed, starting from the $(N+1)$-body truncation.

If we restrict our consideration of the scalar
Yukawa model to calculations of observables inside the three-body
approximation, we may completely get rid of PV particles, assuming
the limit $\mu_1\to \infty$. Once logarithmic divergences coming
from the self-energy and the mass counterterm are mutually
canceled in their
combination~(\ref{cancel}), one can take the limit
$\mu_1\to\infty$ directly in Eq.~(\ref{G2ff}) by omitting all
contributions with either $j=1$ or $j'=1$. The reason is that the
kernel $V^{00}$, Eq.~(\ref{Vker}) at $j=j'=0$,
does not produce new divergences requiring regularization by PV
particles. This does not mean that we would automatically get
$\Gamma_2^{(3)j=1}=0$ in the limit $\mu_1\to\infty$. The PV
components of the vertex functions may tend to a finite nonzero
limit, but they do not affect the physical
components or the
calculated observables, or the Fock sector
norms~(\ref{psi13})--(\ref{I33}). This statement relates to both the
two- and three-body vertices and reasonably simplifies subsequent
numerical calculations. Note that in the spinor Yukawa model,
where divergences are stronger, such a procedure does not work and
one has to retain PV particle contributions till the end of
calculations~\cite{Karmanov12,Karmanov10}.

The inhomogeneous linear integral equation~(\ref{G2ff}) was solved
numerically for various values of the physical coupling constant
$\alpha$ defined by Eq.~(\ref{alpha}) and the physical particle
masses $m=0.94$ and $\mu=0.14$. To find the solution we
employ an iterative method. We first approximate the integrals by
using Gauss-Legendre quadratures. We start with an educated
guess for $\Gamma_2^{{(3)j}}$ and substitute it onto the
right-hand side of Eq.~(\ref{G2ff}). We solve for
$\Gamma_2^{(3)j}$ on the left-hand side on the quadrature grid,
interpolating as needed. The obtained $\Gamma_2^{(3)j}$ then
serves as the input for the next round of iterations. We update
$\Gamma_2^{(3)j}$ until the point-by-point total deviation is
sufficiently small.

\begin{figure}
\centering
\includegraphics[width=0.48\textwidth]{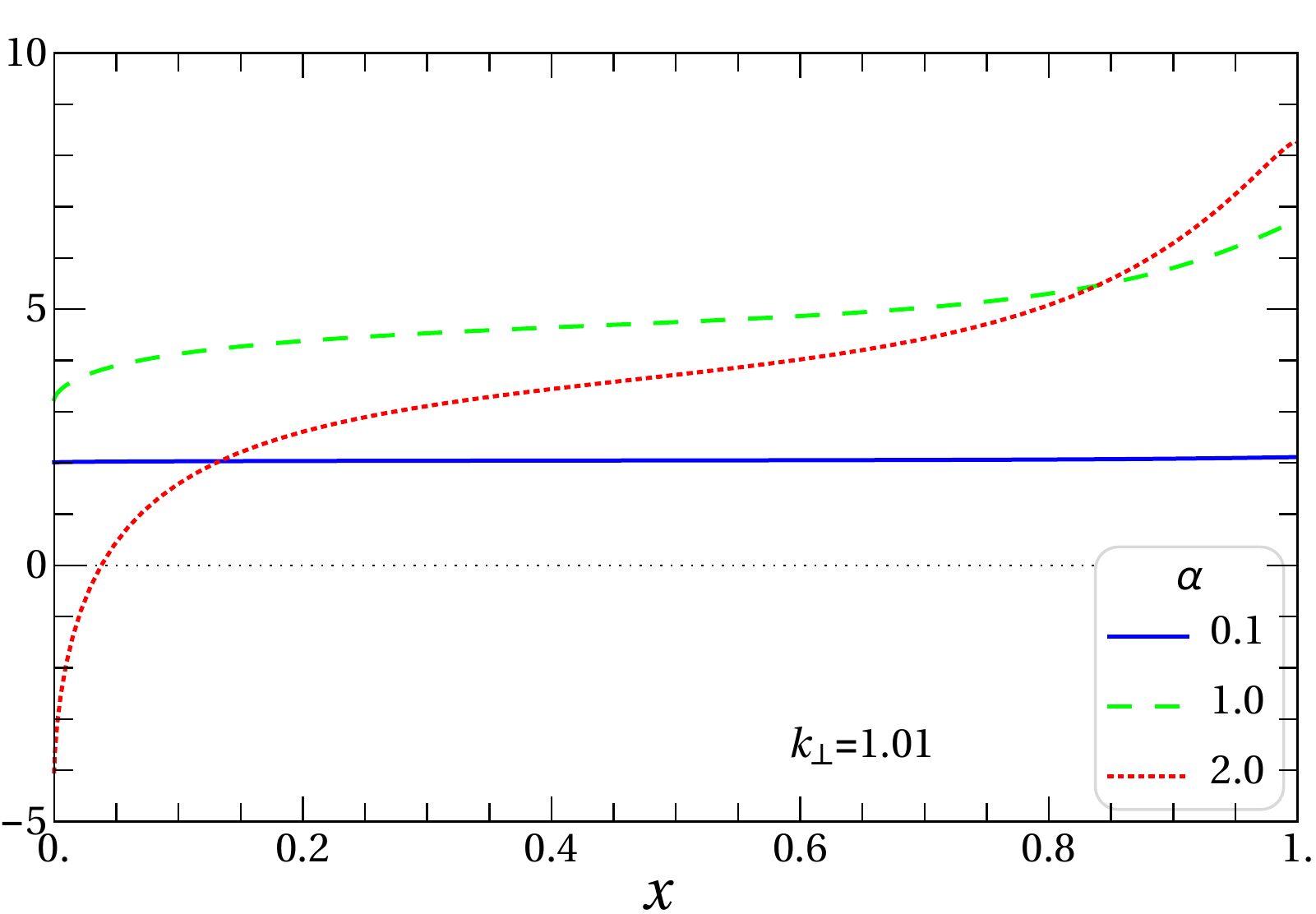}
\raisebox{-0.01\height}{\includegraphics[width=0.475\textwidth]{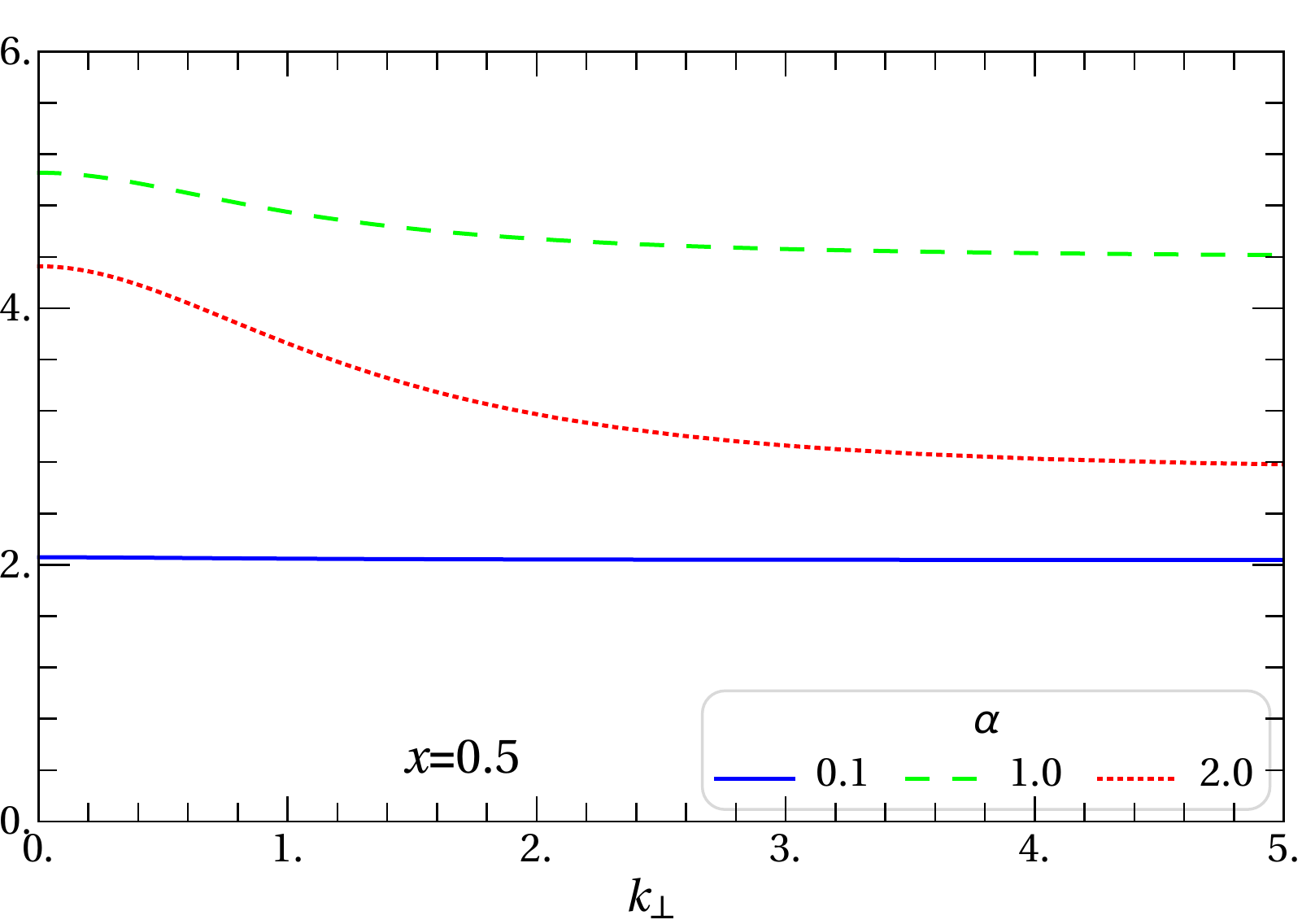}}
\caption{The vertex function $\Gamma_2^{(3){j=0}}(k_\perp, x)$ {as
a function of $x$ at fixed $k_{\perp}$} (left panel), and as a function of $k_{\perp}$ at fixed $x$ (right panel), calculated in the
three-body truncation for several values of the physical coupling
constant $\alpha$. The PV mass has been
removed by taking the limit $\mu_1\to \infty$.  }
 \label{fig:gamma2N3byx}
\end{figure}

Representative solutions for
$\Gamma_2^{(3){j=0}}(k_{\perp},x)$ are shown in
Fig.~\ref{fig:gamma2N3byx}. We removed the PV mass
by taking the limit
$\mu_1\to\infty$.~\footnote{The limiting solution
for $\Gamma_2^{(3)j=0}$ is sufficient for calculations within
the three-body Fock space truncation, but the solution with a
finite PV mass is useful in the
four-body truncation, where the PV mass cannot be easily removed.}
The curves in Fig.~\ref{fig:gamma2N3byx} reflect
typical behavior of $\Gamma_2^{(3){j=0}}(k_{\perp},x)$ as a function of
its arguments.

Our calculations distinctly indicate that the physical coupling
constant $\alpha$ cannot be taken arbitrarily large. If we fix $x$
and consider $\Gamma_2^{(3){j=0}}$ as a function of
$k_{\perp}$, its limiting ($k_{\perp}\to\infty$) value rapidly
increases in magnitude with the increase of $\alpha$. 
The same happens in the limit $x\to 0$ at fixed $k_{\perp}$. At certain
$\alpha=\alpha_c$ it seems that
$\Gamma_2^{(3){j=0}}$ becomes unbounded. Further
increase of $\alpha$ leads to the absence of stable numerical
solutions of Eq.~(\ref{G2ff}). Numerical estimations give
$\alpha_c \simeq 2.630$. In the next section we will
explain the reason why the critical coupling appears in the given
physical model and calculate $\alpha_c$ exactly.

\begin{figure}
 \centering
 \includegraphics[width=.5\columnwidth]{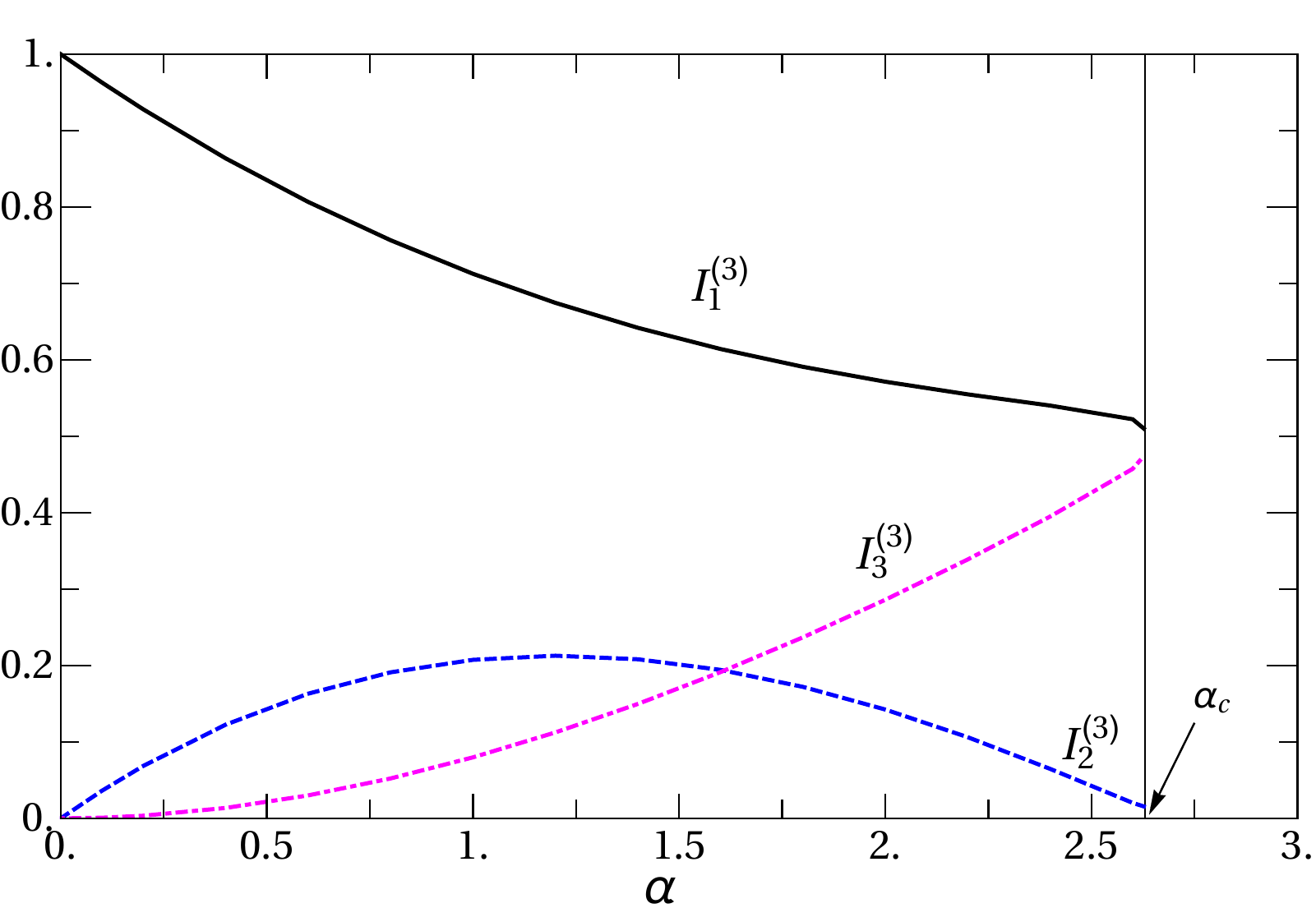}
 \caption{Fock sector norms
 {in the three-body truncation} as a function of the physical coupling constant
 $\alpha$.}
 \label{Inum}
\end{figure}

To estimate relative contributions of different
Fock sectors to the full state vector norm, we calculated the
corresponding sector norms as a function of the coupling constant
which varies from zero up to the critical value. The results are
presented in Fig.~\ref{Inum}. One observes that the one-body sector
always dominates, though its contribution monotonically
decreases with the increase of the coupling constant. The behavior
of the two-body sector contribution looks nontrivial: it increases
to a maximum and then decreases as a function of the coupling
constant. The
three-body sector contribution increases monotonically, but it does
not reach the value of the one-body sector contribution.


\subsection{Renormalization Parameters}
\label{RP3}


Having found $\Gamma_{2}^{(3)j}(k_{\perp},x)$ and $\psi_1^{(3)}$, we
can calculate the bare coupling constant ${g_{03}=}g_{03}(x)$ from
Eq.~(\ref{g03psi1}):
\begin{eqnarray}
g_{03}(x)&=&{\frac{1}{\sqrt{I_1^{(3)}}}\left[\frac{g}{\sqrt{1-g^2\bar{I}_2^{(2)}}}
\right.}\nonumber\\
&&{-\frac{g^2}{8\pi^2\left[1-g^2\bar{I}_2^{(2)}\right]}\sum_{j'=0}^1(-1)^{j'}\int_0^{1-x}dx'\,x'\int_0^{\infty}
\frac{k'_{\perp}dk'_{\perp}}{{k'}^2_{\perp}+\mu_{j'}^2(1-x')+m^2x^{'2}}}\nonumber\\
&&{\left.\times\frac{\Gamma_2^{(3)j'}(k'_\perp,x')}{\sqrt{[{k'}^2_{\perp}(1-x)+m^2{x'}^2(1+x)-\mu^2{x'}^2+
\mu_{j'}^2(1-x-x')]^2+4{k'}^2_{\perp}[m^2x^2+\mu^2(1-x)]{x'}^{2}}}\right].}
\label{g03f}
\end{eqnarray}
Note that the integrand in Eq.~(\ref{g03f}) is not
singular even without PV regularization and the term with $j'=1$
in the sum vanishes in the limit $\mu_1\to\infty$. Therefore, $g_{03}(x)$
does not contain divergences. As outlined above, it does explicitly depend on $x$.

\begin{figure}
\centering
\includegraphics[width=.5\columnwidth]{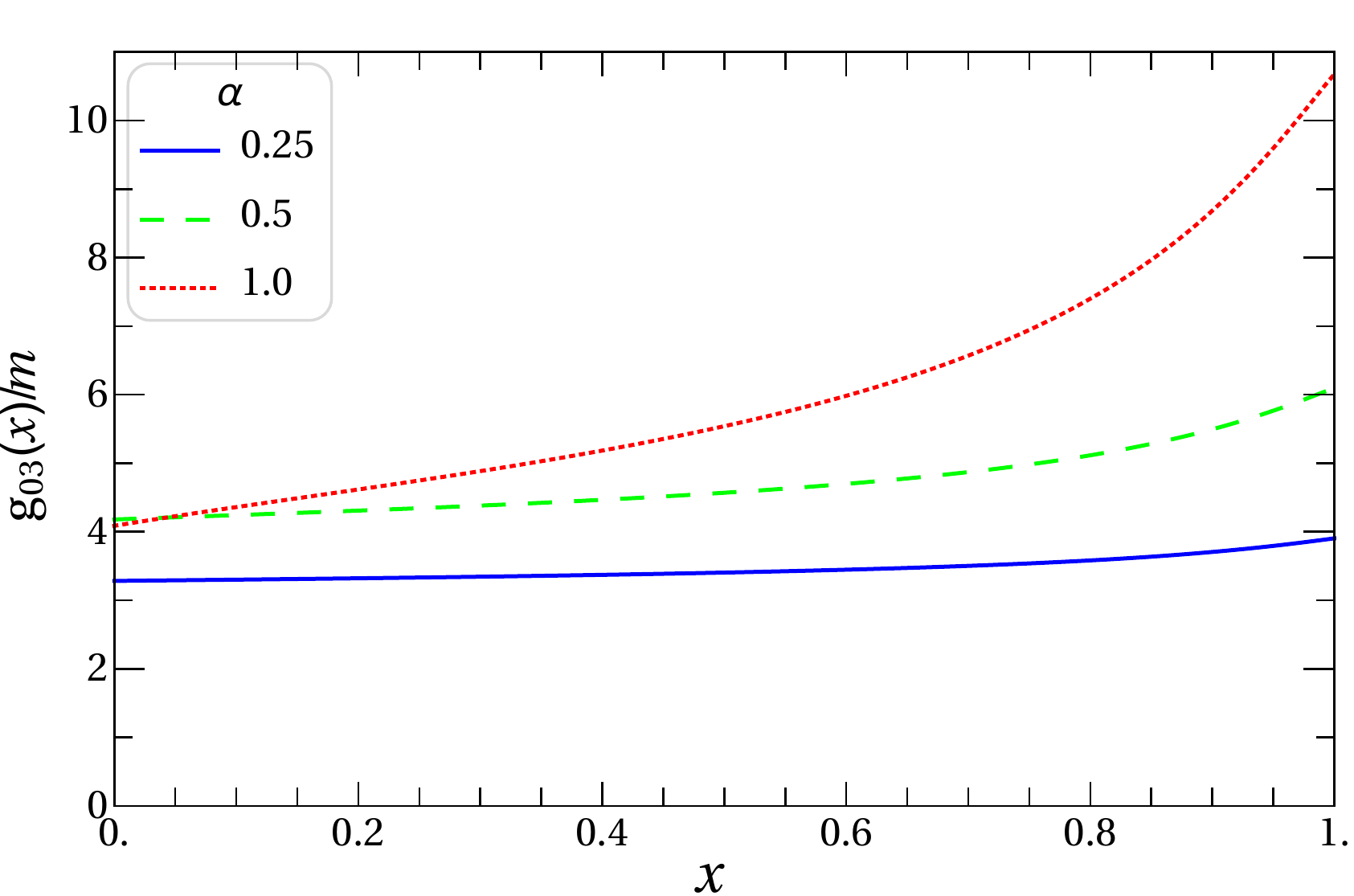}
\caption{
Ratio $g_{03}(x)/m$
as a function of $x$ for a few values of
$\alpha$.} \label{fig_g03}
\end{figure}

In Fig.~\ref{fig_g03} we show the dependence of
$g_{03}$, in units $m$, on the kinematical variable $x$ for
several values of the physical coupling constant $\alpha$.  If
rotational symmetry was not broken by the Fock space truncation,
$g_{03}$ would be a true constant independent of $x$. As is seen
from Fig.~\ref{fig_g03}, this is not the case: $g_{03}$ depends on
$x$; the larger the value of $\alpha$ the stronger is the $x$-dependence. Such a
property is a price we pay to have the renormalization
condition~(\ref{rcN3}) satisfied for arbitrary $x$. The question
of $x$-dependence of $g_{03}$ is discussed below in a special
Sec.~\ref{g03x}.

Similarly, the three-body mass counterterm $\delta m_3^2$ can be
found from Eq.~(\ref{eqn:SoE_N3n1}) in the limit
$M\to m$, taking into account the $x$-dependence of $g_{03}$:
\begin{equation}
\label{dm3f}
 \delta m_3^2 ={-}
{\frac{1}{8\pi^2\sqrt{I_1^{(3)}}}\sum_{j=0}^1(-1)^j\int_0^1dx\,g_{03}(x)\int_0^{\infty}dk_{\perp}\,k_{\perp}\,\frac{\Gamma_2^{(3)j}(k_{\perp
},x)}
{k^2_{\perp}+\mu_j^2(1-x)+m^2x^2}}.
\end{equation}
In contrast to $g_{03}(x)$, the mass counterterm
$\delta m_3^2$ is a true constant independent of kinematical
variables. If $\mu_1\to\infty$, $\delta m_3^2$ diverges like
$\log(\mu_1/m)$, i.e., one cannot avoid PV particle contributions,
when calculating it.

A question may arise, why one should insert
$g_{03}(x)$ into the integrand in Eq.~(\ref{dm3f}), rather than to
leave it as a free factor [like it appears originally in
Eq.~(\ref{eqn:SoE_N3n1})], making $\delta m_3^2$ to be
$x$-dependent as well. An answer can not be found in the framework
of the three-body Fock space truncation, and the above recipe
appears as an ansatz. The rule is however justified in the four-body
truncation~\cite{Li15a,Li15b}, where $g_{03}(x)$ and $\delta
m_3^2$ are necessary to calculate the Fock components. It can be
easily seen that $g_{03}(x)$ enters into amplitudes of light-front
diagrams constructed according to the FSDR requirements, being
integrated over $dx$.

It is
instructive to consider not only
the mass counterterm $\delta m_3^2$, but also the three-body
self-energy
[cf. Eq.~(\ref{SE})]:
\begin{equation}
\label{SE3} \Sigma^{(3)}(p^2) = -\frac{1}{ \sqrt{I_1^{(3)}}}
\sum_{j=0}^1 (-1)^j \int_0^1 \frac{d x\,g_{03}(x)}{2x(1-x)} \int
\frac{d^2 k_\perp}{(2\pi)^3} \frac{\Gamma_2^{(3)j}(k_\perp, x;
p^2)} {s_2^j-p^2}
\end{equation}
with $\delta m^2_3 ={\Sigma^{(3)}(m^2)}$.
$\Gamma_2^{(3)j}(k_\perp, x; p^2)$ is the fully off-energy-shell
two-body vertex function, i.e., that introduced in
Eq.~(\ref{Gamman}) with $M^2=p^2\neq m^2$. It satisfies the same
integral equation~(\ref{eqn:SoE_N3n2}), changing $M^2$ to $p^2$,
with the renormalization condition
$\Gamma_2^{(3)j{=0}}(k_\perp^*(x), x;
m^2)=g\sqrt{I_1^{(3)}}$. After simple transformations, fully
analogous to those made above, one can derive the following
renormalized equation for it:
\begin{eqnarray}
\left[1-\frac{g^2\bar{\Sigma}_r^{(2)}(\ell_p^2)}{\ell_p^2-m^2}\right]
\Gamma_2^{(3)j}(k_{\perp},x;p^2)&=&{g\sqrt{1-g^2\bar{I}_2^{(2)}}}\nonumber\\
&&+\frac{g^2}{8\pi^2}\sum_{j'=0}^1
(-1)^{j'}\int_0^{1-x}dx'\int_0^{\infty}dk_{\perp}'\,k_{\perp}'\nonumber\\
&&\times \left[V^{jj'}(k_{\perp},x,k_{\perp}',x';p^2)
\Gamma_2^{(3)j'}(k_{\perp}',x';p^2)\right.\nonumber\\
&&\left.-V^{{0}j'}(k_{\perp}^*(x),x,k_{\perp}',x';m^2)
\Gamma_2^{(3)j'}(k_{\perp}',x';m^2)\right],
\label{G2ffoff}
\end{eqnarray}
where $\ell_p^2=-\frac{k_{\perp}^2}{x}+p^2(1-x)-\mu_j^2\left(\frac{1-x}{x}\right)$ and
\begin{eqnarray}
V^{jj'}(k_{\perp},x,k_{\perp}',x';p^2)&=&\frac{1}{{k'}_{\perp}^{2}+\mu_{j'}^2(1-x')+m^2x'-p^2x'(1-x')}\nonumber\\
&&\times
\left[(1-x-x')^2\left(\frac{k_{\perp}^2+\mu_j^2}{x}+\frac{{k'}_{\perp}^{2}+\mu_{j'}^2}{x'}
+\frac{k_{\perp}^2+{k'}_{\perp}^{2}+m^2}{1-x-x'}-p^2\right)^2-4k_{\perp}^2{k'}_{\perp}^{2}\right]^{-1/2}.\nonumber\\
\label{Vkeroff}
\end{eqnarray}
The derivative ${\Sigma^{(3)}}'(m^2)$ is related
to the field strength renormalization factor
\begin{equation}
\label{eqn:Z3}
Z^{(3)}_\chi = \bigg[ 1 
{-}{\Sigma^{(3)}}'(m^2)\bigg]^{-1}.
\end{equation}

In spite of both $\Sigma^{(3)}(p^2)$ and
$\Gamma_2^{(3)j}(k_{\perp},x;p^2)$ having a three-body ``origin'', they
are actually not needed within the three-body truncation, like
$\delta m_3^2$ and $g_{03}$. So, without going beyond the $N=3$
case, one may ignore the properties of these off-shell quantities.
The latter quantities however naturally appear, when finding the Fock
components in the four-body truncation, where they affect the
calculated results in full measure. In particular, the fully
off-energy-shell two-body vertex function
$\Gamma_2^{(3)j}(k_{\perp},x;p^2)$
is a source of the critical value
of the coupling constant
for $N=4$. This point is discussed in more detail
in the next section.
\begin{figure}
\centering
\includegraphics[width=.5\columnwidth]{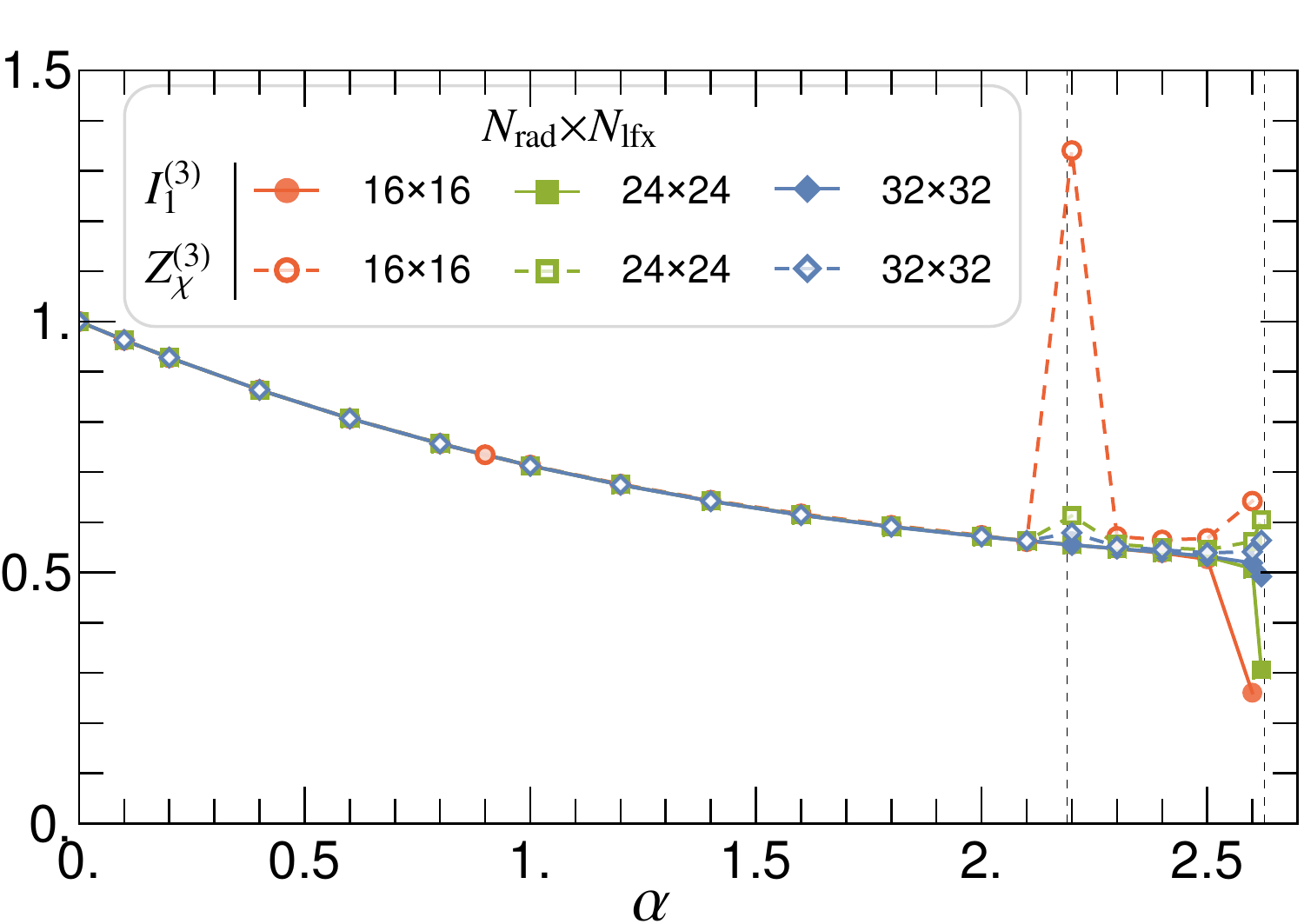}
\caption{Comparison of {the one-body norm}
$I_1^{(3)}$ and {the field strength renormalization
factor} $Z_\chi^{(3)}$.
$N_{\mbox{\scriptsize rad}}$ and $N_{\mbox{\scriptsize lfx}}$ are, respectively, the numbers of Gaussian
integration nodes in the variables $k_{\perp}$ and $x$.
$I_1^{(3)}$ is obtained from the wave
function normalization, Eq.~(\ref{psi13}). $Z_\chi^{(3)}$ is
obtained from the self-energy~{(\ref{SE3}) by means
of} Eq.~(\ref{eqn:Z3}). These two quantities agree within the
numerical precision. The dashed lines mark the positions of the
critical coupling constants $\alpha_c^{{\mbox{\scriptsize nr}}}\simeq
2.190$ (see Sec.~{\ref{critN3}
below}) and $\alpha_{{c}} \simeq 2.630$.
} \label{I1Z3}
\end{figure}

The comparison of the calculated value of the field strength
renormalization factor $Z^{(3)}_\chi$ with the one-body
normalization integral $I_1^{(3)}$ serves as an additional test of
our numerical computations. As is seen from Fig.~\ref{I1Z3}, these
two quantities do coincide with each other within the numerical precision.


\subsection{Critical coupling}
\label{critN3}


The parameters entering into the linear integral
equation~(\ref{G2ff}) --- the coupling constant $g$ and the
particle masses --- should be chosen to allow a physically proper
solution for the two-body vertex function. We will not perform
here the full analysis, but study the behavior of
$\Gamma_2^{(3)j}$ as a function of $g$ for fixed values of the
particle masses $m$ and $\mu$. Some of our conclusions can be proven
analytically, while we will rely on the results
of numerical computations for the remainder.

For simplicity, we consider the case of an infinite
PV mass $\mu_1$. As discussed above, this limit is
reached by omitting the term with $j'=1$ in the sum in
Eq.~(\ref{G2ff}). We thus obtain a single linear integral equation
for $\Gamma_2^{(3)j=0}(k_{\perp},x)$ which we will denote here
simply $\Gamma_2$, for brevity. Then we represent
Eq.~(\ref{G2ff}) in the following operator form:
\begin{equation}
\label{hHmatr} \Gamma_2=f+\hat{\sf A}\Gamma_2,
\end{equation}
where $f=g\sqrt{1-g^2\bar{I}_2^{(2)}}$ is the inhomogeneous part,
and the operator $\hat{\sf A}$ is represented as a sum of the two contributions
\begin{equation}
\label{Asum}
\hat{\sf A}=\hat{\sf A}'+\hat{\sf K},
\end{equation}
where
\begin{eqnarray}
\label{AG1} \hat{\sf A}'\Gamma_2&=& {\cal
F}(\ell^2)\Gamma_2^{(3)j=0}(k_{\perp},x),\\
\label{AG2}
\hat{\sf K}\Gamma_2&=&\frac{g^2}{8\pi^2}\int_0^{1-x}dx'\int_0^{\infty}dk_{\perp}'\,k_{\perp}'\,\Delta
V^{00}(k_{\perp},x,k_{\perp}',x') \Gamma_2^{(3)j=0}(k_{\perp}',x'),
\end{eqnarray}
collect, respectively, all nonintegral and integral terms coming
from the interaction in the three-body states.
The function
\begin{equation}
\label{calF} {\cal
F}(\ell^2){\equiv}\frac{g^2\bar{\Sigma}_r^{(2)}(\ell^2)}{\ell^2-m^2},
\end{equation}
where $\ell^2$ is given by Eq.~(\ref{ell}) with $j=0$, is
generated by the scalar nucleon self-energy. The formal solution
of Eq.~(\ref{hHmatr}), which can be written as
$\Gamma_2=(1-\hat{\sf A})^{-1}f$, is regular, if the operator
$(1-\hat{\sf A})$ is nonsingular. To find out conditions when this
is satisfied, we consider a more general eigenvalue problem for
the operator $\hat{\sf A}$:
\begin{equation}
\label{hHmatrhomo} \lambda\Gamma_2=\hat{\sf A}\Gamma_2.
\end{equation}
Varying the physical coupling constant, we can trace the behavior of the
eigenvalues $\lambda$. As soon as we encounter at least one
eigenvalue $\lambda=1$, the solution of
Eq.~(\ref{hHmatr}) becomes singular.

For numerical analysis, we represent the operator
$\hat{\sf A}$ in a matrix form. It can be achieved by discretizing
the integrals in Eq.~(\ref{AG2}) by means of the Gaussian procedure. The same is done for the operator
$\hat{\sf A}'$ which is reduced to a diagonal matrix. We thus approximate the operator
$\hat{\sf A}$ by a finite $n_A\times n_A$ matrix with the
dimension $n_A=n_kn_x$, where $n_k$ and $n_x$ are the numbers of
the integration nodes in the variables $k'_{\perp}$ and $x'$,
respectively.
After this transformation, we calculate all the $n_A$ eigenvalues
$\lambda$. Gradually increasing $n_A$, we analyze the spectrum
each time till the eigenvalues which are interesting for us become
stable.

It is more convenient to work with the dimensionless coupling
constant $\alpha$ related to $g^2$ by Eq.~(\ref{alpha}).
Evidently, at $\alpha=0$ we have a trivial result $\hat{\sf A}=0$
and the only eigenvalue is $\lambda=0$. Once $\alpha$ starts
increasing, the eigenvalues are concentrated in a region of a
finite size. We are interested in the maximal real eigenvalue
$\lambda_{\mbox{\scriptsize max}}$. Varying $\alpha$, we get a
function $\lambda_{\mbox{\scriptsize max}}(\alpha)$. The minimal
positive root of the equation $\lambda_{\mbox{\scriptsize
max}}(\alpha)=1$ just gives the critical coupling constant
$\alpha_c$.

\begin{figure}
\centering
\includegraphics[width=.7\columnwidth]{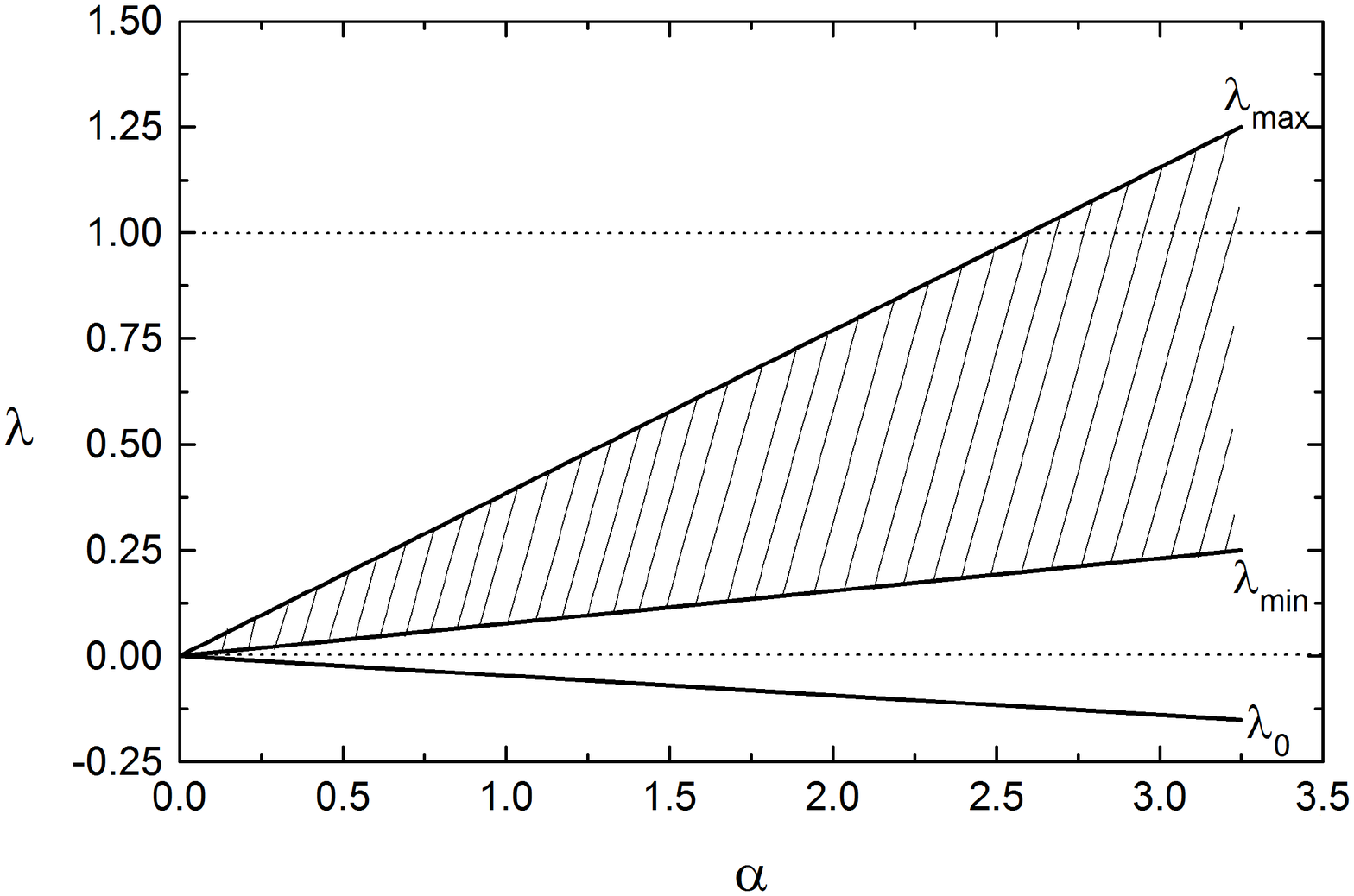}
\caption{Eigenvalue spectrum of Eq.~(\ref{hHmatrhomo}) as a function of
the physical coupling constant $\alpha$. The cross-hatched region represents the continuous part
of the spectrum, while $\lambda_0$ is a discrete eigenvalue.}
\label{spectrum}
\end{figure}

The calculated spectrum includes one discrete eigenvalue
$\lambda_0(\alpha)$ and a set of $(n_A-1)$ eigenvalues distributed, almost uniformly, in the interval
\begin{equation}
\label{lambdasp} \lambda_{\mbox{\scriptsize
min}}(\alpha)<\lambda<\lambda_{\mbox{\scriptsize max}}(\alpha).
\end{equation}
$\lambda_0(\alpha)$ is always negative and therefore has no
relation to the critical coupling. Note that all the three
functions $\lambda_0(\alpha)$, $\lambda_{\mbox{\scriptsize
min}}(\alpha)$, and $\lambda_{\mbox{\scriptsize max}}(\alpha)$ are
very stable as $n_A$ increases, while the density
of $\lambda$'s between $\lambda_{\mbox{\scriptsize min}}(\alpha)$
and $\lambda_{\mbox{\scriptsize max}}(\alpha)$ grows. This provides
a hint that the exact spectrum consists of two parts: a discrete
one including the only eigenvalue $\lambda_0(\alpha)$ plus a
continuous one given by the interval $[\lambda_{\mbox{\scriptsize
min}}(\alpha),\,\lambda_{\mbox{\scriptsize max}}(\alpha)]$. The
results of the numerical calculation of the spectrum for
$m=0.94$ and $\mu=0.14$ are shown in Fig.~\ref{spectrum}.
Note that the functions $\lambda_0(\alpha)$,
$\lambda_{\mbox{\scriptsize min}}(\alpha)$, and
$\lambda_{\mbox{\scriptsize max}}(\alpha)$ are linear, {because
the operator $\hat{\sf A}$ is proportional to $\alpha$.}
The relative computational precision
is about $10^{-{5}}$, corresponding to $n_A\sim
10^4$. When $\alpha$ increases, $\lambda_{\mbox{\scriptsize
max}}(\alpha)$ reaches unity at $\alpha=\alpha_c
\simeq 2.630$.

Our calculation revealed an interesting fact: the continuous
part~(\ref{lambdasp}) of the spectrum is insensitive
to the integral part~(\ref{AG2}) of the operator $\hat{\sf A}$. In
other words, if we calculate the eigenvalue spectrum of $\hat{\sf
A}'$ by means of the matrix equation
\begin{equation}
\label{hHmatrhomop} \lambda'\Gamma_2'=\hat{\sf A}'\Gamma_2',
\end{equation}
then we see that all $n_A$ eigenvalues are confined into
an interval
 with the same boundaries
$\lambda_{\mbox{\scriptsize min}}(\alpha)$ and
$\lambda_{\mbox{\scriptsize max}}(\alpha)$, in spite of the fact that the
eigenvectors $\Gamma_2$ and $\Gamma_2'$ are different. The
coincidence of the continuous parts of the spectra $\lambda$ and
$\lambda'$ is not caused by chance but originates from some common
property of Eqs.~(\ref{hHmatrhomo}) and~(\ref{hHmatrhomop}), which
will be clear now. The merit of Eq.~(\ref{hHmatrhomop}) consists
in that it can be trivially solved analytically. Indeed, because
of the diagonal form of the matrix $\hat{\sf A}'$ the eigenvalues
$\lambda'$ are simply the values of the function ${\cal
F}(\ell^2)$ at the node points. If $n_A\to\infty$, the spectrum
becomes continuous:
\begin{equation}
\label{lsp} \lambda_{\mbox{\scriptsize
min}}(\alpha)<\lambda'<\lambda_{\mbox{\scriptsize max}}(\alpha),
\end{equation}
where
\begin{eqnarray}
\lambda_{\mbox{\scriptsize
min}}(\alpha)&=&\mbox{min}_{\ell^2}{\cal F}(\ell^2),
\label{lmin}\\
\lambda_{\mbox{\scriptsize
max}}(\alpha)&=&\mbox{max}_{\ell^2}{\cal F}(\ell^2). \label{lmax}
\end{eqnarray}
It is easy to check that for $0\leq k_{\perp}<\infty$ and $0\leq
x\leq 1$ we have $-\infty<\ell^2\leq (m-\mu)^2$.

The condition~(\ref{lsp}) has very simple meaning.
It means that there always exists such a point $(k_{\perp},x)$ where the solution of the inhomogeneous equation
\begin{equation}
\label{hHmatrp} \lambda'\Gamma_2'=f+\hat{\sf A}'\Gamma_2',
\end{equation}
which is
\begin{equation}
\label{Gamma2spp} \Gamma_2'=\frac{f}{\lambda'-{\cal F}(\ell^2)},
\end{equation}
becomes singular. The corresponding equation with the whole operator $\hat{\sf A}$,
Eq.~(\ref{Asum}),
\begin{equation}
\label{hHmatrp1} \lambda\Gamma_2=f+\hat{\sf A}\Gamma_2,
\end{equation}
which is a generalization of our initial equation~(\ref{hHmatr}),
can not be solved in a similar trivial way, but its formal
``solution'' can be written
\begin{equation}
\label{Gamma2sp} \Gamma_2=\frac{f+\hat{\sf K}\Gamma_2}{\lambda-{\cal F}(\ell^2)}.
\end{equation}
Both expressions on the right-hand sides of Eqs.~(\ref{Gamma2spp})
and~(\ref{Gamma2sp}) have denominators of the same type. Assume we
take some value $\lambda$ inside the interval~(\ref{lambdasp})
with the boundaries defined by Eqs.~(\ref{lmin}) and~(\ref{lmax}).
Then the equation ${\cal F}(\ell^2)=\lambda$ determines a point
$\ell^2$ [or a set of points $(k_{\perp},x)$] where the
denominator in Eq.~(\ref{Gamma2sp}) vanishes. The solution
is singular, unless
\begin{equation}
\label{nosing}
\hat{\sf K}\Gamma_2=-f
\end{equation}
at the same point. As our analysis shows, this condition is not
satisfied. Hence, the stability of the solution of
Eq.~(\ref{G2ff}) relates to the function ${\cal F}(\ell^2)$ only.
The critical coupling constant is derived from the equation
\begin{equation}
\label{c1}
\mbox{max}_{\ell^2}{\cal F}(\ell^2)=1,
\end{equation}
where $-\infty<\ell^2\leq (m-\mu)^2$.
To find the maximum, we make use of the explicit form of ${\cal F}(\ell^2)$:
$$
{\cal
F}(\ell^2)={-g^2{\bar{\Sigma}^{(2)}}{'}(m^2)+g^2\,\frac{\bar{\Sigma}^{(2)}(\ell^2)-\bar{\Sigma}^{(2)}(m^2)}{\ell^2-m^2}}.
$$
Since ${\bar{\Sigma}^{(2)}}{'}(\ell^2){<}0$ [this is
distinctly seen, e.g., from Eq.~(\ref{SE})], the difference
$\bar{\Sigma}^{(2)}({\ell}^2)-\bar{\Sigma}^{(2)}({m}^2)$
is always positive, while the difference
${\ell}^2-{m}^2$ is negative. So, the quantity
$[\bar{\Sigma}^{(2)}({\ell}^2)-\bar{\Sigma}^{(2)}({m}^2)]
/({\ell}^2-{m}^2)$
is negative. Its maximal (asymptotic) value equals zero, being
achieved at $\ell^2\to -\infty$. Hence,
$$
\mbox{max}_{\ell^2}{\cal
F}(\ell^2)={-g^2{\bar{\Sigma}^{(2)}}{'}(m^2)}
$$
and
\begin{equation}
\label{c2} \alpha_c=-\left[16\pi
m^2{{\bar\Sigma}^{(2)}}{'}(m^2)\right]^{-1}.
\end{equation}
{Substituting here the explicit form~(\ref{A8}) of the derivative
of the self-energy, it is easy to see that $\alpha_c$ is identical
to} the critical coupling constant $\alpha_\textsc{l}$ associated
with the Landau pole~(\ref{alphaL2}). For $m=0.94$ and
$\mu=0.14$ we obtain $\alpha_c \simeq 2.630$, in full agreement with the value
found numerically.

Note that $\alpha_\textsc{l}$ naturally appears within the
two-body approximation, where it nevertheless does not impose any
restrictions on the calculated {renormalized} Fock components. In the three-body
case discussed here it appears again but now it substantially
affects the behavior of the Fock components. Indeed, for coupling
constants above $\alpha_{{c}}$ the equation~(\ref{G2ff}) has no physically acceptable solutions.
An attempt to calculate the vertex functions
numerically for
$\alpha>\alpha_{{c}}$ fails: the calculated results
oscillate strongly, when the number of integration nodes $n_A$
increases, without any tendency to converge.
At $\alpha=\alpha_c$ the solution of Eq.~(\ref{G2ff})
is stable.

We emphasize that the result
$\alpha_c=\alpha_\textsc{l}$ obtained above should be considered
as a feature of the Yukawa model rather than a fundamental
property of FSDR in the given approximation. The value of the
critical coupling constant depends on the particular form of the
interaction and can hardly be predicted before analyzing the
equations for the Fock components. Indeed, the full eigenvalue
spectrum of the equation~(\ref{hHmatrhomo}) is determined by the
behavior of the self-energy $\bar{\Sigma}^{(2)}$ and the kernel
$\Delta V$ as a function of their arguments. In the Yukawa model,
the integral part~(\ref{AG2}) of the operator $\hat{\sf A}$
generates the only eigenvalue $\lambda_0(\alpha)$
[in addition to the continuous spectrum governed
by the self-energy contribution~(\ref{AG1})] having no
influence on the stability of the solution. As a
result, $\alpha_c$ is fully determined by the two-body
self-energy, which just makes $\alpha_c$ identical to
$\alpha_\textsc{l}$. For another dynamical model,
different from the Yukawa model, the
situation may be different.

Going over to a finite PV particle mass does not
change the qualitative conclusions, but the numerical value of the critical coupling
constant increases.
This is not surprising due the fact that each PV
subtraction effectively reduces the interaction strength. So, the
case $\mu_1\to\infty$ analyzed above brings the tightest
limitations on admissible values of the coupling constant.

We thus establish that the solution of the renormalized
equation~(\ref{G2ff}) is nonsingular at $\alpha\leq\alpha_c$. This
statement however does not concern the initial, nonrenormalized,
equation~(\ref{G2f}), if the inhomogeneous part
$g_{03}\psi_1^{(3)}[1-g^2\bar{I}_2^{(2)}]$ is considered as a free
parameter (or an independent function of kinematical variables).
Numerical computations show that its solution becomes singular at
a lower value of the coupling constant
$\alpha=\alpha_c^{\mbox{\scriptsize nr}}$, where
$\alpha_c^{\mbox{\scriptsize nr}}<\alpha_c$. The new critical
coupling constant now essentially depends on the kernel $V$ of the
integral term in Eq.~(\ref{G2f}). If we perform the eigenvalue
analysis of Eq.~(\ref{G2f}) (more precisely, of the corresponding
homogeneous equation), we will see the following. The self-energy
contribution which is the same as in the renormalized
equation~(\ref{G2ff}) generates the continuous part of the
eigenvalue spectrum~(\ref{lambdasp}), as previously, but the
discrete eigenvalue $\lambda_0(\alpha)$ now is different from that
found for the renormalized equation. Moreover, it is positive and
always exceeds the upper boundary of the continuous spectrum
$\lambda_{\mbox{\scriptsize max}}(\alpha)$, in contrast to the
situation shown in Fig.~\ref{spectrum}.  The critical coupling
constant $\alpha_c^{\mbox{\scriptsize nr}}$ is found as a root of
the equation $\lambda_0(\alpha_c^{\mbox{\scriptsize nr}})=1$.
Numerical calculations performed for $m=0.94$, $\mu=0.14$,
and an infinite PV mass $\mu_1$ give $\alpha_c^{\mbox{\scriptsize
nr}}\simeq 2.190$. One may conclude that the renormalization
removes the singularity of the solution for the two-body vertex
function, which appears in the original nonrenormalized equation
at $\alpha=\alpha^{\mbox{\scriptsize nr}}_c$.

Considering the renormalized
equation~(\ref{G2ffoff}) for the fully off-energy-shell two-body
vertex $\Gamma_2^{(3)j}(k_{\perp},x;p^2)$, we encounter
a critical coupling constant $\alpha=\alpha_c^{\mbox{\scriptsize off}}(p^2)$,
depending on $p^2$, which makes $\Gamma_2^{(3)j}(k_{\perp},x;p^2)$ singular.
This singularity exists only if $p^2\neq m^2$.
On the mass shell, when we take $p^2=m^2$, the singularity of
$\Gamma_2^{(3)j}(k_{\perp},x;m^2)$ vs. $\alpha$ is absent.
Without discussing all technical details, we briefly explain below the origin of
$\alpha_c^{\mbox{\scriptsize off}}(p^2)$ and reveal its role in the
calculation of Fock components within the FSDR scheme.

Eq.~(\ref{G2ffoff}) can be solved in two steps.
First, we pay attention that setting $p^2=m^2$ returns us to the
renormalized equation~(\ref{G2ff}), because
$\Gamma_2^{(3)j}(k_{\perp},x;m^2)\equiv \Gamma_2^{(3)j}(k_{\perp},x)$. In
the second step, on finding the latter function, we can reduce
Eq.~(\ref{G2ffoff}) to
\begin{eqnarray}
\left[1-\frac{g^2\bar{\Sigma}_r^{(2)}(\ell_p^2)}{\ell_p^2-m^2}\right]
\Gamma_2^{(3)j}(k_{\perp},x;p^2)&=&G_0(x)
+\frac{g^2}{8\pi^2}\sum_{j'=0}^1
(-1)^{j'}\int_0^{1-x}dx'\nonumber\\
&&\times \int_0^{\infty}dk_{\perp}'\,k_{\perp}'
V^{jj'}(k_{\perp},x,k_{\perp}',x';p^2)
\Gamma_2^{(3)j'}(k_{\perp}',x';p^2), \label{G2ffoff1}
\end{eqnarray}
where the inhomogeneous part given by
\begin{equation}
\label{G2ffoffinh}
G_0(x)=g\sqrt{1-g^2\bar{I}_2^{(2)}}-\frac{g^2}{8\pi^2}\sum_{j'=0}^1
(-1)^{j'}\int_0^{1-x}dx'\int_0^{\infty}dk_{\perp}'\,k_{\perp}'
V^{0j'}(k_{\perp}^*(x),x,k_{\perp}',x';m^2)
\Gamma_2^{(3)j'}(k_{\perp}',x';m^2)
\end{equation}
is already known. As advocated above, the function
$\Gamma_2^{(3)j}(k_{\perp},x;m^2)$ is nonsingular at
$\alpha\leq \alpha_c$. Under this condition, the function $G_0(x)$
is also a finite quantity. Eq.~(\ref{G2ffoff1}) now has the same
shape as the nonrenormalized equation~(\ref{G2f}) for the
half-off-shell two-body vertex function $\Gamma_2^{(3)j}(k_{\perp},x)$,
excepting the fact that the kernel parametrically depends on
$p^2$. Applying the same eigenvalue analysis, as for
Eq.~(\ref{G2f}) above, we calculate the critical coupling
constants $\alpha^{\mbox{\scriptsize off}}_c(p^2)$. Note that
\begin{equation}
\label{anr} \alpha_c^{\mbox{\scriptsize
nr}}=\alpha^{\mbox{\scriptsize off}}_c(m^2).
\end{equation}
The right-hand side of Eq.~(\ref{anr}) should be understood as a
limit $\alpha^{\mbox{\scriptsize off}}_c(p^2\to m^2)$, because at $p^2=m^2$, as has been mentioned
above, the two-body vertex function is smooth, even at
$\alpha=\alpha_c^{\mbox{\scriptsize off}}(m^2)$. Then, since the
critical coupling constant $\alpha_c$ defined by Eq.~(\ref{c2}) always
exists for any of Eqs.~(\ref{G2f}), (\ref{G2ff}),
and~(\ref{G2ffoff}), the coupling constant $\alpha_c^{\mbox{\scriptsize off}}(p^2)$
brings new information, only if
$\alpha_c^{\mbox{\scriptsize off}}(p^2)<\alpha_c$. We emphasize that if
$\alpha\leq \alpha_c$ and $\alpha\neq \alpha_c^{\mbox{\scriptsize
off}}(p^2)$, the solutions of Eqs.~(\ref{G2f}), (\ref{G2ff}),
and~(\ref{G2ffoff}) are nonsingular.

The critical coupling constants considered above
may generate some peculiarities in $\alpha$-dependence of numerically calculated quantities,
especially when using rough computational
grids.  Indeed, while the exact result is nonsingular,
the cancellation of pole contributions may not occur in full measure,
due to approximate character of numerical calculations. For instance, sharp behavior
of the field strength renormalization factor $Z^{(3)}_\chi$ in the vicinity of
the point $\alpha=\alpha_c^{\mbox{\scriptsize nr}}=2.190$  for a relatively small number
of Gaussian integration nodes (see Fig.~\ref{I1Z3})
is a probable manifestation of this effect.

The situation with the simultaneous existence of several types of critical coupling constants
looks, at first glance, rather confusing. In order to make it more transparent, in Appendix~\ref{cc}
we discuss an explicitly solvable toy model
which mimics relevant features of the scalar Yukawa model in the three-body truncation. This
illustrates our conclusions in a very simple and clear manner.

Within the three-body truncation, all calculated observables are
expressed through the renormalized two-body Fock component found
for $p^2=m^2$, while the fully off-shell two-body Fock component
$\Gamma_2^{(3)j}(k_{\perp},x;p^2)$ with $p^2\neq m^2$ is not
needed for this purpose. Thus, it may seem that a set of critical
constants $\alpha_c^{\mbox{\scriptsize off}}(p^2)$ is a sort of
peculiarity having no relation to practical computations of
physical quantities, while all actual restrictions imposed on the
value of the coupling constant reduces to the requirement
$\alpha\leq \alpha_c$. The importance of the function
$\Gamma_2^{(3)j}(k_{\perp},x;p^2)$ becomes evident as one
goes  to the four-body truncation, where the former enters, as an
internal block, into the system of equations for the Fock
components~\cite{Li15a}. In this sense, the fully off-shell
two-body vertex function serves as a ``bridge'' between the three-
and four-body truncations. Respectively, the critical coupling
constants $\alpha_c^{\mbox{\scriptsize off}}(p^2)$ propagate, together with
$\Gamma_2^{(3)j}(k_{\perp},x;p^2)$, to the four-body problem as
well. The parameter $p^2$ in the four-body truncation varies
continuously from $-\infty$ up to $(m-\mu)^2$. Our computations
for $m=0.94$, $\mu=0.14$, and an infinite PV mass $\mu_1$
show that $\alpha_c^{\mbox{\scriptsize off}}(p^2)$ is a decreasing
function of $p^2$. It reaches its minimal value at the maximal
available $p^2$, i.e., $p^2=(m-\mu)^2$. Hence, in the four-body
truncation, one may expect the critical coupling constant to be
not greater than $\alpha_c^{\mbox{\scriptsize
off}}((m-\mu)^2)\simeq 2.382$. At the same time, one cannot
exclude the possibility of appearance of a new, purely ``four-body'',
critical coupling constant. Numerical
estimations~\cite{Li15a,Li15b} based on an iterative procedure
show that the iterations stop converging at $\alpha$ about 2.14 or
larger. Exact calculation of the critical coupling constant in the
four-body truncation however goes beyond the scope of the present
paper. The above example with the hierarchy of the critical
coupling constants is given to demonstrate that some of them
originated from a given order truncation as mathematical
peculiarities may then propagate to higher order truncations and
then introduce further physical restrictions on the parameters of the model.


\section{Calculation of the electromagnetic form factor}
\label{ff}


Form factors are fundamental for the study of hadron structures. They are defined from the
electromagnetic vertex (EMV) $G^{\rho}(p,p')$ which is expressed through the matrix element of the current
operator $\hat{J}_\mathrm{em}^\rho(x)$:
\begin{equation}
\label{EMV}
G^{\rho}(p,p')=e_0\langle p'|\hat{J}^{\rho}_\mathrm{em}(0)|p\rangle,
\end{equation}
where $p$ and $p'$ are the initial and final particle
four-momenta, $e_0$ is the  bare electromagnetic coupling
constant, and $\rho$ is an arbitrary Lorentz index. The
bra and ket vectors here are the same as the state vectors
$\phi^{\dag}(p')$ and $\phi(p)$, respectively. The elastic
electromagnetic form factor $F(Q^2)$ for a scalar particle is
defined as
\begin{equation}
\label{FEMV}
G^{\rho}(p,p')=e(p+p')^{\rho}F(Q^2),
\end{equation}
where $e$ is the physical electromagnetic coupling constant
(physical charge), $Q^2=-q^2$, $q=p'-p$ is the four-momentum
transfer.
 The necessity to distinguish the physical and bare electromagnetic coupling constants
follows from the fact that the elementary electromagnetic vertex,
generally speaking, is renormalized due to
its ``dressing'' by scalar pion lines. The standard
renormalization condition known from QED demands that the
renormalized EMV at zero momentum transfer must coincide with that
for the free particle:
\begin{equation}
\label{Re}
G^{\rho}(p,p)=2ep^{\rho}.
\end{equation}
This condition yields a relation between $e_0$ and $e$.

The structure of the EMV~(\ref{FEMV}) is a consequence of general
physical symmetries of the interaction. In approximate
nonperturbative calculations in the framework of LFD these
symmetries may be broken because of the rotational symmetry
violation. This fact may lead to appearance, in the EMV, of
nonphysical contributions explicitly depending on the light front
orientation~\cite{Karmanov92}. In the spinless case the problem is however absent for
the plus-component of the EMV, provided an additional requirement
$q^+=0$ is  imposed on the momentum transfer. After that, the form
factor can be expressed through the EMV by
\begin{equation}
\label{FEMV1}
eF(Q^2)=\frac{G^+(p,p')}{2p^+}.
\end{equation}
The renormalization condition~(\ref{Re}) is implied to refer to the plus-component of the EMV as well. With Eq.~(\ref{FEMV1}),
it can be written in a very simple form
\begin{equation}
\label{F0}
F(0)=1.
\end{equation}

\begin{figure}
\centering
\includegraphics[width=.3\columnwidth]{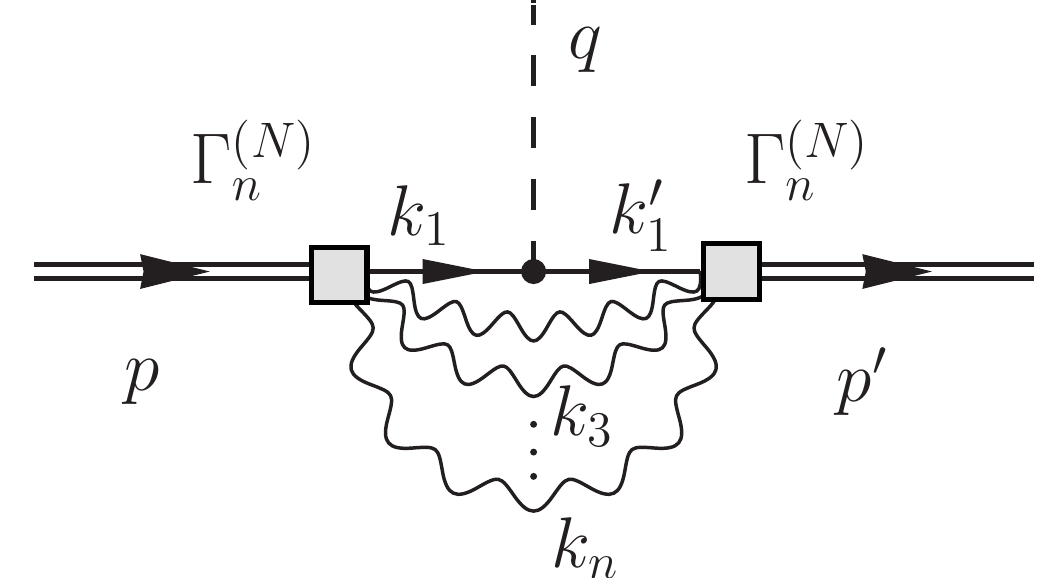}
\caption{$n$-body electromagnetic vertex in the truncation of
order $N$. The dashed line corresponds to a photon. The
nucleon-photon interaction vertex is given by
$e_{0(N-n+1)}({k}_1+{k}_1')^{\rho}$.}
\label{figEMV}
\end{figure}

With the Fock representation of the state vector in $N$-body
truncated Fock space, the total EMV is a sum of $n$-body
contributions ($n=1,\,2,\,\ldots ,\,N$) shown in
Fig.~\ref{figEMV}. According to the FSDR rules, the bare
electromagnetic coupling constant $e_0$ must be a sector-dependent
quantity
$$
e_0\to e_{0l},\quad (l=1,\,2,\ldots ,\,N),
$$
similar to the bare coupling constant $g_0$ which determines the
interaction between the constituents of the state vector [see
Eq.~(\ref{bareq})]. However, in contrast to $g_0$ treated
nonperturbatively, $e_0$ is considered as being small, so that the
EMV is calculated in the leading order in $e_0$ (at the same time,
the renormalization of $e_0$ due to its ``dressing'' by scalar pion
lines is nonperturbative!). Then, since $e_0$ has no relation to
the interactions ``inside'' the state vector, we will refer to it as
an external bare coupling constant~\cite{Karmanov08}. Now the
photon as an external particle should be excluded from the
particle counting, and the Fock sector content is fully
determined by the number of scalar pion-spectators plus one
scalar nucleon. We thus have the following rule to calculate the
index $l$ for the $n$-body Fock sector: $l=N-n_s$, where
$n_s=n-1$ is the number of pion-spectators. The lowest order
external bare coupling constant $e_{01}=e$, because the trivial
case $N=1$ describes the interaction of a photon with a point-like
scalar nucleon.

Applying the LFD graph techniques rules to the
diagram in Fig.~\ref{figEMV} and using Eq.~(\ref{FEMV}), we obtain
for the form factor within the $N$-body Fock space truncation:
\begin{equation}
\label{FFe} eF^{(N)}(Q^2)=\sum_{n=1}^Ne_{0(N-n+1)}F^{(N)}_n(Q^2),
\end{equation}
where
\begin{eqnarray}
F_n^{(N)}(Q^2)&=&\frac{2}{(2\pi)^{3(n-1)}(n-1)!}\int\prod_{i=1}^n
\frac{d^2k_{i\perp}dx_i}{2x_i}\nonumber\\
&&\times \left[\frac{\Gamma_n^{(N)}({\bf k}_{2\perp},x_2,\ldots ,
{\bf k}_{n\perp},x_n)\Gamma_n^{(N)}({\bf k}'_{2\perp},x_2,\ldots ,
{\bf k}'_{n\perp},x_n)}{(s_n-m^2)(s_n'-m^2)}\right]\,
\delta^{(2)}\left(\sum_{i=1}^n{\bf
k}_{i\perp}\right)\delta\left(\sum_{i=1}^nx_i-1\right).\nonumber\\
\label{FFn}
\end{eqnarray}
The primed transverse momenta are defined as
$$
{\bf k}'_{i\perp}
={{\bf k}_{i\perp}-x_i{\bf q}_{\perp}},
$$
$s'_n$ is given by Eq.~(\ref{snlf}), changing ${\bf k}_{i\perp}$
by ${\bf k}'_{i\perp}$. Note that, due to the condition $q^+=0$,
we have $Q^2=q_{\perp}^2$. Comparison of Eqs.~(\ref{FFn})
and~(\ref{Inexpl}) gives
\begin{equation}
\label{FI} F_n^{(N)}(0)=I_n^{(N)},
\end{equation}
i.e., the $n$-body contribution to the form factor at zero
momentum transfer coincides with the $n$-body Fock sector norm.
Setting $Q^2=0$ in Eq.~(\ref{FFe}) and making use of the
renormalization condition~(\ref{F0}) which writes simply
$F^{(N)}(0)=1$ in truncated Fock space, we get
\begin{equation}
\label{FFe0} e=\sum_{n=1}^Ne_{0(N-n+1)}I^{(N)}_n.
\end{equation}
This formula, together with the normalization
condition~(\ref{normN}) leads to the following result:
\begin{equation}
\label{e0e} e_{0l}=e
\end{equation}
for arbitrary $l$. So, the electromagnetic coupling constant in
the framework of FSDR is not renormalized at all. Eq.~(\ref{FFe})
now becomes
\begin{equation}
\label{FFef} F^{(N)}(Q^2)=\sum_{n=1}^NF^{(N)}_n(Q^2).
\end{equation}
The one-body contribution is
\begin{equation}
\label{FF1b}
F^{(N)}_1(Q^2)=\left[\psi_1^{(N)}\right]^2=I_1^{(N)}=1-\sum_{n=2}^NI_n^{(N)}.
\end{equation}
It does not depend on $Q^2$. With Eq.~(\ref{FI}), the final
expression for the form factor reads
\begin{equation}
\label{FFeff}
F^{(N)}(Q^2)=1+\sum_{n=2}^N\left[F^{(N)}_n(Q^2)-F_n^{(N)}(0)\right].
\end{equation}

Note that the general formula~(\ref{FFn}) for the
$n$-body Fock sector contribution to the form factor does not take
into account PV particles, since the corresponding integrals do
not need regularization in the scalar Yukawa model without
antiparticles. One may thus consider Eq.~(\ref{FFn}) to be related to
the limiting case $\mu_1\to\infty$.

In the two-body truncation the form factor is
\begin{equation}
\label{FF2a}
F^{(2)}(Q^2)=1+F_2^{(2)}(Q^2)-F_2^{(2)}(0).
\end{equation}
Substituting the solution~(\ref{eqn:Gamma2_N2}) into Eq.~(\ref{FFn}) for $n=N=2$,
we arrive at
\begin{equation}
\label{F22b}
F_2^{(2)}(Q^2)=\frac{g^2}{16\pi^3}\int_0^1dx\,x(1-x)\int\frac{d^2k_{\perp}}
{[k_{\perp}^2+\mu^2(1-x)+m^2x^2][({\bf k}_{\perp}-x{\bf q}_{\perp})^2+\mu^2(1-x)+m^2x^2]}.
\end{equation}
The integrals can be expressed in terms of elementary functions. It is interesting to note that
the formulas~(\ref{FF2a}) and~(\ref{F22b}) exactly reproduce the familiar perturbative result,
though our approach does not rely on perturbation theory. This rather surprising fact can be
explained by the simplicity of the two-body approximation. Already in the three-body truncation
both two- and three-body vertex functions have rather complex dependence on the physical coupling constant,
determined by Eqs.~(\ref{G2ff}) and~(\ref{eqn:SoE_N3n3}).

\begin{figure}
\centering
\subfigure[\ $\alpha = 1.0$]{
\includegraphics[width=.48\columnwidth]{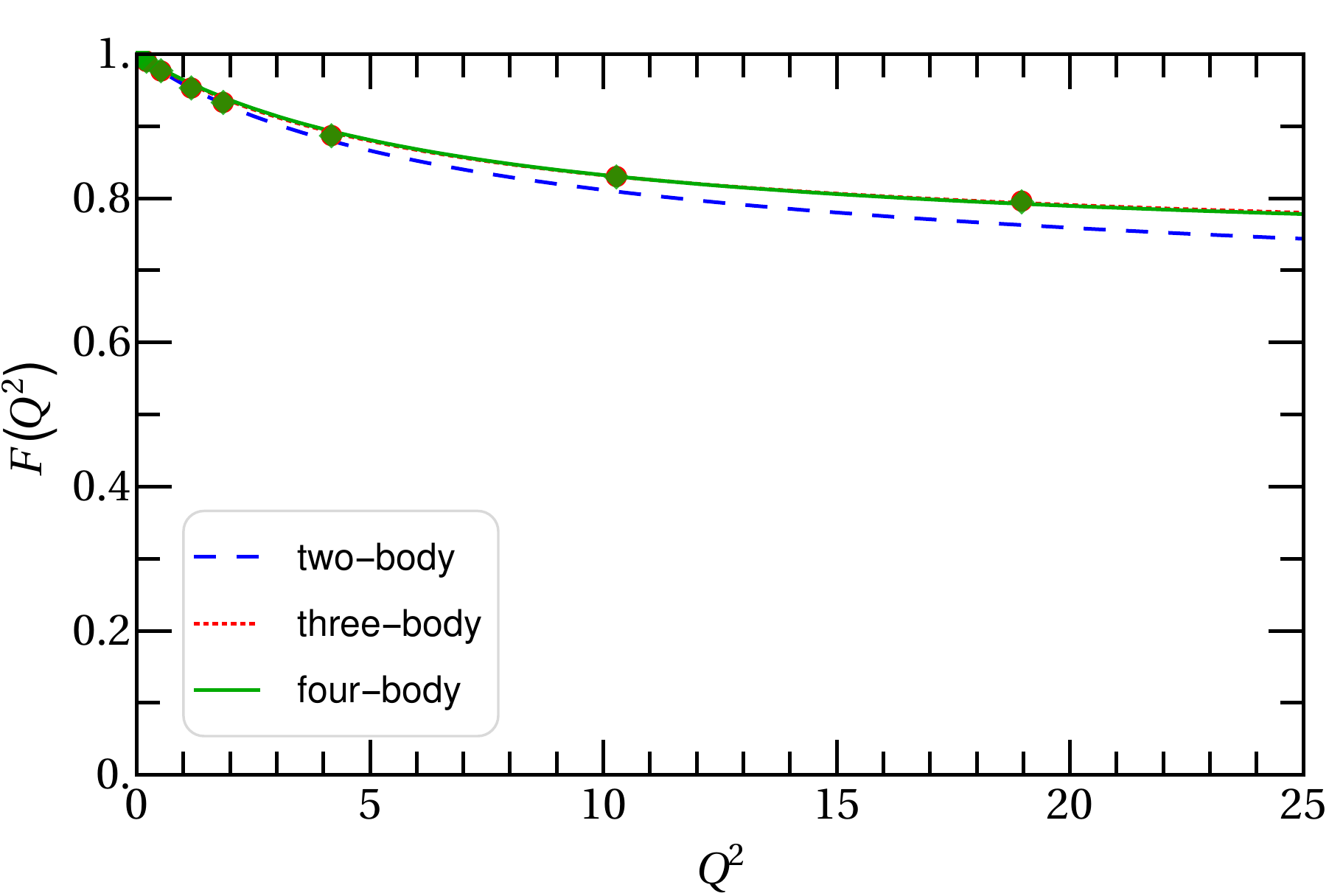}
}
\subfigure[\ $\alpha = 2.0$]{
\includegraphics[width=.48\columnwidth]{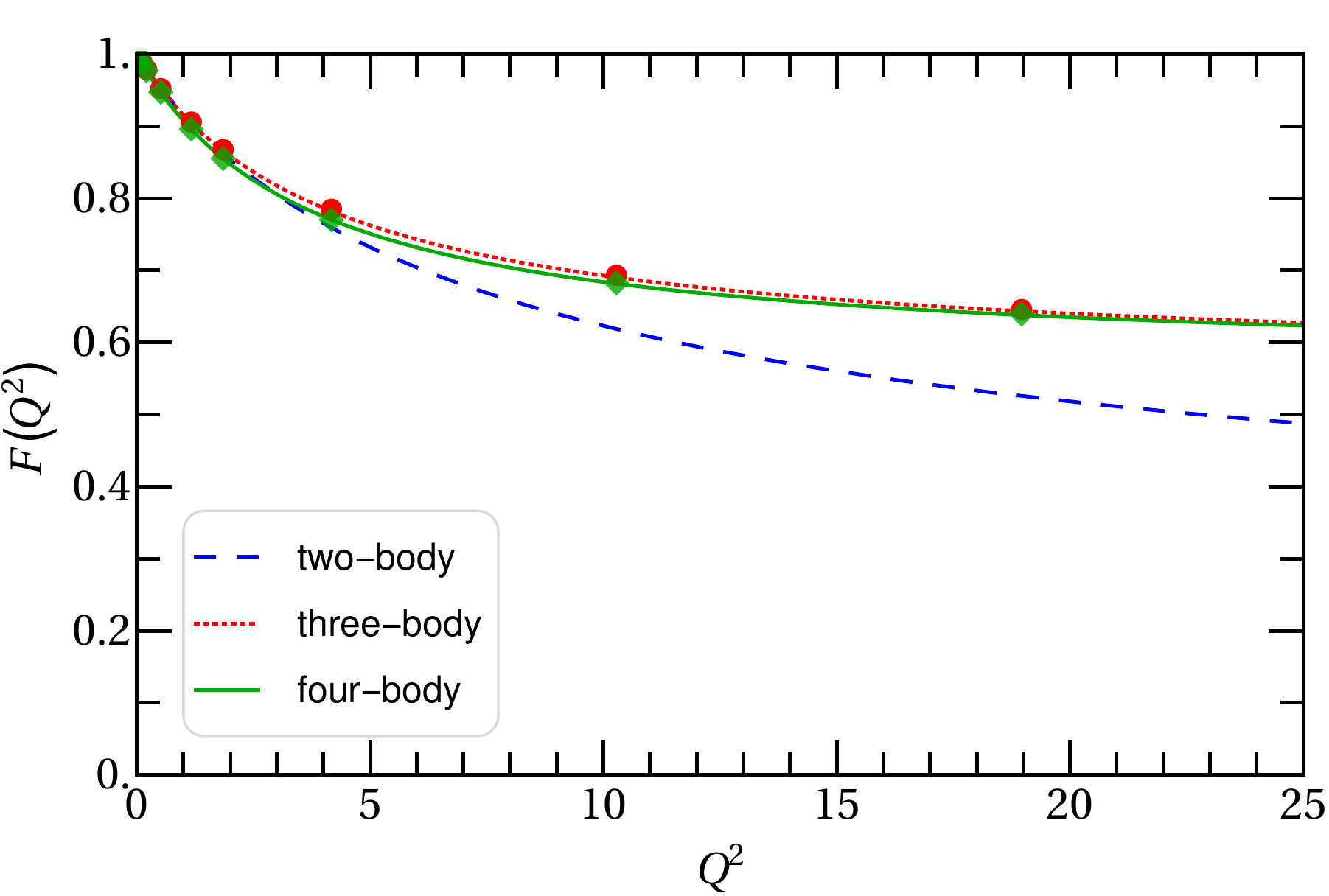}
} \caption{Electromagnetic form factor calculated in the two-,
three- and four-body truncations for $\alpha=1.0$ (left panel) and
$\alpha=2.0$ (right panel), at $m=0.94$, $\mu=0.14$, and $\mu_1=15$.
 The results for the four-body
truncation are adopted from Ref.~{\cite{Li15b}}. The form factor
in the two-body truncation admits an analytic expression. For the
three- and four-body truncations, the obtained results (symbols)
are fitted to a function (lines) $f(Q^2) = I_1 +
(1-I_1)/(1+c_1Q^2)/(1+c_2Q^2)$, where $c_1$ and $c_2$ are
some constants depending on $\alpha$ and on the order of
truncation.
} \label{fig:formfactor}
\end{figure}

The form factor $F^{(N)}(Q^2)$ calculated numerically for $N=2$,
$N=3$, and $N=4$ is shown in Fig.~\ref{fig:formfactor}. The
calculations were carried out for $m=0.94$, $\mu=0.14$, $\mu_1=15$,
and two different values of the coupling
constant: $\alpha=1.0$ and $\alpha=2.0$. Here we retain a finite PV 
mass in the two- and three-body truncations in order to compare with the 
results in the four-body truncation obtained with the same PV mass \cite{Li15b}.
This value of $\mu_1$ is large enough to make the calculated results almost 
insensitive (in the scale of the plots) to its further increase. In principle, in the
three-body truncation, one might take a bigger $\alpha$, up to
$\alpha_{{c}}{\simeq 2.630}$, inclusive. However, keeping in mind
stronger limitations on the coupling constant in the four-body
truncation (see the end of Sec.~\ref{critN3}), we took a lower
value of $\alpha$, in order to have the possibility to compare
with each other the results for the form factor, obtained in the successive
$N=2$, $N=3$, and $N=4$ truncations.

Note that the functions $F_n^{(N)}(Q^2)$,
Eq.~(\ref{FFn}), with $n\geq 2$ fall rapidly in the asymptotic
region $Q^2\gg \{m,\,\mu\}$ and tend to zero if $Q^2\to\infty$.
The limiting value of the form factor thus coincides with the
one-body Fock sector norm:
\begin{equation}
\label{FFas}
F^{(N)}(Q^2\to\infty)=I_1^{(N)},
\end{equation}
that is, generally speaking, a finite nonzero quantity.

%
%
%


\section{Discussion for the $x$-dependence of $g_{03}(x)$}
\label{g03x}


The $x$-dependent bare coupling constant $g_{03}(x)$ was
introduced in Ref.~\cite{Karmanov12}. This $x$-dependence which,
at first glance,  seems to be an oddity, is a consequence of
truncation. As it was already mentioned,  by truncating Fock space, we replace the initial
light-front Hamiltonian~(\ref{Hamiltonian}) by a
finite matrix. Finding, with this  finite matrix, the vertex function $\Gamma^{(N)}_2(k_{\perp},x)$ and solving
the renormalization condition (\ref{rencgN2}) relative to $g_{0N}$,  we find that the latter becomes  dependent on $x$:
$g_{0N}=g_{0N}(x)$ [see Eq.~(\ref{g03f}) for the $N=3$ case].

More precisely, the mechanism for the appearance of the
$x$-dependent $g_{0N}(x)$ is the following. The renormalization
condition (\ref{rencgN2}) contains the vertex function
$\Gamma^{(N)}_2(k_{\perp},x)$ calculated in $N$-body truncated
Fock space and dependent on the two kinematical variables
$k_{\perp}$ {and} $x$. The on-energy-shell condition $s_2=m^2$
does not fix both variables, but gives a relation between them. We
express from this relation the value $k_{\perp}=k^{*}_{\perp}(x)$
which is given by Eq.~(\ref{kstar}). So, the value of the vertex
function which appears on the left-hand side of the
renormalization condition (\ref{rencgN2})  is
$\Gamma^{(N)}_2(k^*_{\perp}(x),x)$, i.e., it depends on $x$ via
$k^*_{\perp}(x)$ and also via its ``own'' argument $x$.  Since the
right-hand side of Eq.~(\ref{rencgN2}) is a constant, this
condition can be satisfied identically only if the bare coupling
constant $g_{0N}$, which the two-body vertex depends on,  becomes
a function of $x$ as well.

\begin{figure}[ht!]
\centering
\subfigure[\ $\Gamma_{2,a}$\label{Unt3a}]{
  \includegraphics[width=0.25\textwidth]{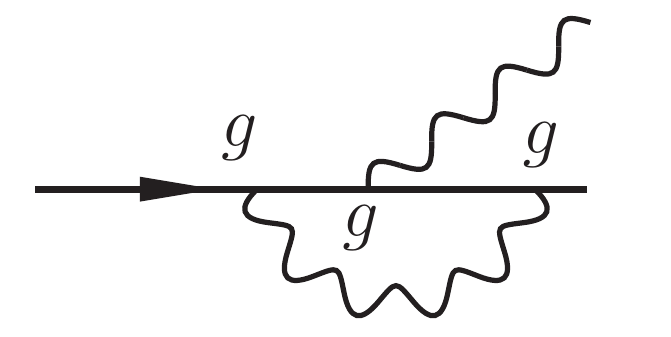}
}
\subfigure[\ $\Gamma_{2,b}$\label{Unt3b}]{
  \includegraphics[width=0.25\textwidth]{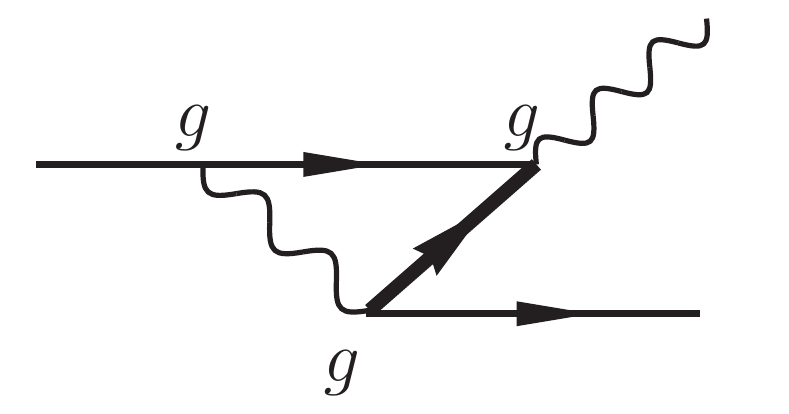}
}
\caption{Full {set of} perturbative contributions
to
the two-body vertex function at order $\mathrm O(g^3)$.
The thinner straight lines represent the
scalar nucleon. The thicker straight line represents the
antinucleon.
The wavy lines represent the
scalar pions.} \label{Unt3}
\end{figure}

As discussed in Sec.~\ref{sect3}, the $x$-dependence
of the on-energy-shell two-body vertex function must disappear, if
the latter was calculated in full (i.e., not truncated) Fock space.
This general property is based on fundamental physical symmetries.
To illustrate how the cancellation of $x$-dependence happens in
practice, within LFD, there is no need to perform nonperturbative
calculations of $\Gamma_2^{(N)}(k^*_{\perp}(x),x)$ involving
contributions from all possible Fock sectors (i.e., for
$N\to\infty$). One may use the perturbative expansion of the
two-body vertex function, which can be written as
\begin{equation}
\label{G2exp}
\Gamma_2^{(N\to\infty)}(k^*_{\perp}(x),x)=\sum_{n=1}^{\infty} g^n
\Gamma_2^{(g^n)}(k^*_{\perp}(x),x).
\end{equation}
If $\Gamma_2^{(N\to\infty)}(k^*_{\perp}(x),x)=\mbox{const}$, then
any coefficient $\Gamma_2^{(g^n)}(k^*_{\perp}(x),x)$ of the
perturbation series is also $x$-independent. We emphasize that
$\Gamma_2^{(g^n)}$ involves contributions from all possible Fock
sectors at order $g^n$ of perturbation theory.

The simplest nontrivial case is the third order of perturbation
theory. All the contributions to
$\Gamma_2^{(g^3)}(k^*_{\perp}(x),x)$ are exhausted by the two
shown in Fig.~\ref{Unt3}. The graph (a) generated by the
three-body Fock sector (one scalar nucleon plus two scalar pions)
represents a contribution incorporated in our nonperturbative
three-body calculations of $\Gamma_2^{(3)}$ in Sec.~\ref{sec61}.
The graph (b) represents a contribution from another three-body
Fock sector (one scalar nucleon plus one nucleon-antinucleon
pair), which was omitted in the truncation we used. Below we will
demonstrate that the full $\Gamma_2^{(g^3)}(k^*_{\perp}(x),x)$, determined by
the sum of two contributions (a) and (b), is indeed a constant with respect to $x$.

The amplitude of the diagram in Fig.~\ref{Unt3a} reads
%
\begin{widetext}
\begin{equation}\label{Ga}
\Gamma_{2,a}^{{(g^3)}}(
{k}_{\perp},x)=\frac{g^3}{(2\pi)^3}\int\limits_0^{1-x}\frac{dx'}{2x'(1-x')(1-x-x')}
 \int \frac{d^2k'_{\perp}}{
 ({s_2'}-m^2)({s_3}-m^2)}.
\end{equation}
\end{widetext}
%
The
amplitude of the diagram in
Fig.~\ref{Unt3b} reads
\begin{widetext}
\begin{equation}\label{Gb}
\Gamma_{2,b}^{{(g^3)}}(
k_{\perp},x)=\frac{g^3}{(2\pi)^3}\int\limits_{1-x}^1\frac{dx'}{2x'(1-x')(x+x'-1)}
\int \frac{d^2k'_{\perp}}{
({s_2'}-m^2)({\bar{s}_3}-m^2)}.
\end{equation}
\end{widetext}
The quantities $s_2'$ and $s_3$ defined by
Eqs.~(\ref{s2}), changing $k_{\perp}\to k'_{\perp}$, $x\to x'$,
and ~(\ref{s3jjp}) with $j=j'=0$, respectively, are the invariant
mass squared of each of the intermediate states of Fig.~\ref{Unt3a}: nucleon plus pion and
nucleon plus two pions. The quantity $\bar{s}_3$ is the invariant
mass squared of the three-body state of Fig.~\ref{Unt3b}, i.e., the nucleon plus
nucleon-antinucleon pair:
\begin{equation}
\label{s3bar} \bar{s}_3=\frac{k^2_\perp+m^2}{1-x}+\frac{{
k'}^2_\perp+m^2}{1-x'}+\frac{({\bf k}_\perp+{\bf
k}'_\perp)^2+m^2}{x+x'-1}.
\end{equation}
Since each of the amplitudes~(\ref{Ga})
and~(\ref{Gb}) converge, we omit the PV
particle contributions. To calculate
the integrals,
it is convenient to use the Feynman
parametrization:
%
\[\frac{1}{ab}=\int_0^1\frac{dv}{[va+(1-v)b]^2}.\]
%
Then both integrals over $d^2k'_{\perp}$ can be calculated
analytically. The integrals over $dv$ are also calculated
analytically. We substitute $k_{\perp}= k^*_{\perp}(x)$ with the
imaginary value $k^*_{\perp}{(x)}$ from {Eq.~}(\ref{kstar}) and
calculate the residual one-dimensional integrals over $dx'$
numerically.

\begin{figure}
 \centering
\includegraphics[width=0.5\columnwidth]{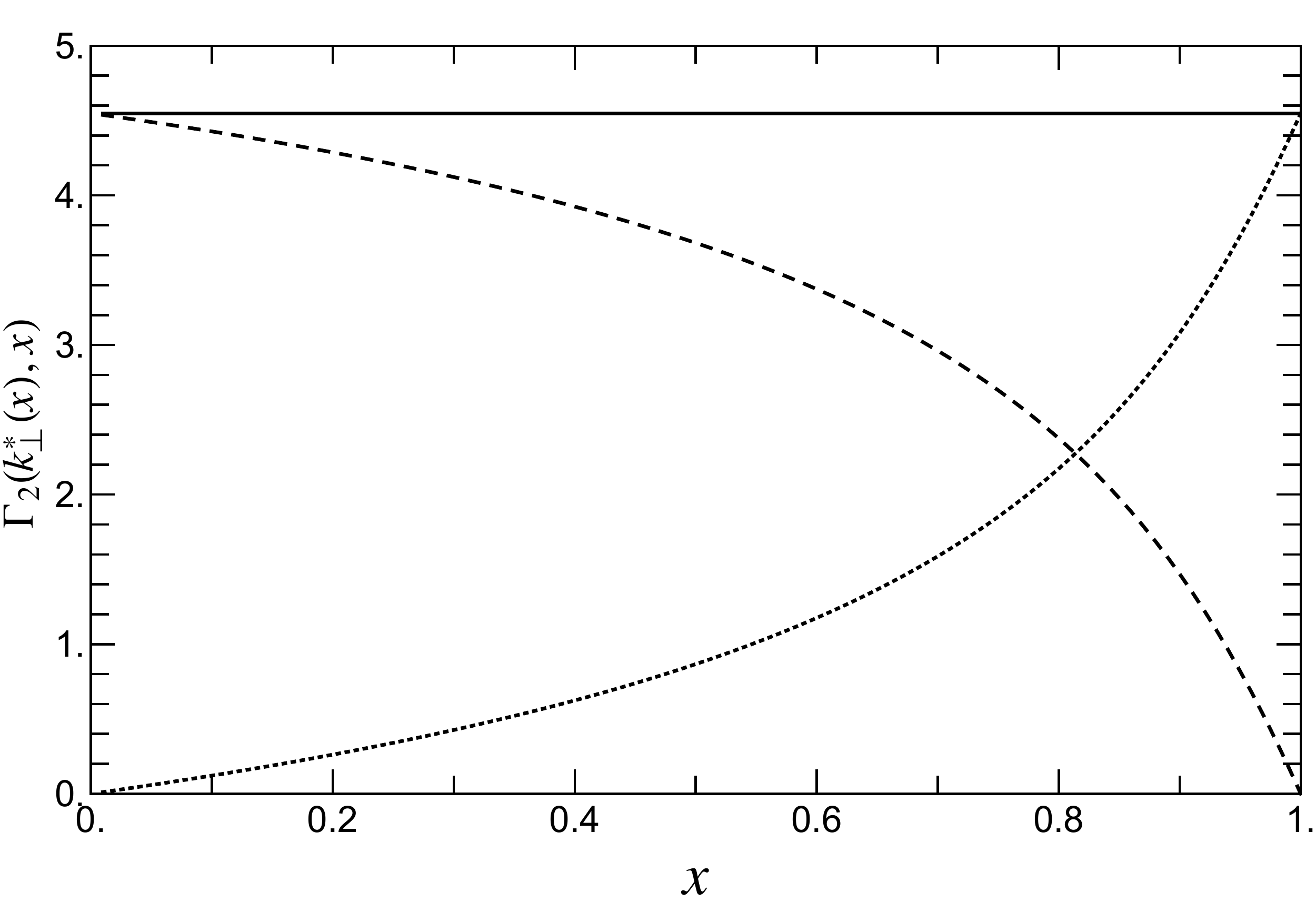}
\caption{Dependence of the on-energy-shell two-body vertex
function in the $g^3$-order of perturbation theory on the
kinematical variable $x$. The dashed curve is
$\Gamma_{2,a}^{{(g^3)}}(k^*_{\perp}(x),x)$,
Fig.~\protect{\ref{Unt3a}}; the dotted curve is
$\Gamma_{2,b}^{{(g^3)}}(k^*_{\perp}(x),x)$,
Fig.~\protect{\ref{Unt3b}}; the solid curve represents their
sum which is a constant.} \label{xc}
\end{figure}

The calculated results are shown in
Fig.~\ref{xc}.
The dashed curve is
$\Gamma_{2,a}^{{(g^3)}}(k^*_{\perp}(x),x)$, the contribution
shown in Fig.~\ref{Unt3a}.
It depends on $x$. This $x$-dependence generates the
$x$-dependence of $g_{03}(x)$, Eq.~(\ref{g03f}).
The dotted curve is $\Gamma_{2,b}^{{(g^3)}}(k^*_{\perp}(x),x)$, the contribution
shown in Fig. \ref{Unt3b}. It also depends on
$x$. The solid line is the sum
$$
{\Gamma_2^{{(g^3)}}(k^*_{\perp}(x),x)}=\Gamma_{2,a}^{{(g^3)}}(k^*_{\perp}(x),x)+\Gamma_{2,b}^{{(g^3)}}(k^*_{\perp}(x),x).
$$
It does not depend on $x$. In the Yukawa model with spin, also in
the perturbative framework, the same result was found in
Ref.~\cite{Karmanov12}.

This example clearly shows that the origin of the $x$-dependence
of the bare coupling constant $g_{03}(x)$ is the Fock space
truncation. Taking into account the previously omitted
contribution with an antinucleon we restore the constant value of
$g_{03}$.

In principle, antiparticle degrees of freedom can be included into
Fock space within the nonperturbative approach based on FSDR. This was
done in Refs.~\cite{Mathiot11} (within the scalar Yukawa model)
and~\cite{Karmanov12} (within the spinor Yukawa model in the
quenched approximation, i.e., neglecting fermion-antifermion loop
contributions). The results of numerical
{nonperturbative} calculations of the on-energy-shell 
two-body vertex function $\Gamma_2^{(3)}(k^*_{\perp}(x),x)$
or the bare coupling constant $g_{03}(x)$ in the three-body
truncation with the nucleon-nucleon-antinucleon Fock sector
included show that the latter makes the $x$-dependence of the
calculated quantities much weaker, even for rather large coupling
constant values.


\section{Conclusion}
\label{concl}


With the interaction Hamiltonian {$\mathcal H_\text{int}{(x)} = -
g \, \chi^\dagger \chi \varphi$}, where {$\chi$} and {$\varphi$}
are spinless fields referred as a ``scalar nucleon'' and ``scalar
pion'', respectively, in the framework of light-front dynamics, we
found nonperturbatively the Fock components of the state vector in
truncated Fock space including one-body ($\chi$), two-body
($\chi$+$\varphi$), and three-body ($\chi$ +$2\varphi$) states
(Fock sectors). The sector dependent renormalization of the
coupling constant and the scalar nucleon mass was
used. In this transparent example, we exposed the general
principles of nonperturbative renormalization in truncated Fock
space and demonstrated, by practical application, the main steps
required to solve the problem. The procedure contains the
principal ingredients of more general applications, and,
especially, the main features of the sector dependent renormalization
-- appearance of the sector dependent renormalization parameters,
i.e., the bare coupling constants like $g_{02}$, $g_{03}$ and the
mass counterterms like $\delta m^2_2$, $\delta m_3^2$, related to
different Fock sectors, simultaneously in one system of equations
for the Fock components. Though the constant $g_{03}$ is not a
true constant -- it depends on the kinematical variable $x$, --
this and other constants do not contain any uncertainties and are
found unambiguously.

The case of the true Yukawa model (or other field theories),
incorporating spin, differs from the example considered here by
technical details only (the form of propagators, the spin
structure of the wave functions, etc.), but contains the same
steps. The case of higher order truncation  is
more complicated technically, since it requires the solution of a
more complicated system of equations, but it uses the same
solution procedure.

This work presents  the detailed theoretical framework that underlines the successful
solution of the scalar Yukawa model in four-body truncation when the ($\chi$+$3\varphi$)
Fock sector is added to the three listed above \cite{Li15a, Li15b}. Comparison of
results in the three-body truncation with those in four-body truncation \cite{Li15a, Li15b}
shows that convergence
with respect to the number of Fock sectors involved
is achieved. 

 \section*{Acknowledgements}
The authors thank P. Maris for valuable remarks and constructive criticisms.
One of the authors (V.A.K.) thanks the Nuclear
Theory Group at Iowa State University, where a part of this paper was
prepared, for kind hospitality during his visits.
This work was supported in part by the Department of Energy under
Grant Nos. DE-FG02-87ER40371 and DESC0008485 (SciDAC-3/NUCLEI).
Computational resources were provided by the National Energy
Research Supercomputer Center (NERSC), which is supported by the
Office of Science of the U.S. Department of Energy under Contract
No. DE-AC02-05CH11231.

\appendix


\section{Two-body self-energy}
\label{SE2b}


The two-body scalar nucleon self-energy is given
by Eq.~(\ref{SE}) which can be written as
\begin{equation}
\bar{\Sigma}^{(2)}(p^2)={-}
\frac{1}{16\pi^2}\sum_{j=0}^1(-1)^j\int_0^1dx\int_0^{\infty}\frac{dk_{\perp}^2}{k_{\perp}^2+\mu_j^2(1-x)+m^2x-p^2x(1-x)}.
\label{A1}
\end{equation}
Without PV regularization, the integral over $dk_{\perp}^2$
diverges logarithmically at the upper limit. It is convenient to
define the regular function
\begin{equation}
\label{A2} a(p^2,m_1,m_2)\equiv
\int_0^1dx\int_0^{\infty}dk_{\perp}^2\,\left[\frac{1}{k_{\perp}^2+m_1^2(1-x)+m_2^2x-p^2x(1-x)}
-\frac{1}{k_{\perp}^2+m^2}\right].
\end{equation}
Then
\begin{equation}
\label{A3}
\bar{\Sigma}^{(2)}(p^2)={-}\frac{1}{16\pi^2}\left[a(p^2,\mu,m)-a(p^2,\mu_1,m)\right].
\end{equation}
The integrals in Eq.~(\ref{A2})
are easily calculated. We introduce the notation
$$
D\equiv p^4-2(m_1^2+m_2^2)p^2+(m_1^2-m_2^2)^2.
$$
 Then
\begin{equation}
\label{A4}
a(p^2,m_1,m_2)=2-\log\frac{m_1m_2}{m^2}+\frac{m_1^2-m_2^2}{p^2}\log\frac{m_2}{m_1}+
\frac{\sqrt{|D|}}{p^2}\,\Phi(p^2,m_1,m_2),
\end{equation}
where
\begin{equation}
\label{A5} \Phi(p^2,m_1,m_2)= \left\{
\begin{array}{lll}
\log\left(\frac{m_1^2+m_2^2-p^2+\sqrt{D}}{2m_1m_2}\right),&
\mbox{if}& D\geq 0,\\
- \arctan\left(\frac{\sqrt{|D|}}{m_1^2+m_2^2-p^2}\right),&
\mbox{if}& D<0.
\end{array}
\right.
\end{equation}
The function $a(p^2,m_1,m_2)$ is symmetric with respect to the
permutation of $m_1$ and $m_2$. At $p^2<(m_1+m_2)^2$ it is real.

In the limit of infinite PV mass $\mu_1$ the
difference
$\bar{\Sigma}^{(2)}_c(p^2)=\bar{\Sigma}^{(2)}(p^2)-\bar{\Sigma}^{(2)}(m^2)$
tends to a finite value:
\begin{equation}
\label{A6}
\bar{\Sigma}^{(2)}_c(p^2)={-}
\frac{1}{16\pi^2}\left[a(p^2,\mu,m)-a(m^2,\mu,m)\right].
\end{equation}
At $p^2\to -\infty$ we get the following asymptotic behavior:
\begin{equation}
\label{A7} \bar{\Sigma}^{(2)}_c(p^2)\approx
{\frac{1}{16\pi^2}\log\frac{|p^2|}{m^2}}+\ldots,
\end{equation}
where the dots designate finite terms.

In contrast to the self-energy, its derivative over $p^2$  does
not need regularization, so, ${\bar{\Sigma}^{(2)}}{'}(m^2)$
is finite in the limit of infinite PV mass. Its limiting value is
calculated as
$$
{\bar{\Sigma}^{(2)}}{'}(m^2)={-}
\frac{1}{16\pi^2}\,\left. \frac{\partial a(p^2,\mu,m)}{\partial
p^2}\right|_{p^2=m^2}.
$$
The calculation of the derivative is straightforward. It yields
\begin{equation}
\label{A8} {\bar{\Sigma}^{(2)}}{'}(m^2)=
{-}
\frac{1}{16\pi^2m^2}\left[\frac{\xi(3-\xi^2)}{\sqrt{4-\xi^2}}\,\arctan\left(
\frac{\sqrt{4-\xi^2}}{\xi}\right)-1+(1-\xi^2)\log\frac{1}{\xi}\right],
\end{equation}
where $\xi=\mu/m$. Eq.~(\ref{A8}) is valid for $\mu<2m$.


\section{Critical coupling constant in explicitly solvable model}
\label{cc}


In this section we consider, as an illustration, an explicitly solvable model which reflects
all important properties of Eqs.~(\ref{G2f}), (\ref{G2ff}), and~(\ref{G2ffoff}), related to the existence
of the critical coupling constant. We will not analyze the ``Landau pole'' type critical coupling constant~(\ref{alphaL1}) caused by
the two-body self-energy contribution but, instead, focus on the critical coupling originating from the kernel of the
integral term in each of the equations discussed. According to the terminology of Sec.~\ref{critN3}, we are interested in
the critical coupling constant coming from the discrete eigenvalue $\lambda_0$ which strongly
depends on the particular form of the integration kernel.

Each of the equations is a Fredholm integral equation of
the second kind, which can be written schematically
in the form~(\ref{hHmatr}). 
The integral operator $\hat{\sf A}$ depends on the coupling constant $\alpha$. The critical coupling
constant $\alpha_c$ is a solution of the matrix equation $\det(\hat{\sf A}-{\sf I})=0$,
where ${\sf I}$ is a unity matrix. If $\alpha=\alpha_c$,
the solution $\Gamma_2$ as a function of $\alpha$ becomes singular: it has a pole $\sim 1/(\alpha-\alpha_c)$.

Let us first summarize what is already known about, concerning the
critical coupling in the equations mentioned above. The
equation~(\ref{G2f}) for the nonrenormalized
$\Gamma_2^{(3)j}(k_{\perp},x)$, where the inhomogeneous part
$g_{03}\psi_1^{(3)}[1-g^2\bar{I}_2^{(2)}]$ is treated as an
independent quantity (a constant or a function of $x$), has a
critical coupling constant. For the physical particle masses
$m=0.94$, $\mu=0.14$, and an infinite PV mass $\mu_1$ its
value is $\alpha_c^{\mbox{\scriptsize nr}}\simeq 2.190$. After the
renormalization leading to Eq.~(\ref{G2ff}), this critical
coupling disappears, i.e., the renormalized
$\Gamma_2^{(3)j}(k_{\perp},x)$  is smooth at
$\alpha=\alpha_c^{\mbox{\scriptsize nr}}$. We see that the role
of renormalization in deleting infinities is two-fold: the
renormalization not only deletes {field theoretical} divergences
(e.g., the logarithmic ultraviolet divergence in the two-body
self-energy $\Sigma^{(2)}$),
but also removes the pole singularity of the Fock components at
$\alpha=\alpha_c^{\mbox{\scriptsize nr}}$. In the
generalized equation~(\ref{G2ffoff}) for the fully off-shell
two-body vertex function $\Gamma_2^{(3)j}(k_{\perp},x;p^2)$ the
critical coupling constant arises again, now as a function of
$p^2$. We will show that all these properties can be easily
traced and explained by using a toy model which admits analytic
solution.

We start with the equation
 \begin{equation}\label{eq1b}
F(k,x)=g_0(x)\psi_1+\int K(k,x,k',x')F(k',x')dk'dx'
\end{equation}
with the separable kernel
\begin{equation}\label{eq2b}
K(k,x,k',x')=\alpha h(k,x)h(k',x').
\end{equation}
Here $F$ is an unknown function to be found, $g_0$ and $h$ are smooth bounded functions
of their arguments, and
$\psi_1$ is a constant. We do not specify the limits of integration, assuming that all integrals
hereafter are convergent (e.g., due to proper regularization). Eq.~(\ref{eq1b}) is an analog of the equation~(\ref{G2f})
for the nonrenormalized two-body vertex function. Its solution is easily
found and has the form
\begin{equation}\label{eq5c}
F(k,x)=g_0(x)\psi_1+ \frac{\alpha
h(k,x)}{1-\frac{\alpha}{\alpha_c}}\int h(k',x')g_0(x')\psi_1dk'dx',
\end{equation}
where
\begin{equation}\label{alphac}
\frac{1}{\alpha_c}=\int h^2(k',x')dk'dx'.
\end{equation}
It is seen that $F(k,x)$ is singular at
$\alpha=\alpha_c$. The latter quantity is a full analog of the
critical coupling constant $\alpha_c^{\mbox{\scriptsize nr}}$ (see
Sec.~\ref{critN3}).

Now we apply a ``renormalization'' procedure to the function $F(k,x)$.
In the Yukawa model we
imposed the renormalization condition on the function $\Gamma_2^{(N)}(
k_{\perp},x)$ at $s_{{2}}=m^2$
which corresponds to an $x$-dependent
point $k_{\perp}=k^*_{\perp}(x)$ given by Eq.~(\ref{kstar}).  We
will keep this analogy and impose the renormalization condition on
$F(k,x)$  in some point $k=k^*(x)$:
\begin{equation}\label{eq6b}
F(k^*(x),x)=g
\end{equation}
and demand its fulfillment for all values of $x$.
Now the function $g_0(x)$ can not be considered as being fixed
{\em a priori} and should be found along with the renormalized
$F(k,x)$. Substituting  $F(k,x)$ from Eq.~(\ref{eq5c})
into Eq.~(\ref{eq6b}),  we get
%
\begin{equation}\label{eq7b}
g_0(x)\psi_1+ \frac{\alpha
h(k^*(x),x)}{1-\frac{\alpha}{\alpha_c}}\int
h(k',x')g_0(x')\psi_1dk'dx'=g.
\end{equation}
%
This equation is easily solved relative to $g_0(x)$.
After that, we find the relation between
the ``bare'' ($x$-dependent) coupling constant $g_0(x)$ and the ``physical'' {one},
$g$:
%
\begin{equation}\label{g0r}
g_0(x)\psi_1= g
\left(\frac{1-\frac{\alpha}{\alpha_c}+\alpha\int
\bigl[h(k^*(x'),x')-h(k^*(x),x)\bigr]
h(k',x')dk'dx'}{1-\frac{\alpha}{\alpha_c}+\alpha \int
h(k^*(x'),x') h(k',x') dk'dx' }\right).
\end{equation}
%
Substituting this expression for $g_0(x)\psi_1$ into
Eq.~(\ref{eq5c}), we obtain
the renormalized solution
%
 \begin{equation}\label{eq10b}
F(k,x)=g+\frac{g\alpha [h(k,x)-h(k^*(x),x)]\int
h(k',x')dk'dx'}{1-\frac{\alpha}{\alpha_c}+\alpha  \int
h(k^*(x'),x') h(k',x') dk'dx'}.
\end{equation}
%
It satisfies the equation
\begin{equation}
\label{eq10bb}
F(k,x)=g+\int \left[K(k,x,k',x')-K(k^*(x),x,k',x')\right]F(k',x')dk'dx'
\end{equation}
analogous to the equation~(\ref{G2ff}) for the renormalized
two-body vertex function. Note that after the renormalization
$F(k,x)$ has no any singularity at $\alpha=\alpha_c$.

The mechanism for the removal of the singularity at
$\alpha=\alpha_c$ is nontrivial, since it is not reduced to
the cancellation of factors like $(\alpha-\alpha_c)$. Such a
cancellation would take place if
$g_0(x)$ did not depend on $x$. To explain this point,
consider, for a moment, another form of the
renormalization condition, as compared to Eq.~(\ref{eq6b}).
Namely, we impose it not for all values of $x$ simultaneously, but
for a particular value $x=x^*$ only:
$F(k^*(x^*),x^*)=g$. Now, expressing
the product $g_0\psi_1$ via $\alpha$, we obtain
that the former is a true constant proportional to
$(\alpha-\alpha_c)$.
It cancels the $ 1/(\alpha-\alpha_c)$ singularity
of the nonrenormalized solution. At $\alpha=\alpha_c$
the inhomogeneous term vanishes and the renormalized $F(k,x)$ becomes
a solution of the homogeneous equation, which
exists just at this critical value of $\alpha$.
After the $x$-dependence of $g_0(x)$ is taken into account (as it should be),
the function $g_0(x)$ does not turn into zero at
$\alpha=\alpha_c$, in contrast to the case $g_0(x)=const$.
The cancellation of the singularity
$1/(\alpha-\alpha_c)$ in Eq.~(\ref{eq5c}) takes place only after
the calculation of the double integral. Due to this fact
the renormalized solution
becomes nonsingular and smooth at $\alpha=\alpha_c$.

It is also instructive to find the corresponding ``off-shell''
solution satisfying the equation {[cf. with
Eq.~(\ref{eq1b})]}:
 \begin{equation}\label{eq4da}
F(k,x;p)=g_0(x)\psi_1(p)+\int K(k,x,k',x';p)F(k',x';p)dk'dx',
\end{equation}
where
\begin{equation}\label{eq2ba}
K(k,x,k',x';p)=\alpha h(k,x;p)h(k',x';p).
\end{equation}
The ``off-shell'' continuation is given by the
additional dependence
of all parts of the equation on the
parameter $p$. The ``on-shell'' equation,
completely equivalent to Eq.~(\ref{eq1b}), is obtained at $p=m$. The
renormalization condition is still imposed on the mass shell $p=m$:
$$
F(k^*(x),x;m)=g,
$$
whereas Eq.~(\ref{eq4da}) determines the solution $F(k,x;p)$
for arbitrary $p$.
After the renormalization, Eq.~(\ref{eq4da}) becomes an analog
of Eq.~(\ref{G2ffoff}).
Its solution can be found in a similar fashion to the on-shell
solution~(\ref{eq10b}) and
has the form
\begin{widetext}
\begin{eqnarray}
F(k,x;p)&=&g\frac{\psi_1(p)}{\psi_1(m)}\left\{1-\alpha
\frac{
J_1(m)h(k^*(x),x;m)}{1-\frac{\alpha}{\alpha_c(m)}+\alpha
J_2(m)}
+\alpha
\left[\frac{1-\frac{\alpha}{\alpha_c(m)}}{1-\frac{\alpha}{\alpha_c(p)}}
\right] \frac{
J_1(p)h(k,x;m)}{1-\frac{\alpha}{\alpha_c(m)}+\alpha
J_2(m)}\right.
\nonumber\\
&+&\left.\alpha
\left[\frac{
J_1(p)
J_2(m)-
J_1(m)
J_2(p)}{1-\frac{\alpha}{\alpha_c(p)}}\right]
\frac{h(k,x;p)}{1-\frac{\alpha}{\alpha_c(m)}+\alpha
J_2(m)}\right\},
\label{eq1ca}
\end{eqnarray}
\end{widetext}
where
\begin{eqnarray*}
J_1(p)&=&\int h(k',x';p)dk'dx',
\\
J_2(p)&=&\int h(k',x';p) h(k^*(x'),x';m)dk'dx',
\end{eqnarray*}
and
$$
\frac{1}{\alpha_c(p)}=
\int h^2(k',x';p)dk'dx'.
$$
A peculiarity of
the solution~(\ref{eq1ca})
consists in the fact that, in contrast to the
on-shell renormalized solution~(\ref{eq10b}), it
is singular at $\alpha=\alpha_c(p)$, in spite of renormalization.
A similar peculiarity happens with the fully off-shell two-body
vertex function $\Gamma_2^{(3)j}(k_{\perp},x;p^2)$ which is
singular at $\alpha=\alpha_c^{\mbox{\scriptsize off}}(p^2)$.

On the mass shell $p=m$, we find that
\begin{eqnarray*}
&&\left[\frac{1-\frac{\alpha}{\alpha_c(m)}}{1-\frac{\alpha}{\alpha_c(p)}}
\right]_{p=m}\Rightarrow 1,
\\
&&\left[\frac{
{J}_1(p)
{J}_2(m)-
{J}_1(m)
{J}_2(p)}{1-\frac{\alpha}{\alpha_c(p)}}\right]_{p=m}\Rightarrow
0.
\end{eqnarray*}
Then the solution~(\ref{eq1ca}) coincides with the on-shell
solution~(\ref{eq10b}) and it is nonsingular at
$\alpha=\alpha_c(m)$.

{It is interesting to trace how the singularity at
$\alpha=\alpha_c(p)$ disappears, when $p\to m$. Extracting the
pole term $\sim [\alpha-\alpha_c(p)]^{-1}$ from Eq.~(\ref{eq1ca})
in this limit, we get, up to terms of order $(p-m)$, inclusive:
\begin{equation}
\label{eqlim} F(k,x;p\to
m)=c\cdot\frac{h(k,x;m)(p-m)}{\alpha-\alpha_c(p)} +\ldots,
\end{equation}
where
$$
c=\frac{g\left\{\alpha_c(m)[J_1(m)J_2'(m)-J_1'(m)J_2(m)]+\alpha_c'(m)J_1(m)\right\}}{J_2(m)},
$$
the primes denote the corresponding derivatives and the dots
designate all nonpole contributions. At $\alpha=\alpha_c(p)$ the
solution~(\ref{eqlim}) vs. $\alpha$ is singular for arbitrary $p\neq m$, though
the residue at the pole reduces when $p$ approaches $m$. The
singularity disappears only if $p$ exactly equals $m$. So, the
cancellation of the singularity in the on-shell solution
$F(k,x;m)$ happens due to subtle balance between different terms
in the equation, caused by the renormalization.}

{The above analysis distinctly shows that there
exists one-to-one correspondence between the properties of this
toy model and the scalar Yukawa model, concerning the behavior of
their solutions as a function of the coupling constant. Thus, in
each of the two models}
\newline
(i) the nonrenormalized solution has a pole at a certain
(critical) value of the coupling constant $\alpha=\alpha_c$;
\newline
(ii) the singularity $1/(\alpha-\alpha_c)$ disappears after the
renormalization and the renormalized solution is smooth at
$\alpha=\alpha_c$;
\newline
(iii) both the ``off-shell'' solution like the two-body vertex
$\Gamma_2^{(3)j}(k_{\perp},x;p^2)$ introduced in Sec.~\ref{RP3} and
the function $F(k,x;p)$ satisfying Eq.~(\ref{eq4da}) are singular at
some (critical) value of the coupling constant
$\alpha=\alpha_c(p)$ depending on the parameter $p$, even after
renormalization;
\newline
(iv) the pole $1/[\alpha-\alpha_c(p)]$ of the ``off-shell'' solution
exists for an arbitrary value of $p$, not equal to its on-mass-shell
value $m$, and disappears identically for $p=m$ only.


\end{document}